\documentclass[review]{elsarticle}
\pdfoutput=1
\usepackage{lineno,hyperref}
\usepackage{textcomp,amsmath}
\usepackage{pdflscape}
\usepackage{graphicx}
\usepackage{epsfig}

\begin{document}

\begin{frontmatter}

\title{Transmission stability and Raman-induced amplitude dynamics in 
multichannel soliton-based optical waveguide systems}

\author[AP_address]{Avner Peleg\corref{correspondingauthor}}
\cortext[correspondingauthor]{Corresponding author}
\ead{avpeleg@gmail.com}

\author[QMN_address]{Quan M. Nguyen}
\author[TPT_address]{Thinh P. Tran}

\address[AP_address]{Department of Exact Sciences, Afeka College of Engineering, 
Tel Aviv 69988, Israel}

\address[QMN_address]{Department of Mathematics, International University, 
Vietnam National University-HCMC, Ho Chi Minh City, Vietnam}

\address[TPT_address]{Department of Theoretical Physics, University of Science, 
Vietnam National University-HCMC, Ho Chi Minh City, Vietnam}

\date{\today}

\begin{abstract}
We study transmission stability and dynamics of pulse amplitudes in 
$N$-channel soliton-based optical waveguide systems, taking into account 
second-order dispersion, Kerr nonlinearity, delayed Raman response, 
and frequency dependent linear gain-loss.   
We carry out numerical simulations with systems of $N$ coupled 
nonlinear Schr\"odinger (NLS) equations and compare the results with 
the predictions of a simplified predator-prey model 
for Raman-induced amplitude dynamics.  
Coupled-NLS simulations for single-fiber transmission with $2 \le N \le 4$ frequency channels show  
stable oscillatory dynamics of soliton amplitudes at short-to-intermediate distances, 
in excellent agreement with the predator-prey model's predictions. 
However, at larger distances, we observe transmission destabilization 
due to resonant formation of radiative sidebands, which is caused by Kerr nonlinearity. 
The presence of linear gain-loss in a single fiber leads 
to a limited increase in transmission stability.    
Significantly stronger enhancement of transmission stability 
is achieved in a nonlinear $N$-waveguide coupler due to efficient 
suppression of radiative sideband generation by the linear gain-loss.  
As a result, the distances along which stable Raman-induced dynamics 
of soliton amplitudes is observed are significantly larger 
in the waveguide coupler system compared with the single-fiber system.   
\end{abstract}

\begin{keyword}
\texttt{Optical solitons, multichannel optical waveguide transmission, Raman crosstalk, 
transmission stability}
\PACS{42.81.Dp, 42.65.Dr, 42.65.Tg, 42.81.Qb}
\end{keyword}


\end{frontmatter}

\section{Introduction}
\label{Introduction}
Transmission of information in broadband optical waveguide links 
can be significantly enhanced by launching many pulse sequences 
through the same waveguide \cite{Agrawal2001,Tkach97,Mollenauer2006,Gnauck2008,Essiambre2010}. 
Each pulse sequence propagating through the waveguide 
is characterized by the central frequency of its pulses, and is therefore called a frequency channel. 
Applications of these multichannel systems, which are also known 
as wavelength-division-multiplexed (WDM) systems, include fiber optics transmission 
lines \cite{Tkach97,Mollenauer2006,Gnauck2008,Essiambre2010}, 
data transfer between computer processors through silicon waveguides \cite{Agrawal2007a,Dekker2007,Gaeta2008}, 
and multiwavelength lasers \cite{Chow96,Shi97,Zhang2009,Liu2013}. Since pulses from different frequency channels 
propagate with different group velocities, interchannel pulse collisions are very 
frequent, and can therefore lead to error generation and cause severe transmission 
degradation \cite{Agrawal2001,Tkach97,Mollenauer2006,Gnauck2008,Essiambre2010,Iannone98,MM98}.

In the current paper, we study pulse propagation in broadband multichannel 
optical fiber systems with $N$ frequency channels, 
considering optical solitons as an example for the pulses. 
The two main processes affecting interchannel soliton collisions in these systems  
are due to the fiber's instantaneous nonlinear response (Kerr nonlinearity) 
and delayed Raman response. The only effects of Kerr nonlinearity on a single interchannel collision between 
two isolated solitons in a long optical fiber are a phase shift and a position shift, 
which scale as $1/\Delta\beta$ and $1/\Delta\beta^{2}$, respectively, 
where $\Delta\beta$ is the difference between the frequencies of the colliding solitons \cite{Zakharov72,MM98,CP2005}. 
Thus, in this long fiber setup, the amplitude, frequency, 
and shape of the solitons do not change due to the collision.   
However, the situation changes, once the finite length of the 
fiber and the finite separation between the solitons are taken into account \cite{CPN2016}. 
In this case, the collision leads to emission of 
small amplitude waves (continuous radiation) 
with peak power that is also inversely proportional to $\Delta\beta$. 
The emission of continuous radiation in many collisions in an $N$-channel transmission 
system can eventually lead to pulse-shape distortion and as a result, 
to transmission destabilization \cite{CPN2016}.     
The main effect of delayed Raman response on single-soliton propagation 
in an optical fiber is an $O(\epsilon_{R})$ frequency downshift, 
where $\epsilon_{R}$ is the Raman coefficient \cite{Mitschke86,Gordon86,Kodama87}. 
This Raman-induced self frequency shift is a result of 
energy transfer from high frequency components of the pulse to 
lower frequency components. The main effect of delayed Raman 
response on an interchannel two-soliton collision is an $O(\epsilon_{R})$ 
amplitude shift, which is called Raman-induced crosstalk
\cite{Chraplyvy84,Tkach95,Chi89,Malomed91,Kumar98,
Kaup99,P2004,CP2005,NP2010b}. It is a result of energy transfer
from the high frequency pulse to the low frequency one.  
The amplitude shift is accompanied by 
an $O(\epsilon_{R}/\Delta\beta)$ collision-induced frequency downshift 
(Raman cross frequency shift) and by emission of continuous radiation 
\cite{Chi89,Agrawal96,Kumar98,Kaup99,P2004,CP2005,NP2010b}. 
Note that the Raman-induced amplitude shift in a single collision 
is independent of the magnitude of the frequency difference between the colliding solitons. 
Consequently, the cumulative amplitude shift experienced 
by a given pulse in an $N$-channel transmission line is proportional to $N^{2}$, 
a result that is valid for linear transmission \cite{Chraplyvy84,Tkach95,Jander96,Ho2000}, 
conventional soliton transmission \cite{Chi89,Malomed91,Kumar98,P2004}, 
and dispersion-managed soliton transmission \cite{Kaup99}.  
Thus, in a 100-channel system, for example, Raman crosstalk 
effects are larger by a factor of $2.5\times 10^3$ compared with a 
two-channel system operating at the same bit rate per channel.          
For this reason, Raman-induced crosstalk is considered to be one of the most 
important processes affecting the dynamics of optical pulse amplitudes 
in broadband fiber optics transmission lines 
\cite{Agrawal2001,Chraplyvy84,Tkach95,Tkach97,Ho2000,Yamamoto2003,
P2007,CP2008,PC2012B}.

The first studies of Raman crosstalk in multichannel fiber optics transmission 
focused on the dependence of the energy shifts on the total number of channels 
\cite{Chraplyvy84}, as well as on the impact of energy depletion 
and group velocity dispersion on amplitude dynamics \cite{Jander96,Cotter84}. 
Later studies turned their attention to the interplay between 
bit-pattern randomness and Raman crosstalk in on-off-keyed (OOK) transmission, 
and showed that this interplay leads to lognormal statistics of pulse amplitudes 
\cite{Tkach95,Ho2000,P2004,CP2005,Yamamoto2003,Kumar2003}. 
This finding means that the $n$th normalized moments of the probability 
density function (PDF) of pulse amplitudes grow exponentially with both propagation 
distance and $n^{2}$. Furthermore, in studies of soliton-based multichannel transmission,  
it was found that the $n$th normalized moments of the PDFs of the Raman self and cross frequency shifts 
also grow exponentially with propagation distance and $n^{2}$ \cite{P2007,CP2008}.     
The exponential growth of the normalized moments of pulse parameter PDFs
can be interpreted as intermittent dynamics, in the sense that the statistics of 
the amplitude and frequency is very sensitive to bit-pattern randomness \cite{P2007,CP2008,P2009}.  
Moreover, it was shown in Refs. \cite{P2007,CP2008,PC2012B} that this intermittent 
dynamics has important practical consequences in massive multichannel transmission, 
by leading to relatively high bit-error-rate values at intermediate and large propagation distances.  
Additionally, the different scalings and statistics of Raman-induced and 
Kerr-induced effects lead to loss of scalability in these systems \cite{PC2012B}.

One of the ways to overcome the detrimental effects of Raman crosstalk 
on massive OOK multichannel transmission is by employing encoding schemes, 
which are less susceptible to these effects. 
The phase shift keying (PSK) scheme, in which 
the information is encoded in the phase difference 
between adjacent pulses, is among the most promising encoding methods, 
and has thus become the focus of intensive research \cite{Xu2004,Gnauck2005}.  
Since in PSK transmission the information is encoded in the phase, 
the amplitude patterns are deterministic, and as a result, the Raman-induced 
amplitude dynamics is also approximately deterministic. A key question about 
this deterministic dynamics concerns the possibility to achieve stable steady-state 
transmission with nonzero predetermined amplitude values in all channels. 
In Ref. \cite{Jander96}, it was demonstrated that this is not possible in 
unamplified optical fiber lines. 
However, the experiments in Refs. \cite{Golovchenko2000,Kim2008}  
showed that the situation is very different in amplified multichannel transmission. 
More specifically, it was found that the introduction of amplification 
enables transmission stabilization and significant reduction of the cumulative Raman crosstalk effects.
In Ref. \cite{NP2010}, we provided a dynamical explanation for the stabilization 
of PSK soliton-based multichannel transmission, by demonstrating that the       
Raman-induced amplitude shifts can be balanced by an appropriate choice 
of amplifier gain in different channels. 
Our approach was based on showing that the collision-induced dynamics 
of soliton amplitudes in an $N$-channel system can be described by 
a relatively simple $N$-dimensional predator-prey model. 
Furthermore, we obtained the Lyapunov function for the predator-prey model 
and used it to show that stable transmission with nonzero amplitudes in all channels 
can be realized by overamplification of high frequency channels and underamplification 
of low frequency channels.

All the results in Ref. \cite{NP2010} were obtained with the $N$-dimensional 
predator-prey model, which is based on several simplifying assumptions, whose validity might 
break down with increasing number of channels or at large propagation distances. 
In particular, the predator-prey model neglects high-order effects due to  
radiation emission, intrasequence interaction, and temporal inhomogeneities. 
These effects can lead to pulse shape distortion and eventually 
to transmission destabilization (see, for example, Ref. \cite{CPN2016}). 
The distortion of the solitons shapes can also lead to the breakdown of the predator-prey 
model description at large distances. 
For example, the relation between the onset of pulse pattern distortion 
and the breakdown of the simplified model for dynamics of pulse amplitude 
was noted (but not quantified) in studies of crosstalk induced by 
nonlinear gain or loss \cite{PNC2010,PC2012,CPJ2013}.     
In contrast, the complete propagation model, which consists of a system of $N$ perturbed 
coupled nonlinear Schr\"odinger (NLS) equations, fully incorporates the effects 
of radiation emission, intrachannel interaction, and temporal inhomogeneities. 
Thus, in order to check whether stable long-distance multichannel transmission  
can indeed be realized by a proper choice of linear amplifier gain, 
it is important to carry out numerical simulations with the full coupled-NLS model.

In the current paper, we take on this important task. For this purpose, 
we employ perturbed coupled-NLS models, which take into account 
the effects of second-order dispersion, Kerr nonlinearity, delayed Raman response, 
and frequency dependent linear gain-loss. We perform numerical simulations 
with the coupled-NLS models with $2 \le N \le 4$ frequency channels
for two main transmission setups. In the first setup, 
the soliton sequences propagate through a single optical fiber, 
while in the second setup, the sequences propagate through a waveguide coupler.
We then analyze the simulations results in comparison with the predictions of the predator-prey 
model of Ref. \cite{NP2010}, looking for processes leading to transmission stabilization and destabilization. 
The coupled-NLS simulations for single-fiber transmission show 
that at short-to-intermediate distances soliton amplitudes exhibit stable oscillatory dynamics, 
in excellent agreement with the predator-prey model's predictions. 
These results mean that radiation emission and intrachannel interaction effects 
can indeed be neglected at short-to-intermediate distances. 
However, at larger distances, we observe transmission destabilization 
due to formation of radiative sidebands, which is caused by the effects of Kerr 
nonlinearity on interchannel soliton collisions.  
We also find that the radiative sidebands for the $j$th soliton sequence form near 
the frequencies $\beta_{k}(z)$ of the solitons in the neighboring frequency channels.
Additionally, we find that the presence of frequency dependent 
linear gain-loss in a single fiber leads to a moderate increase in
the distance along which stable transmission is observed. 
The limited enhancement of transmission stability in a single fiber is explained 
by noting that in this case one cannot employ strong linear loss at 
the frequencies of the propagating solitons, and therefore, 
one cannot efficiently suppress the formation of the radiative sidebands.

A stronger enhancement of transmission stability might be achieved 
in a nonlinear waveguide coupler, consisting of $N$ nearby waveguides. 
Indeed, in this case one might expect to achieve a more efficient suppression 
of radiative sideband generation by employing relatively strong linear loss 
outside of the central amplification frequency interval for each of the $N$ 
waveguides in the waveguides coupler. To test this prediction, we   
carry out numerical simulations with the coupled-NLS model for 
propagation in the waveguide coupler. 
The coupled-NLS simulations show that transmission stability 
and the validity of the predator-prey model's predictions in the waveguide coupler system    
are extended to significantly larger distances compared with the distances in the single-fiber system. 
Furthermore, the simulations for the waveguide coupler show that 
no radiative sidebands form throughout the propagation. 
Based on these observations we conclude that the enhanced transmission stability 
in the waveguide coupler is a result of the efficient suppression of radiative sideband generation 
by the frequency dependent linear gain-loss in this setup.

We consider optical solitons as an example for the pulses carrying the information 
for the following reasons. First, due to the integrability of the unperturbed NLS equation 
and the shape-preserving property of NLS solitons, derivation of the predator-prey 
model for Raman-induced amplitude dynamics is done in a rigorous manner \cite{NP2010}. 
Second, the soliton stability and shape-preserving property make soliton-based transmission 
in broadband fiber optics links advantageous compared with other transmission 
methods \cite{Agrawal2001,Mollenauer2006,Iannone98,Kodama95}.   
 Third, as mentioned above, the Raman-induced energy exchange in pulse collisions 
is similar in linear transmission, conventional soliton transmission, and dispersion-managed 
soliton transmission. Thus, even though pulse dynamics in these different transmission systems 
is different, analysis of soliton-based transmission stabilization and destabilization might give 
a rough idea about the processes leading to stabilization and destabilization 
of the optical pulse sequences in other transmission setups.

The remainder of the paper is organized as follows. 
In Section \ref{models}, we present the coupled-NLS model for 
$N$-channel transmission in a single fiber 
together with the $N$-dimensional predator-prey model for 
Raman-induced dynamics of pulse amplitudes. 
We then review the results of Ref. \cite{NP2010}   
for stability analysis of the equilibrium states of the predator-prey model.  
In Section \ref{simu}, we present the results of numerical simulations with the coupled-NLS model for 
single-fiber multichannel transmission and analyze these results in comparison with the predictions 
of the predator-prey model. In Section \ref{coupler}, we present the coupled-NLS model 
for pulse propagation in a nonlinear $N$-waveguide coupler. We then analyze the 
results of numerical simulations with this model and compare the results with the predator-prey model's 
predictions. Our conclusions are presented in Section \ref{conclusions}. 
In \ref{appendA}, we discuss the method for determining the stable propagation distance 
from the results of the numerical simulations.

\section{The propagation model for single-fiber transmission and the predator-prey model 
for amplitude dynamics}
\label{models} 
We consider propagation of pulses of light in a single-fiber $N$-channel transmission link, 
taking into account second-order dispersion, Kerr nonlinearity, delayed Raman response,   
and frequency-dependent linear loss or gain. The net linear gain-loss is the difference 
between amplifier gain and fiber loss, where we assume that the gain is provided by 
distributed Raman amplification \cite{Islam2004,Agrawal2005}. 
In addition, we assume that the frequency difference $\Delta\beta$ 
between adjacent channels is much larger than the spectral width of the pulses, 
which is the typical situation in many soliton-based WDM 
systems \cite{MM98,MMN96,Nakazawa97,Nakazawa99,Nakazawa2000}. 
Under these assumptions, the propagation is described by the following system of 
$N$ perturbed coupled-NLS equations \cite{NLS_model}: 
\begin{eqnarray} &&
i\partial_z\psi_{j}+\partial_{t}^2\psi_{j}+2|\psi_{j}|^2\psi_{j}
+4\sum_{k=1}^{N}(1-\delta_{jk})|\psi_{k}|^2\psi_{j}=
i{\cal F}^{-1}(g(\omega) \hat\psi_{j})/2
\nonumber \\&&
-\epsilon_{R}\psi_{j}\partial_{t}|\psi_{j}|^2 
-\epsilon_{R}\sum_{k=1}^{N}(1-\delta_{jk})
\left[\psi_{j}\partial_{t}|\psi_{k}|^2
+\psi_{k}\partial_{t}(\psi_{j}\psi_{k}^{\ast})\right],  
\label{raman1}
\end{eqnarray}          
where $\psi_{j}$ is proportional to the envelope of the electric field of the $j$th sequence, 
 $1 \le j \le N$, $z$ is propagation distance, and $t$ is time \cite{dimensions}. 
 In Eq. (\ref{raman1}), $\epsilon_{R}$ is the Raman coefficient,   
$g(\omega)$ is the net frequency dependent linear gain-loss function \cite{gain_loss},  
$\hat\psi$ is the Fourier transform of $\psi$ with respect to time, 
${\cal F}^{-1}$ stands for the inverse Fourier transform, and $\delta_{jk}$ is 
the Kronecker delta function. The second term on the left hand side 
of Eq. (\ref{raman1}) describes second-order dispersion effects, 
while the third and fourth terms represent intrachannel and 
interchannel interaction due to Kerr nonlinearity.  
The first term on the right hand side of Eq. (\ref{raman1}) describes  
the effects of frequency dependent linear gain or loss,  
the second corresponds to Raman-induced intrachannel interaction, 
while the third and fourth terms describe Raman-induced interchannel interaction.

\begin{figure}[ptb]
\begin{tabular}{cc}
\epsfxsize=10.0cm  \epsffile{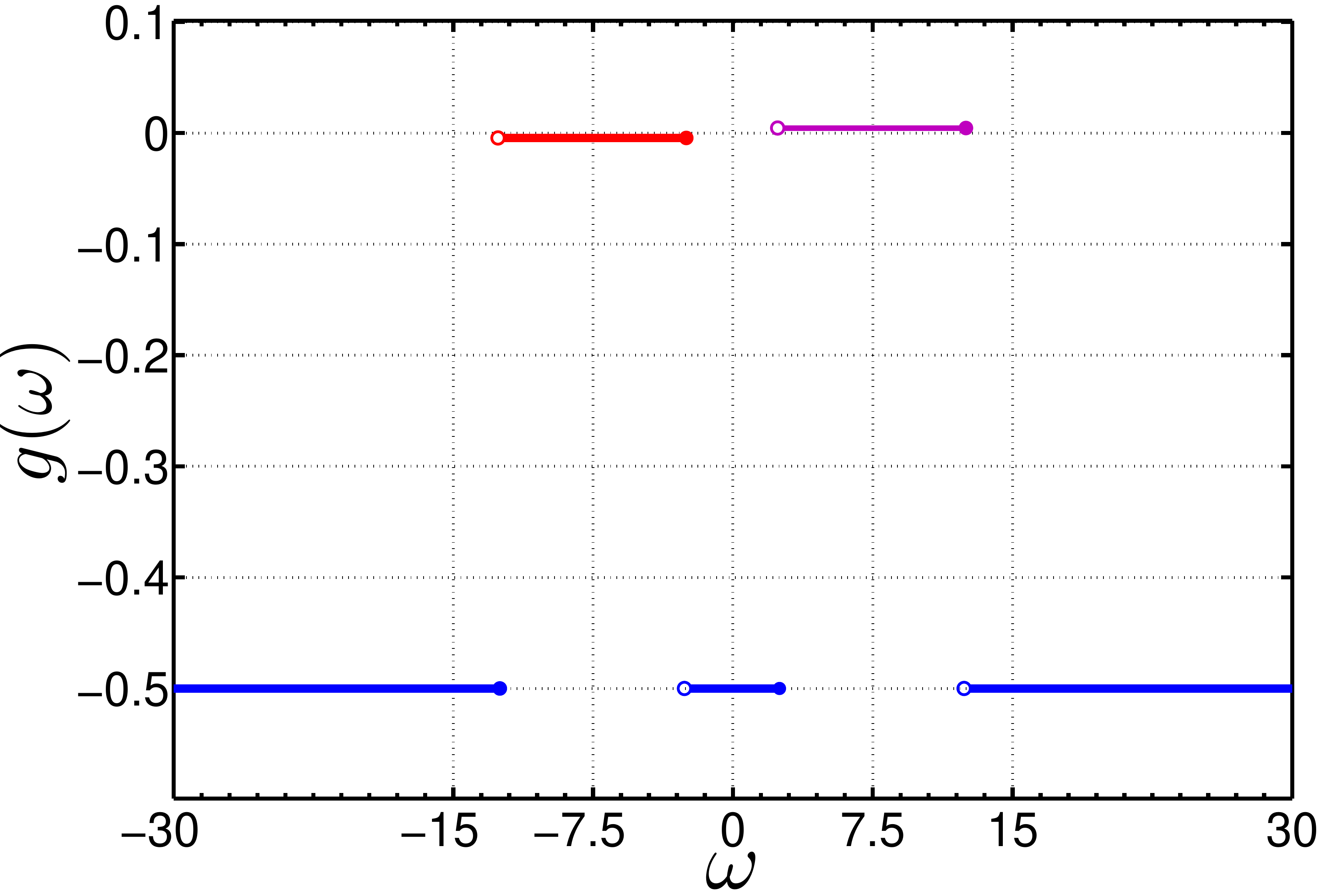} 
\end{tabular}
\caption{(Color online) An example for the frequency-dependent linear gain-loss 
function $g(\omega)$, described by Eq. (\ref{raman2}), in a two-channel system.}
 \label{fig1}
\end{figure}

The form of the net frequency dependent linear gain-loss function $g(\omega)$ is 
chosen so that Raman crosstalk and radiation emission effects are suppressed. 
More specifically, $g(\omega)$ is equal to a value $g_{j}$, required to balance Raman-induced 
amplitude shifts, inside a frequency interval of width $W$ centered about the 
initial frequency of the $j$th-channel solitons $\beta_{j}(0)$, and is equal 
to a negative value $g_{L}$ elsewhere. Thus, $g(\omega)$ is given by: 
\begin{eqnarray} &&
g(\omega) =
\left\{ \begin{array}{l l}
g_{j} &  \mbox{if} \;\;\beta_{j}(0)-W/2 < \omega \le \beta_{j}(0)+W/2 \;\; \mbox{for} \;\; 1\le j \le N,\\
g_{L} &  \mbox{elsewhere,}\\
\end{array} \right. 
\label{raman2}
\end{eqnarray}     
where $g_{L}<0$.
The width $W$ in Eq. (\ref{raman2}) satisfies  $1 < W \le \Delta\beta$, 
where $\Delta\beta=\beta_{j+1}(0)-\beta_{j}(0)$ for $1 \le j \le N-1$.  
Note that the actual values of the $g_{j}$ coefficients are determined 
by the predator-prey model for collision-induced amplitude dynamics, 
such that amplitude shifts due to Raman crosstalk are compensated for 
by the linear gain-loss. The value of $g_{L}$ is determined 
such that instability due to radiation emission is mitigated. 
In addition, the value of $W$ is determined by the following 
two factors. First, we require $W \gg 1$, such that the effects 
of the strong linear loss $g_{L}$ on the soliton patterns and on 
the collision-induced amplitude dynamics are relatively small even 
at large distances. Second, we typically require $W < \Delta\beta$, 
such that instability due to radiation emission is effectively mitigated. 
In practice, we determine the values of $g_{L}$ and $W$ by carrying out 
numerical simulations with the coupled-NLS model (\ref{raman1}), while  
looking for the set of values, which yields the longest stable propagation distance. 
Our simulations show that the optimal $g_{L}$ value is around 0.5, 
while $W$ should satisfy $W \ge 10$.    
Figure \ref{fig1} illustrates a typical linear gain-loss function $g(\omega)$ 
for a two-channel system with $g_{1}=-0.0045$, $g_{2}=0.0045$, $g_{L}=-0.5$,   
$\beta_{1}(0)=-7.5$, $\beta_{2}(0)=7.5$, and $W=10$. 
These parameter values are used in the numerical simulations, 
whose results are shown in Fig. \ref{fig3}(a).

In the current paper we study soliton-based transmission systems, 
and therefore the optical pulses in the $j$th frequency channel are 
fundamental solitons of the unperturbed NLS equation 
$i\partial_z\psi_{j}+\partial_{t}^2\psi_{j}+2|\psi_{j}|^2\psi_{j}=0$. 
The envelopes of these solitons are given by 
$\psi_{sj}(t,z)=\eta_{j}\exp(i\chi_{j})\mbox{sech}(x_{j})$,
where $x_{j}=\eta_{j}\left(t-y_{j}-2\beta_{j} z\right)$, 
$\chi_{j}=\alpha_{j}+\beta_{j}(t-y_{j})+
\left(\eta_{j}^2-\beta_{j}^{2}\right)z$, 
and the four parameters $\eta_{j}$, $\beta_{j}$, $y_{j}$, and $\alpha_{j}$ 
are related to the soliton amplitude, frequency (and group velocity), 
position, and phase, respectively.  
The assumption of a large frequency (and group velocity) difference between 
adjacent channels, means that $|\beta_{j}-\beta_{k}|\gg 1$ for 
$1 \le j \le N$, $1 \le k \le N$, and $j \ne k$. 
As a result of the large group velocity difference, the solitons undergo a large 
number of intersequence collisions. The Raman-induced crosstalk during these 
collisions can lead to significant amplitude and frequency shifts, which can in turn 
lead to severe transmission degradation.

In Ref. \cite{NP2010}, we showed that the dynamics of soliton amplitudes 
in an $N$-channel system can be approximately 
described by an $N$-dimensional predator-prey model. 
The derivation of the predator-prey model was based on the following simplifying assumptions.  
(1) The soliton sequences are deterministic in the sense 
that all time slots are occupied and each soliton is located at the 
center of a time slot of width $T$, where $T \gg 1$. 
In addition, the amplitudes are equal for all solitons 
from the same sequence, but are not necessarily equal for solitons from 
different sequences. This setup corresponds, for example, 
to return-to-zero PSK transmission.   
(2) The sequences are either (a) infinitely long, or (b) subject to 
periodic temporal boundary conditions. Setup (a) 
is an approximation for long-haul transmission systems, 
while setup (b) is an approximation for closed fiber-loop experiments.  
(3) The linear gain-loss coefficients $g_{j}$ in the frequency intervals 
$(\beta_{j}(0)-W/2 < \omega \le \beta_{j}(0)+W/2]$, 
defined in Eq. (\ref{raman2}), are determined 
by the difference between distributed amplifier gain and fiber loss. 
In particular, for some channels this difference can be slightly positive, 
resulting in small net gain, while for other channels this difference can be 
slightly negative, resulting in small net loss.  
(4) Since $T\gg 1$, the solitons in each sequence are 
temporally well-separated. As a result, intrachannel interaction is 
exponentially small and is neglected. 
(5) The Raman coefficient and the reciprocal of the frequency spacing 
satisfy $\epsilon_{R} \ll 1/\Delta\beta \ll 1$. 
Consequently, high-order effects due to radiation emission are neglected,  
in accordance with the analysis of the single-collision problem 
\cite{CP2005,Chi89,Malomed91,Kumar98,Kaup99,P2004,NP2010b}.

By assumptions (1)-(5), the propagating soliton sequences are periodic, 
and as a result, the amplitudes of all pulses in a given sequence undergo the same dynamics.
Taking into account collision-induced amplitude shifts due to delayed Raman response,  
and single-pulse amplitude changes due to linear gain-loss, we obtain the following 
equation for amplitude dynamics of $j$th-channel solitons \cite{NP2010}: 
\begin{eqnarray} &&
\frac{d \eta_{j}}{dz}=
\eta_{j}\left[g_{j}+ 
C\sum_{k=1}^{N}(k-j)f(|j-k|)\eta_{k}\right],  
\label{raman3}
\end{eqnarray}   
where $C=4\epsilon_{R}\Delta\beta/T$, and $1\le j \le N$. The coefficients $f(|j-k|)$ on the right hand side 
of Eq. (\ref{raman3}) are determined by the frequency dependence of the Raman gain. 
In particular, for the commonly used triangular approximation for the Raman gain 
curve \cite{Agrawal2001,Chraplyvy84}, in which the gain is a piecewise linear function of the frequency,  
$f(|j-k|)=1$ for $1\le j \le N$ and $1\le k \le N$ \cite{NP2010}.

In WDM systems it is often desired to achieve steady state transmission, 
in which pulse amplitudes in all channels 
are equal and constant (independent of $z$) \cite{Agrawal2001}. 
We therefore look for a steady state of the system (\ref{raman3}) 
in the form $\eta^{(eq)}_{j}=\eta>0$ for $1 \le j \le N$,  
where $\eta$ is the desired equilibrium amplitude value.  
This yields the following expression for the $g_{j}$: 
\begin{eqnarray} &&
g_{j}=-C\eta\sum_{k=1}^{N}(k-j)f(|j-k|).  
\label{raman4}
\end{eqnarray}      
Thus, in order to maintain steady-state transmission  
with equal amplitudes in all channels, high-frequency channels should be overamplified 
and low-frequency channels should be underamplified, compared with central  frequency channels. 
Substituting Eq. (\ref{raman4}) into Eq. (\ref{raman3}), 
we obtain the following model for amplitude dynamics \cite{NP2010}: 
 \begin{eqnarray} &&
\frac{d \eta_{j}}{dz}=
C\eta_ {j}\sum_{k=1}^{N}(k-j)f(|j-k|)(\eta_{k}-\eta),   
\label{raman5}
\end{eqnarray}   
which has the form of a predator-prey model for $N$ species \cite{Volterra26}.

The steady states of the predator-prey model (\ref{raman5}) with nonzero 
amplitudes in all channels are determined by solving the following 
system of linear equations:
\begin{eqnarray} &&
\sum_{k=1}^{N}(k-j)f(|j-k|)(\eta_{k}^{(eq)}-\eta)=0, 
\;\;\;\;\;\; 1 \le j \le N.  
\label{raman6}
\end{eqnarray}   
The trivial solution of Eq. (\ref{raman6}), i.e., the solution with 
$\eta^{(eq)}_{k}=\eta>0$ for $1 \le k \le N$, 
corresponds to steady state transmission with equal nonzero amplitudes. 
Note that the coefficients $(k- j)f(|j-k|)$ in Eq. (\ref{raman6}) are antisymmetric 
with respect to the interchange of $j$ and $k$. As a result, 
for WDM systems with an odd number of channels, 
Eq. (\ref{raman6}) has infinitely many nontrivial solutions, which correspond to steady 
states of the predator-prey model (\ref{raman5}) with unequal nonzero amplitudes. 
This is also true for WDM systems with an even number of channels, provided that 
the Raman gain is described by the triangular approximation \cite{NP2010}.

The stability of all the steady states with nonzero amplitudes, 
$\eta_{j}=\eta_{j}^{(eq)}>0, 1 \le j \le N$, was established in Ref. \cite{NP2010}, 
by showing that the function 
\begin{eqnarray} &&
V_{L}(\pmb{\eta})=\sum_{j=1}^{N}
\left[\eta_{j}-\eta_{j}^{(eq)}
+\eta_{j}^{(eq)}\ln\left(\frac{\eta_{j}^{(eq)}}{\eta_{j}}\right)\right],
\label{raman7}
\end{eqnarray}      
where $\pmb{\eta}=(\eta_{1},\dots,\eta_{j}, \dots, \eta_{N})$, 
is a Lyapunov function for the predator-prey model (\ref{raman5}).  
This stability was found to be independent of the $f(|j-k|)$ values, i.e., of 
the specific details of the approximation to the Raman gain curve.  
Furthermore, since $dV_{L}/dz=0$ along trajectories of (\ref{raman5}), 
rather than $dV_{L}/dz<0$, typical dynamics of the amplitudes $\eta_{j}(z)$ 
for input amplitudes that are off the steady state value is oscillatory \cite{NP2010}.     
This behavior also means that the steady states with nonzero amplitudes in 
all channels are nonlinear centers of Eq. (\ref{raman5}) \cite{Smale74}.

\section{Numerical simulations for single-fiber transmission}
\label{simu}
The predator-prey model, described in section \ref{models},   
is based on several simplifying assumptions, whose validity might 
break down with increasing number of channels or at large propagation distances. 
In particular, the predator-prey model neglects 
radiation emission and modulation instability, intrasequence interaction, 
and deviations from the assumed periodic form of the soliton sequences. 
These effects can lead to instabilities and pulse-pattern corruption, 
and also to the breakdown 
of the predator-prey model description (see, for example Refs. \cite{PNC2010,PC2012,CPJ2013}, 
for the case of crosstalk induced by nonlinear gain or loss). 
In contrast, the coupled-NLS model (\ref{raman1}) provides 
a fuller description of the propagation, which includes all these effects. 
Thus, in order to check the predictions of the predator-prey model 
(\ref{raman5}) for stable dynamics of soliton amplitudes and 
the possibility to realize stable long-distance multichannel soliton-based transmission, 
it is important to carry out numerical simulations with the full coupled-NLS model.

In the current section, we first present numerical simulations 
with  the system (\ref{raman1}) without the Raman and the linear gain-loss terms. 
We then present a comparison between simulations 
with the full coupled-NLS model (\ref{raman1}) with the Raman term and the linear gain-loss profile (\ref{raman2}) 
and the predictions of the predator-prey model (\ref{raman5}) 
for collision-induced amplitude dynamics. We conclude the section by analyzing 
pulse-pattern deterioration at large distances, as observed in the full coupled-NLS simulations.

The coupled-NLS system (\ref{raman1}) is numerically 
solved using the split-step method with periodic boundary conditions \cite{Agrawal2001}. 
The use of periodic boundary conditions means that  
the numerical simulations describe pulse dynamics in a closed fiber loop.
The initial condition is in the form of $N$ periodic sequences 
of $2J$ solitons with initial amplitudes $\eta_{j}(0)$, initial frequencies $\beta_{j}(0)$, 
and initial zero phases:  
\begin{eqnarray} &&
\psi_{j}(t,0)\!=\!\sum_{k=-J}^{J-1}
\frac{\eta_{j}(0)\exp\{i\beta_{j}(0)[t-(k+1/2)T-\delta_{j}]\}}
{\cosh\{\eta_{j}(0)[t-(k+1/2)T-\delta_{j}]\}}, 
\label{raman8}
\end{eqnarray}
where $1\le j \le N$. The coefficients $\delta_{j}$ in Eq. (\ref{raman8}) 
correspond to the initial position shift of the pulses in the $j$th 
sequence relative to pulses located at $(k+1/2)T$ for $-J\le k \le J-1$.   
We simulate multichannel transmission with two, three, and four channels 
and two solitons in each channel. Thus, $2 \le N \le 4$ and $J=1$ are 
used in our numerical simulations.  
To maximize the stable propagation distance, we choose  
$\beta_{1}(0)$=-$\beta_{2}(0)$ for a two-channel system; 
$\beta_{1}(0)$=-$\beta_{3}(0)$, $\beta_{2}(0)=0$ for a three-channel system;
and $\beta_{1}(0)$=-$\beta_{4}(0)$, $\beta_{2}(0)$=-$\beta_{3}(0)$ for a four-channel system.   
In addition, we take $\delta_{j}=(j-1)T/N$ for $1\le j \le N$.  
These choices are based on extensive numerical simulations with 
Eq. (\ref{raman1}) and different values of $\beta_j(0)$ and $\delta_{j}$.


In the numerical simulations, we consider as a concrete example transmission at 
a bit-rate $B=12.5$ Gb/s per channel with the following physical parameter values
\cite{parameter_values}. The pulse width and time slot width are $\tau=5$ ps 
and $\tilde T=80$ ps, and the frequency spacing is taken as 
$\Delta \nu=0.48$ THz for $N=2,3$, and 4 channels.   
Thus, the total bandwidth of the system is smaller than 13.2 THz, 
and all channels lie within the main body of the Raman gain curve. 
The values of the dimensionless parameters for this system are 
$\epsilon_{R}=0.0012$, $T=16$, and $\Delta\beta=15$ for $N=2,3,4$. 
Assuming $\tilde{\beta_2} =  - 4$ ps$^2$km$^{-1}$ and 
$\gamma = 4$ W$^{-1}$km$^{-1}$ for the second-order dispersion and 
Kerr nonlinearity coefficients, the soliton peak power is $P_{0}=40$ mW. 
Tables 1 and 2 summarize the values of the dimensionless and dimensional 
physical parameters used in the simulations. In these tables, $W$ and $\tilde W$ 
stand for the dimensionless and dimensional width of the linear gain-loss function $g(\omega)$ 
in Eq. (\ref{raman2}), while $z_{s}$ and $X_{s}$ correspond to the dimensionless 
and dimensional distance along which stable propagation is observed.

\begin{table}[ht]
{\bf \caption{The dimensionless parameters}}                     
\begin{center}
\begin{tabular}{ c c c c c c c c c}
\hline \textnumero & $N$ & $\epsilon_{R}$ & $T$ & $\Delta\beta$ & $W$ & $g_L$ & $z_{s}$  & Figures\\ 
\hline $1$ & $2$ & $0.0012$ & $16$ & $15$ & $10$ & $-0.5$ & $950$  & \ref{fig1}, \ref{fig3}(a),  \ref{fig4}(a)-(b)\\ 
$2$ & $2$ & $0$ & $16$ & $15$ & $0$ & $0$ & $550$  & \ref{fig2}(a)-(b)\\  
$3$ & $2$ & $0.0012$ & $16$ & $15$ & $10$ & $-0.5$ & $11200$ & \ref{fig7}(a), \ref{fig8}(a)-(b)\\
$4$ & $3$ & $0$ & $16$ & $15$ & $0$ & $0$ & $510$ &  \ref{fig2}(c)-(d) \\
$5$ & $3$ & $0.0012$ & $16$ & $15$ & $10$ & $-0.5$ & $620$  & \ref{fig3}(b), \ref{fig4}(c)-(d) \\
$6$ & $3$ & $0.0012$ & $16$ & $15$ & $10$ & $-0.5$ & $12050$  & \ref{fig7}(b), \ref{fig8}(c)-(d)\\
$7$ & $4$ & $0$ & $16$ & $15$ & $0$ & $0$ & $340$  & \ref{fig2}(e)-(f)\\
$8$ & $4$ & $0.0012$ & $16$ & $15$ & $11$ & $-0.5$ & $500$  & \ref{fig3}(c), \ref{fig4}(e)-(f)\\
$9$ & $4$ & $0.0012$ & $16$ & $15$ & $11$ & $-0.5$ & $3600$ & \ref{fig7}(c), \ref{fig8}(e)-(f)\\
$10$ & $4$ & $0.0012$ & $16$ & $15$ & $15$ & $-0.5$ & $1800$ & \ref{fig9} \\
\hline 
\end{tabular} 
\end{center}
\label{table1}
\end{table}

\begin{table}[ht]         
{\bf \caption{The dimensional parameters}}     
\begin{center}
\begin{tabular}{ c c c c c c c }
\hline \textnumero &$N$&$\tau_0$ (ps)&$\tilde{T}$ (ps)&$\tilde W$ (THz)&$X_{s}$ (km)& Figures\\ 
\hline 
$1$ & $2$  & $5$ & $80$ & $0.32$ & $11875$ & \ref{fig1}, \ref{fig3}(a),  \ref{fig4}(a)-(b) \\
$2$ & $2$  & $5$ & $80$ & $0$ & $6875$ & \ref{fig2}(a)-(b)\\
$3$ & $2$  & $5$ & $80$ & $0.32$ & $140000$ & \ref{fig7}(a), \ref{fig8}(a)-(b)\\
$4$ & $3$  & $5$ & $80$ & $0$ & $6375$ & \ref{fig2}(c)-(d)\\
$5$ & $3$  & $5$ & $80$ & $0.32$ & $7750$ & \ref{fig3}(b), \ref{fig4}(c)-(d)\\
$6$ & $3$  & $5$ & $80$ & $0.32$ & $150625$ & \ref{fig7}(b), \ref{fig8}(c)-(d)\\
$7$ & $4$  & $5$ & $80$ & $0$ & $4250$ & \ref{fig2}(e)-(f)\\
$8$ & $4$  & $5$ & $80$ & $0.35$ & $6250$ & \ref{fig3}(c), \ref{fig4}(e)-(f)\\
$9$ & $4$ & $5$ & $80$ & $0.35$ & $45000$ & \ref{fig7}(c), \ref{fig8}(e)-(f)\\
$10$ & $4$ & $5$ & $80$ & $0.48$ & $22500$ & \ref{fig9}\\
\hline 
\end{tabular} 
\end{center}
\label{table2}
\end{table}

\begin{figure}[ptb]
\begin{tabular}{cc}
\epsfxsize=5.8cm  \epsffile{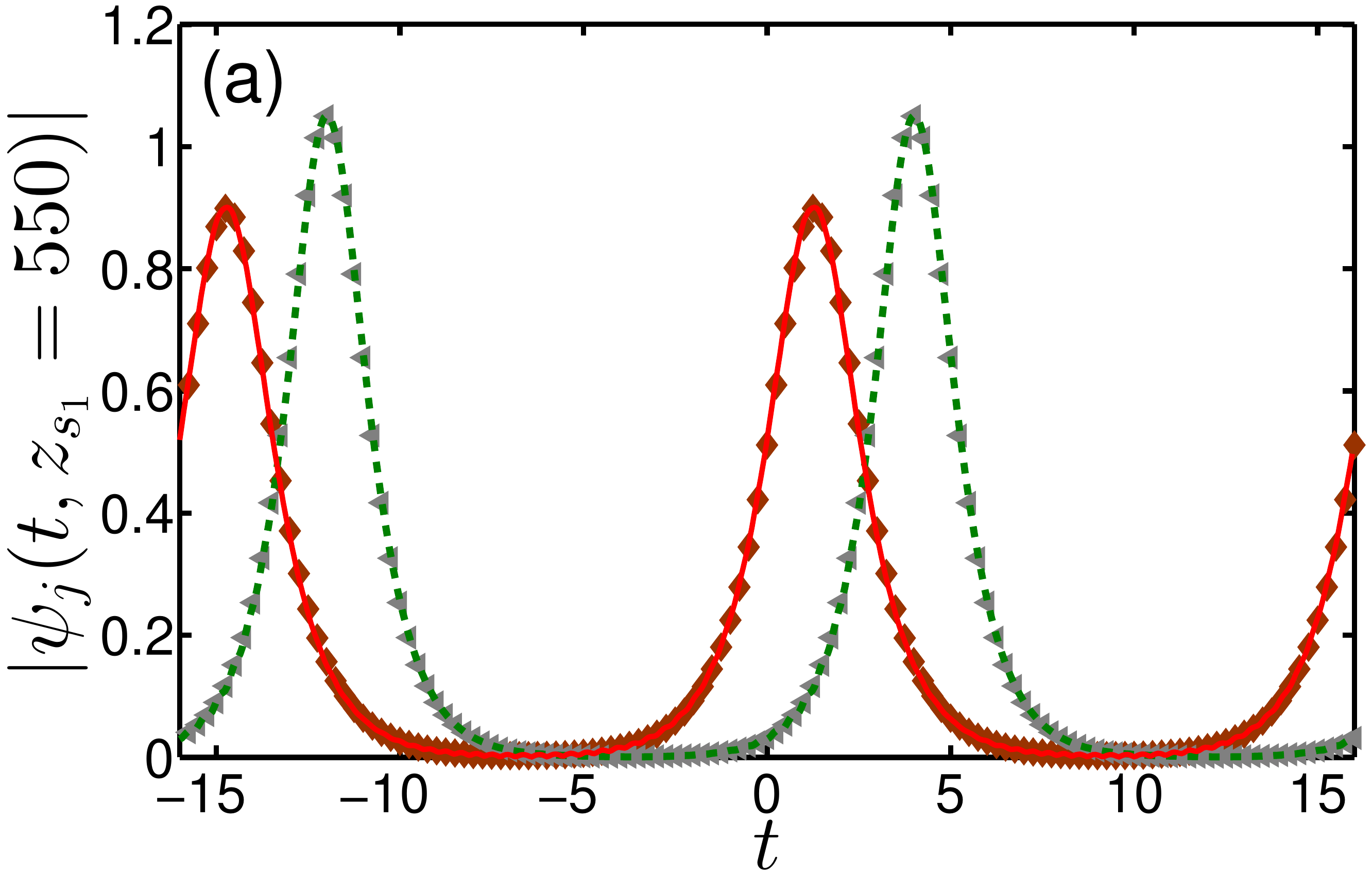} &
\epsfxsize=5.8cm  \epsffile{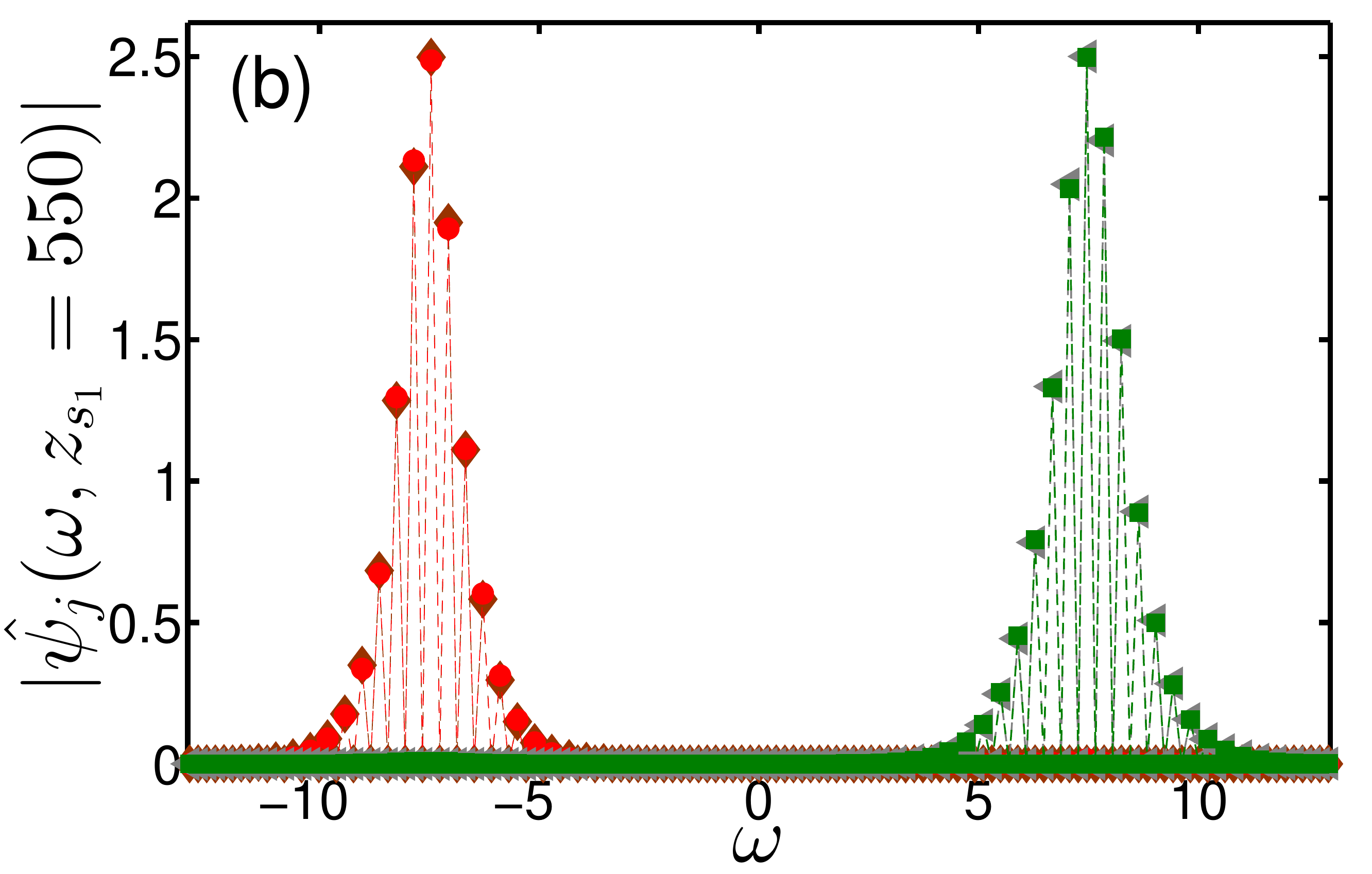} \\
\epsfxsize=5.8cm  \epsffile{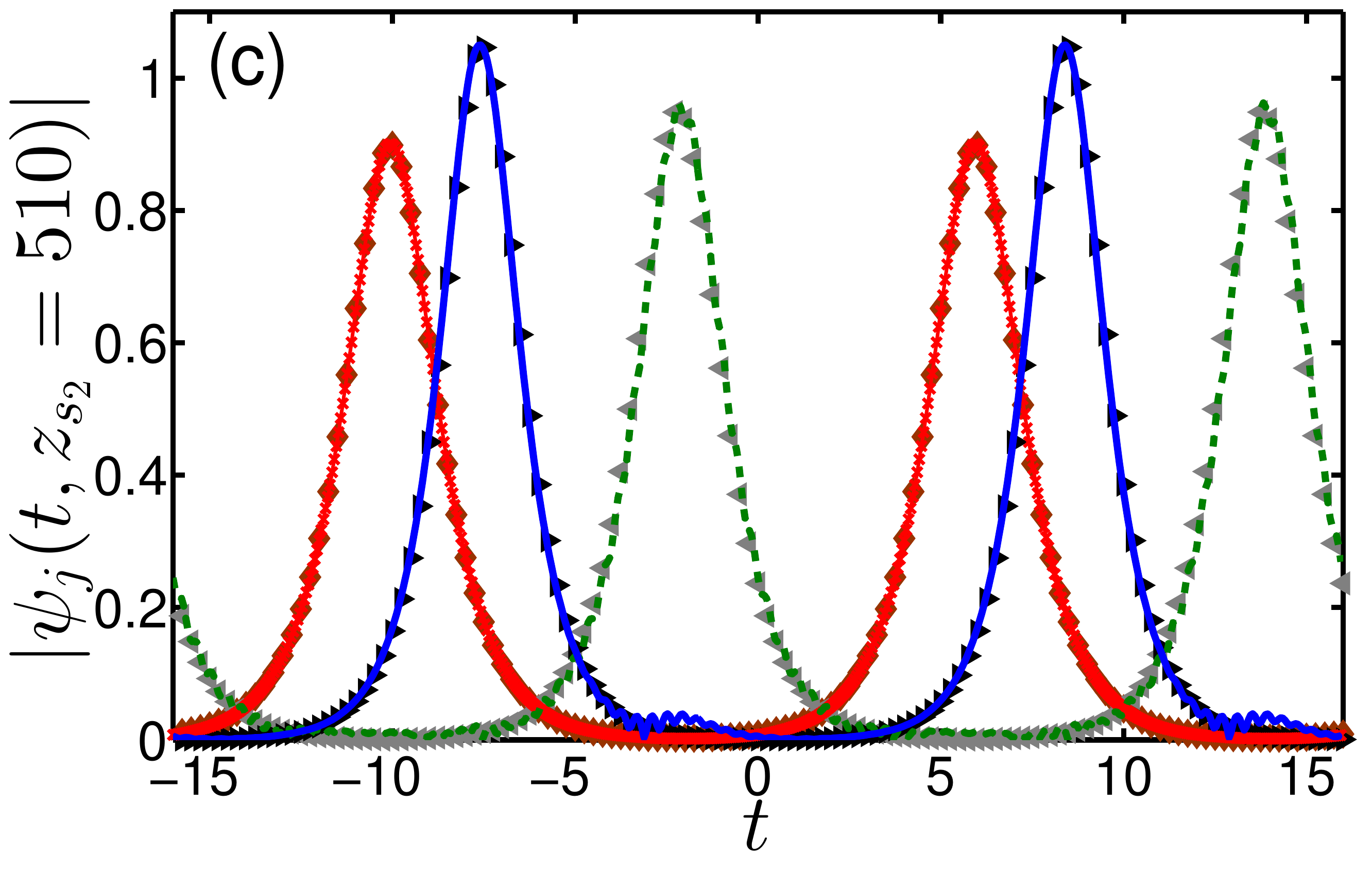} &
\epsfxsize=5.8cm  \epsffile{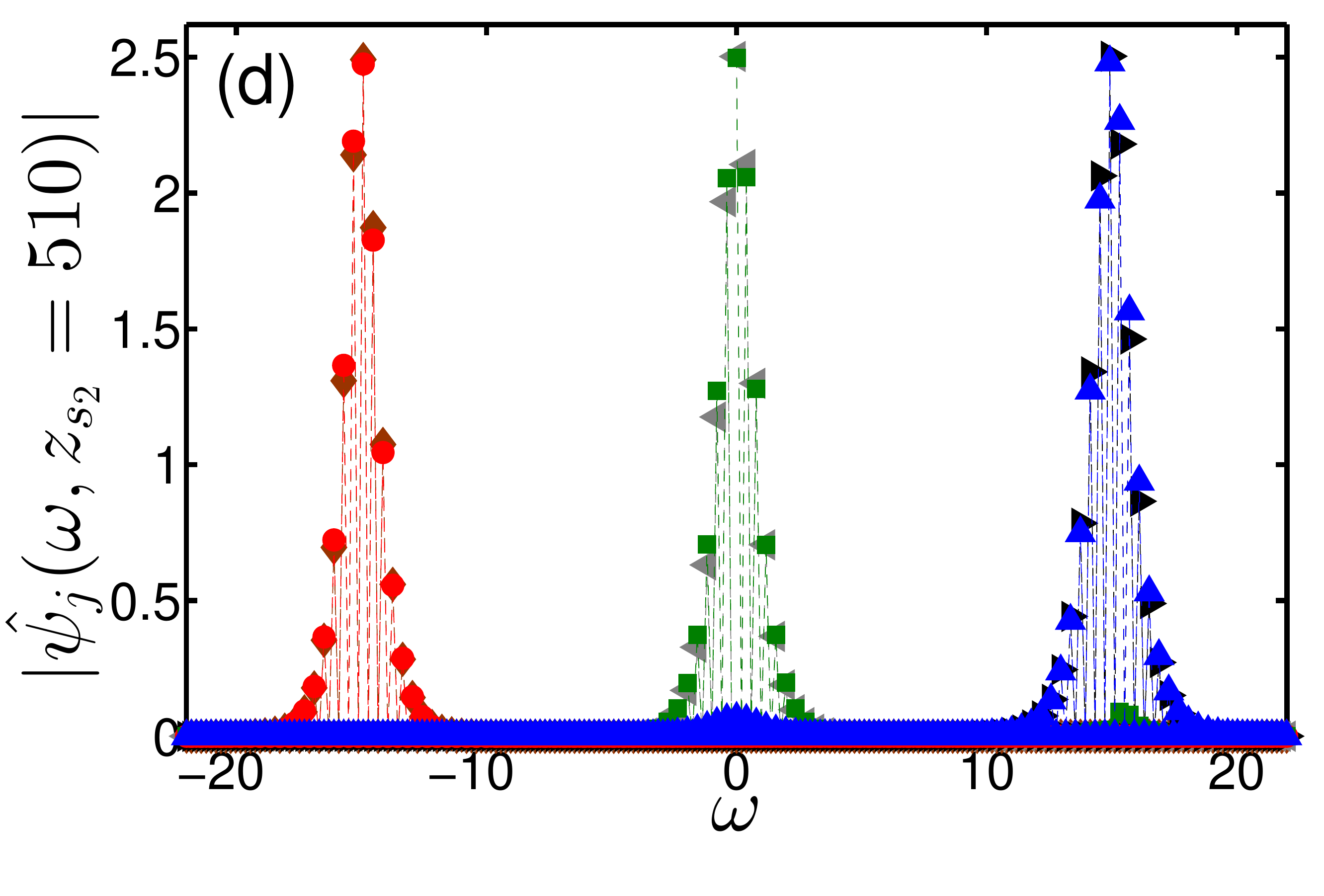}\\
\epsfxsize=5.8cm  \epsffile{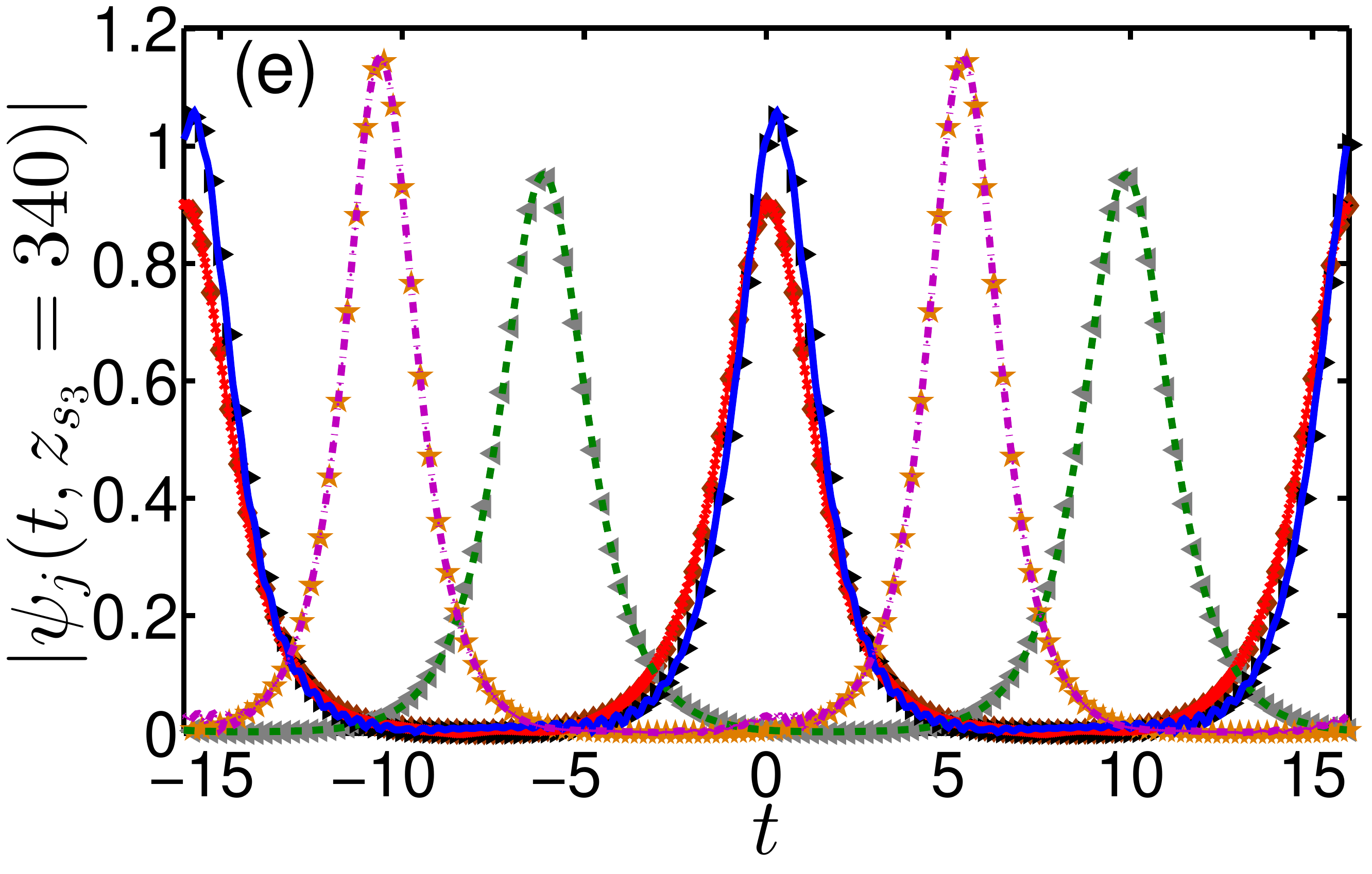} &
\epsfxsize=5.8cm  \epsffile{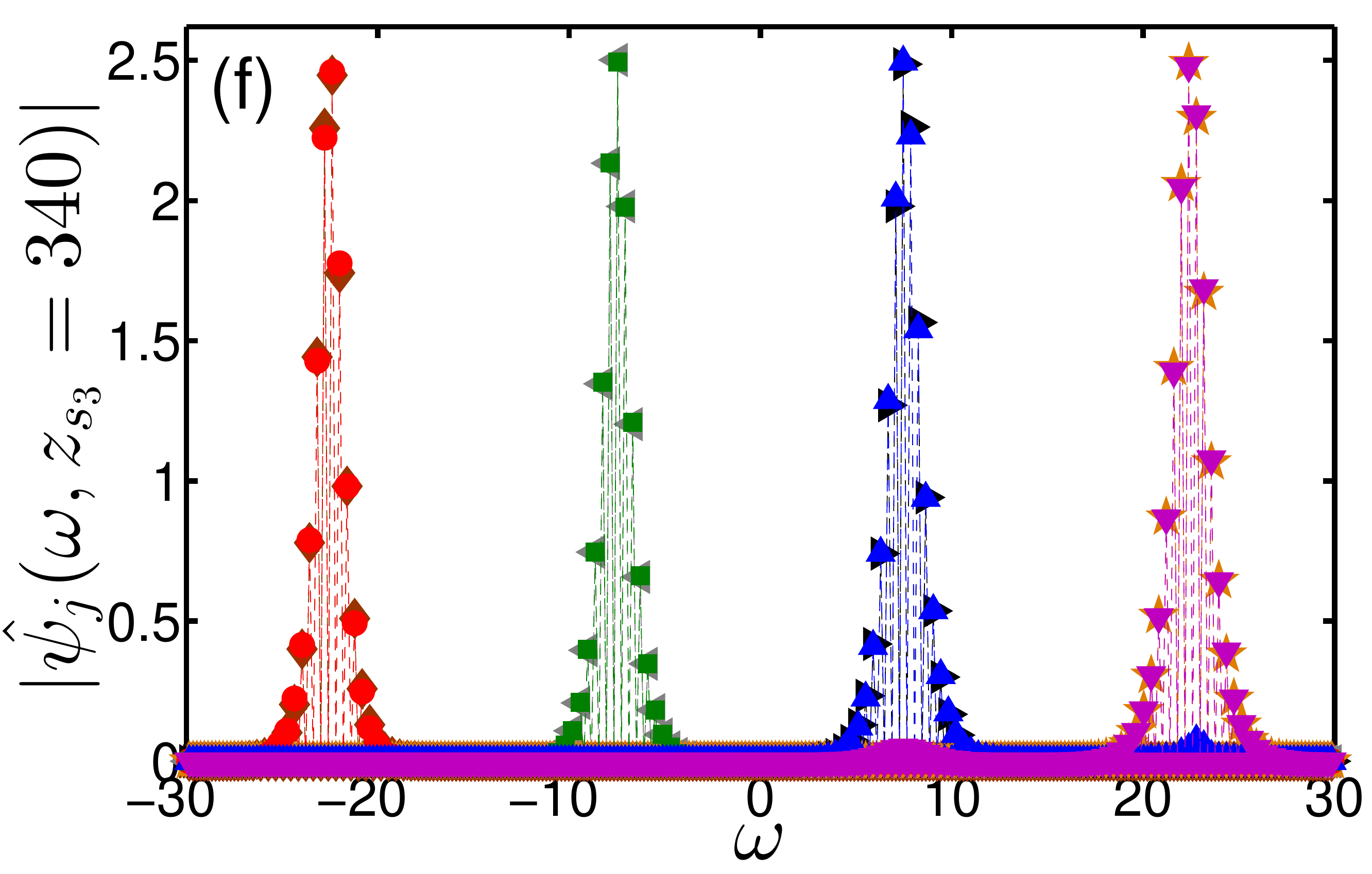}
\end{tabular}
\caption{(Color online) The pulse patterns at the onset of transmission instability  
$|\psi_{j}(t,z_{s})|$ and their Fourier transforms  
$|\hat{\psi_{j}}(\omega,z_{s})|$ for two-channel [(a)-(b)], 
three-channel [(c)-(d)], and four-channel [(e)-(f)] transmission  
in the absence of delayed Raman response and linear gain-loss. 
The physical parameter values are listed in rows 2, 4, and 7 of Table 1. 
The stable transmission distances are $z_{s_1}=550$ for $N=2$, 
$z_{s_2}=510$ for $N=3$, and $z_{s_3}=340$ for $N=4$. 
The solid-crossed red curve [solid red curve in (a)], dashed green curve, 
solid blue curve, and dashed-dotted magenta curve represent 
$|\psi_{j}(t,z_{s})|$ with $j=1,2,3,4$, obtained by 
numerical simulations with Eq. (\ref{raman1B}).  
The red circles, green squares, blue up-pointing triangles, 
and magenta down-pointing triangles represent 
$|\hat{\psi_{j}}(t,z_{s})|$ with $j=1,2,3,4$, obtained by the simulations. 
The brown diamonds, gray left-pointing triangles, black right-pointing triangles, 
and orange stars represent the theoretical prediction for $|\psi_{j}(t,z_{s})|$ 
or $|\hat{\psi_{j}}(\omega,z_{s})|$ with $j=1,2,3,4$, respectively.}
 \label{fig2}
\end{figure}

\begin{figure}[ptb]
\begin{tabular}{cc}
\epsfxsize=5.8cm  \epsffile{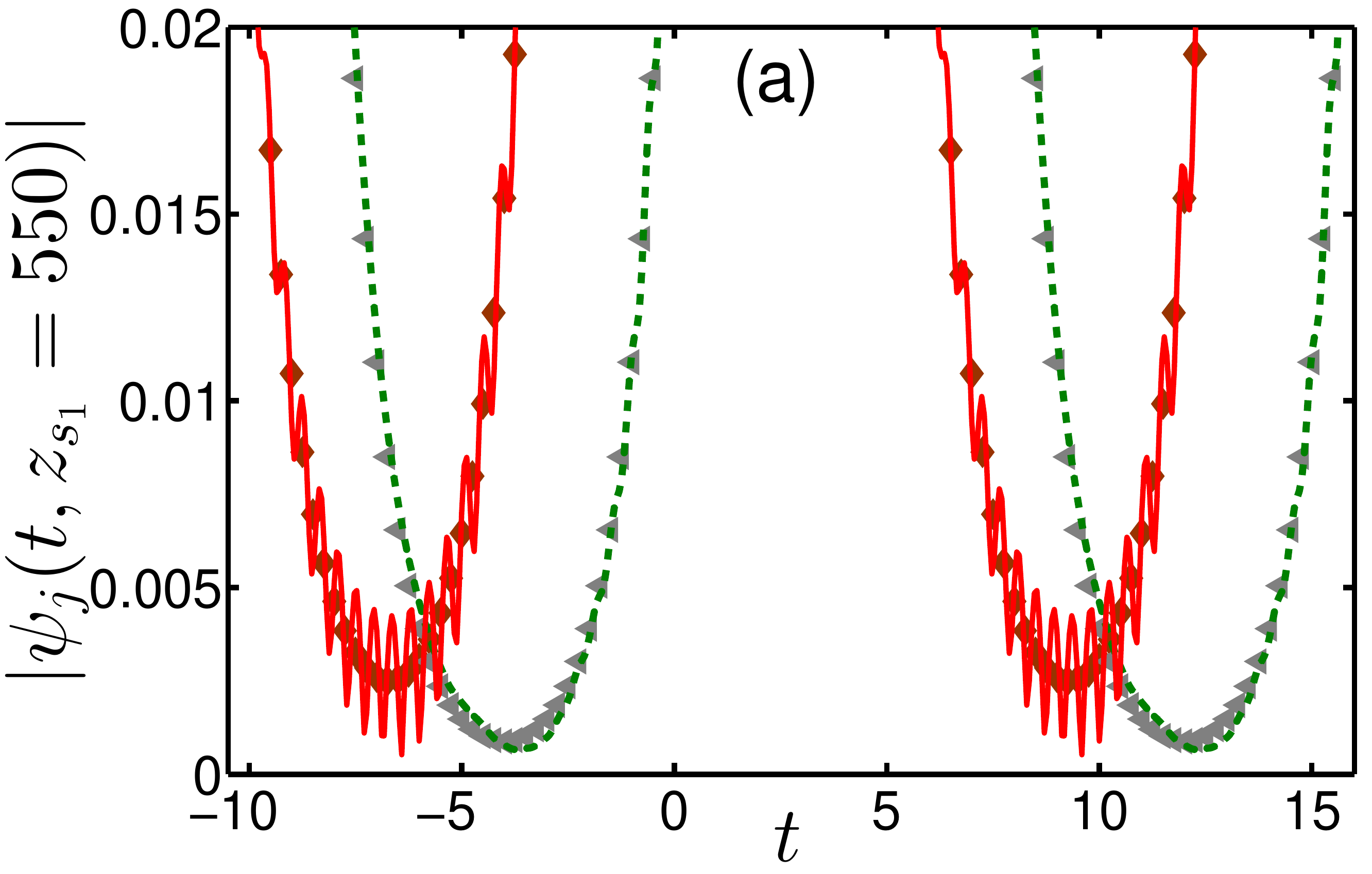} &
\epsfxsize=5.8cm  \epsffile{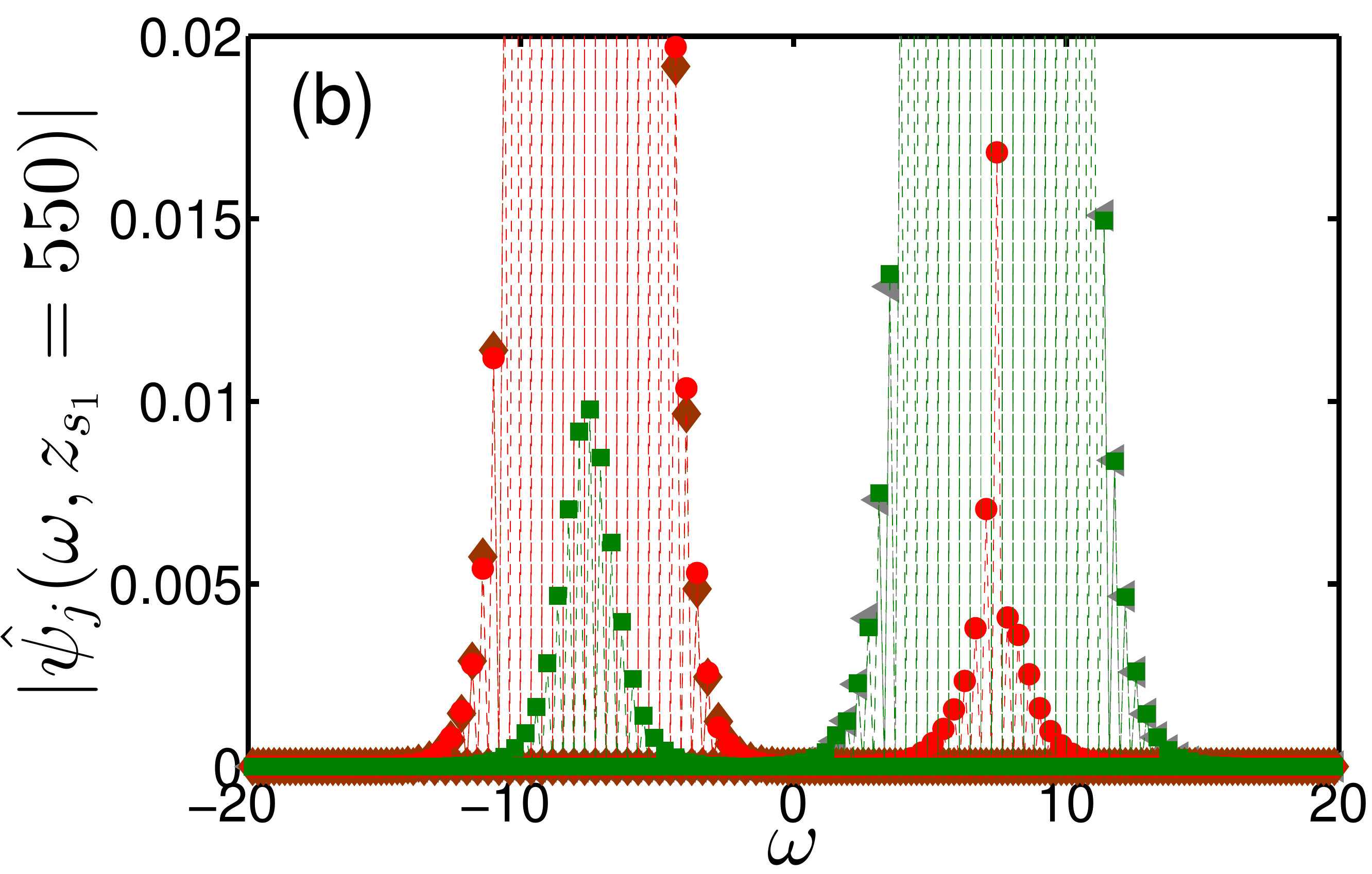} \\
\epsfxsize=5.8cm  \epsffile{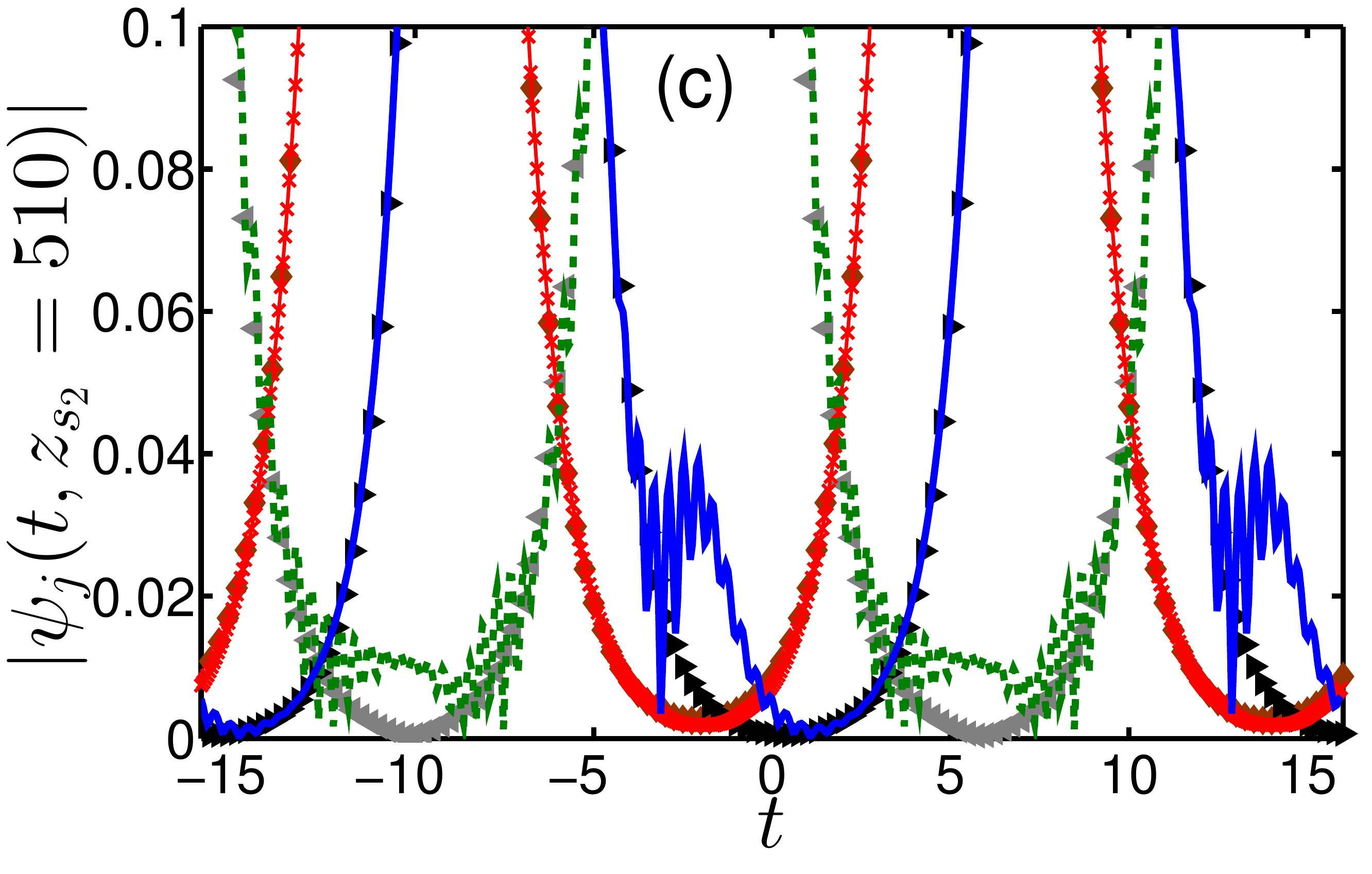}&
\epsfxsize=5.8cm  \epsffile{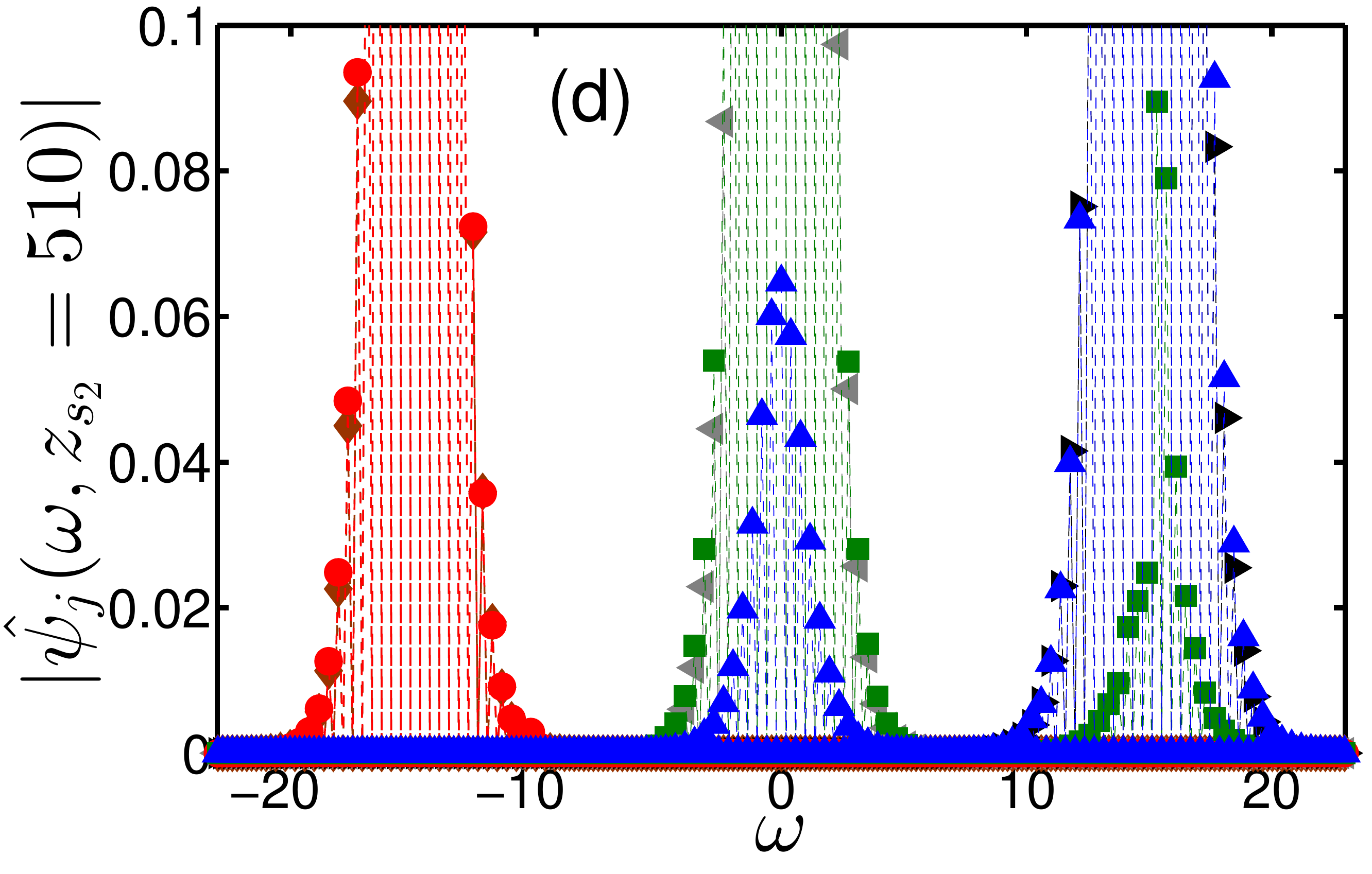} \\
\epsfxsize=5.8cm  \epsffile{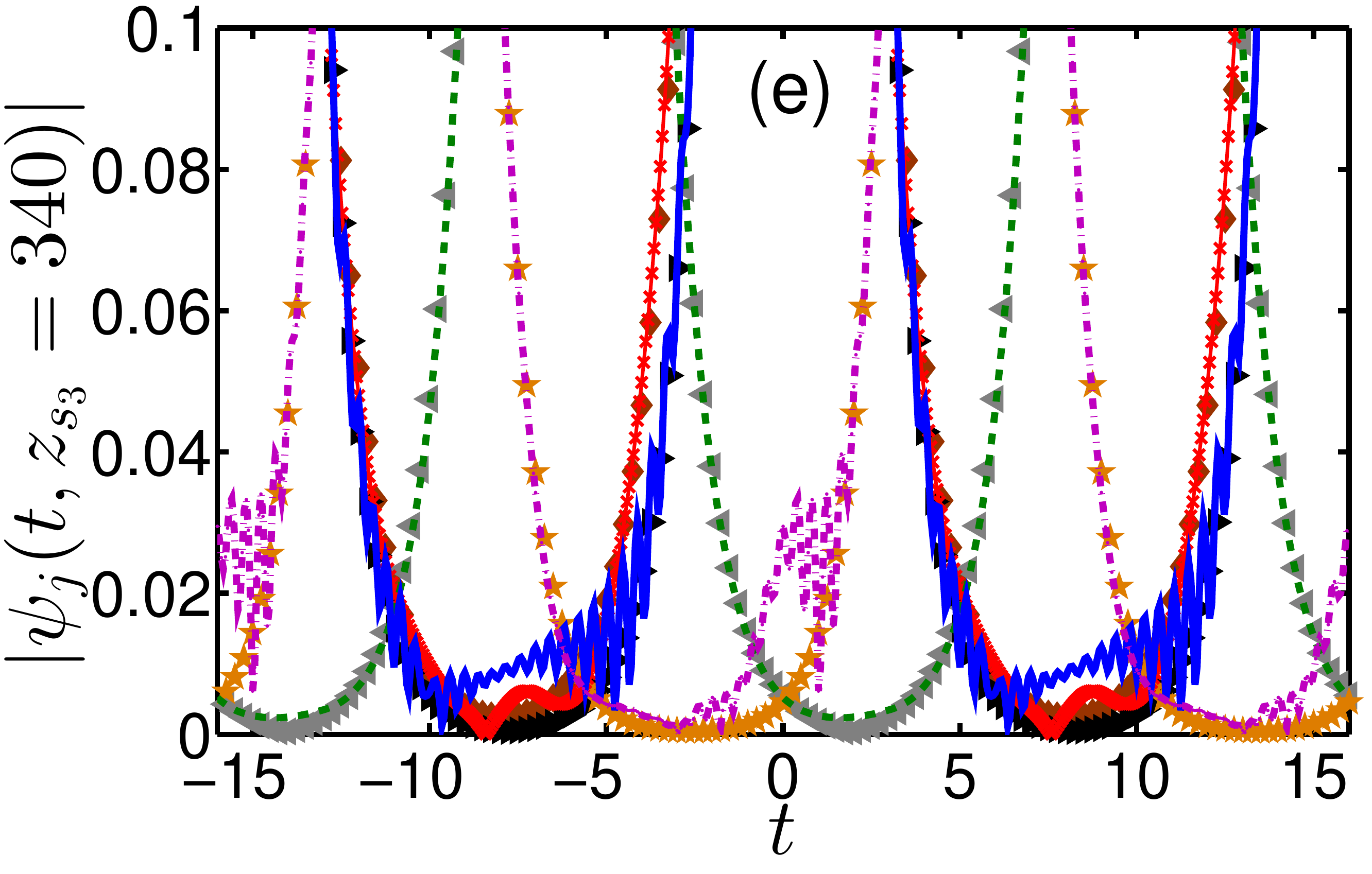}&
\epsfxsize=5.8cm  \epsffile{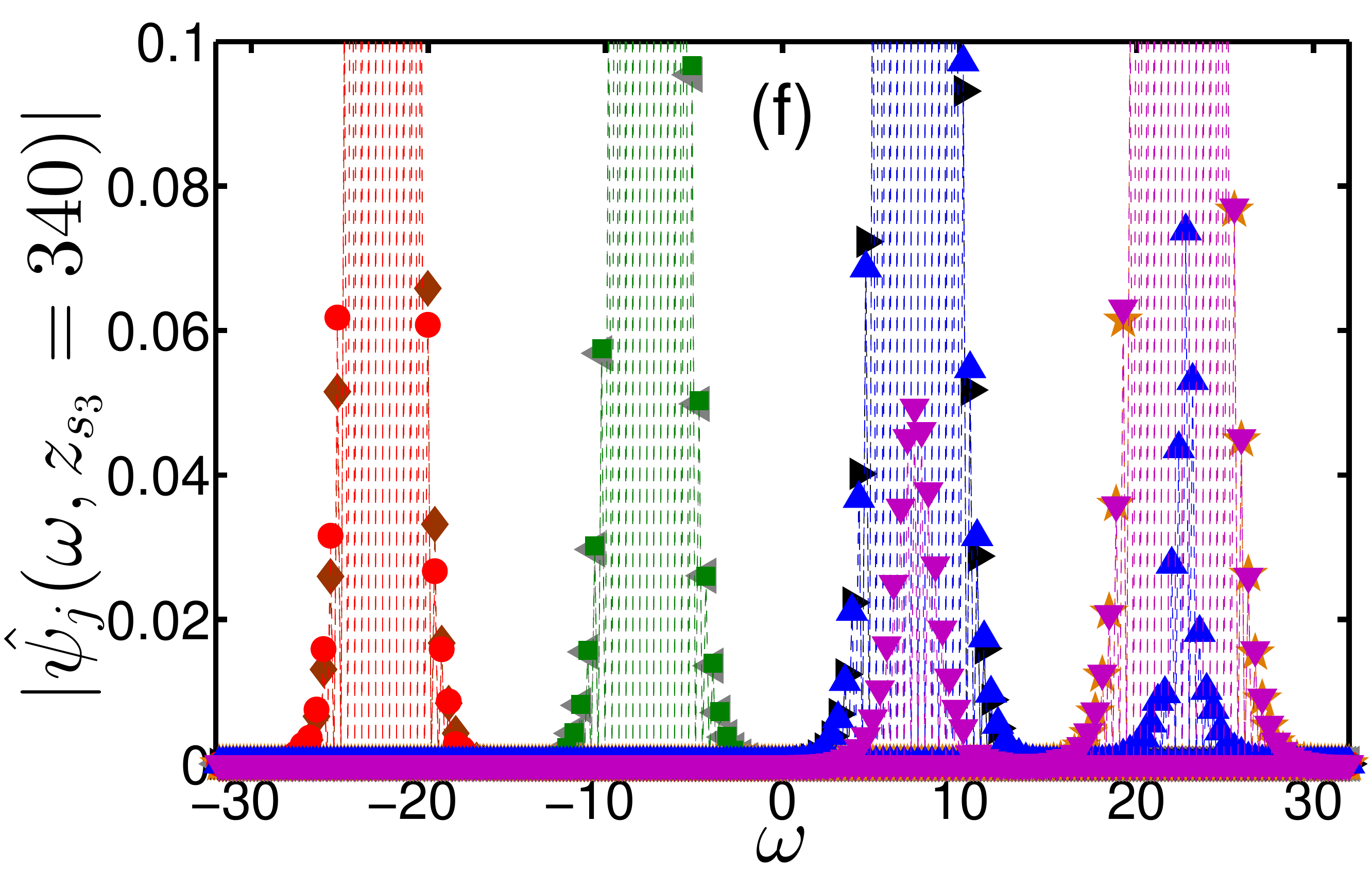} \\
\end{tabular}
\caption{(Color online) Magnified versions of the graphs in Fig. \ref{fig2} for 
small $|\psi_{j}(t,z_{s})|$ and $|\hat{\psi_{j}}(\omega,z_{s})|$ values. 
The symbols are the same as in Fig. \ref{fig2}.}
 \label{mag_fig2}
\end{figure}

Note that the Kerr nonlinearity terms appearing in Eq. (\ref{raman1}) are nonperturbative. 
Even though these terms are not expected to affect the shape, amplitude, and frequency 
of a single soliton, propagating in an ultralong optical fiber, the situation can be very different 
for multiple soliton sequences, circulating in a fiber loop. In the latter case, Kerr-induced 
effects might lead to radiation emission, modulation instability, and eventually to 
pulse-pattern corruption \cite{CPN2016}. 
It is therefore important to first analyze the effects of Kerr 
nonlinearity alone on the propagation. For this purpose, we carry out numerical simulations 
with the following coupled-NLS model, which incorporates second-order dispersion 
and Kerr nonlinearity, but neglects delayed Raman response and linear gain-loss: 
\begin{eqnarray} &&
i\partial_z\psi_{j}+\partial_{t}^2\psi_{j}+2|\psi_{j}|^2\psi_{j}
+4\sum_{k=1}^{N}(1-\delta_{jk})|\psi_{k}|^2\psi_{j}=0,  
\label{raman1B}
\end{eqnarray}               
where $1\le j \le N$. The simulations are carried out for two, three, and four 
frequency channels with the physical parameter values listed in rows 2, 4, and 7 of Table 1. 
As an example, we present the results of the simulations for the 
following sets of initial soliton amplitudes: 
$\eta_{1}(0)=0.9$, $\eta_{2}(0)=1.05$ for $N=2$; 
$\eta_{1}(0)=0.9$, $\eta_{2}(0)=0.95$, $\eta_{3}(0)=1.1$ for $N=3$;
and $\eta_{1}(0)=0.9$, $\eta_{2}(0)=0.95$, $\eta_{3}(0)=1.05$, 
$\eta_{4}(0)=1.15$ for $N=4$. We emphasize, however, that similar 
results are obtained with other choices of the initial soliton amplitudes. 
The numerical simulations are carried out up to a distance $z_{s}$, 
at which instability appears. 
More specifically, we define $z_{s}$ as the largest distance 
at which the values of the integrals $I_{j}(z)$ in Eq. (\ref{criterion3}) 
in \ref{appendA} are still smaller than 0.05 for $1\le j \le N$.        
The actual value of $z_{s}$ depends on the values of the physical parameters 
and in particular on the number of channels $N$.    
For the coupled-NLS simulations with Eq. (\ref{raman1B}) 
and the aforementioned initial amplitude values, 
we find $z_{s_1}=550$ for $N=2$, 
$z_{s_2}=510$ for $N=3$, and $z_{s_3}=340$ for $N=4$. 
Figure \ref{fig2} shows the pulse patterns $|\psi_{j}(t,z_{s})|$ 
and their Fourier transforms $|\hat{\psi_{j}}(\omega,z_{s})|$
at the onset of instability, as obtained by the numerical solution of Eq. (\ref{raman1B}). 
Also shown are the theoretical predictions for the pulse patterns 
and their Fourier transforms at the onset of instability.  
Figure \ref{mag_fig2} shows magnified versions of the graphs in Fig. \ref{fig2} for 
small $|\psi_{j}(t,z_{s})|$ and $|\hat{\psi_{j}}(\omega,z_{s})|$ values. 
The theoretical prediction for $|\psi_{j}(t,z_{s})|$ is obtained by summation 
over fundamental NLS solitons with amplitudes $\eta_{j}(0)$, 
frequencies $\beta_{j}(0)$, and positions $y_{j}(z_{s})+kT$ 
for $-J\le k \le J-1$, which are measured from the simulations (see \ref{appendA}).   
The theoretical prediction for $|\hat{\psi_{j}}(\omega,z_{s})|$ 
is obtained by taking the Fourier transform of the latter sum.      
As can be seen from Fig. \ref{fig2}, the soliton patterns 
are almost intact at $z=z_{s}$ for $N=2,3,4$. 
Additionally, the soliton amplitude and frequency values 
are very close to their initial values. 
Thus, the solitons propagate in a stable manner 
up to the distance $z_{s}$. However, an examination of 
Figs. 3(a), 3(c), and 3(e) reveals that the soliton patterns are 
in fact slightly distorted at $z_{s}$, and that the distortion appears 
as fast oscillations in the solitons tails. 
Furthermore, as seen in Figs. 3(b), 3(d), and 3(f), the distortion 
is caused by resonant generation of radiaitive sidebands, 
where the largest sidebands for the $j$th soliton sequence form 
at frequencies $\beta_{j-1}(0)$ and/or $\beta_{j+1}(0)$ of the 
neighboring soliton sequences. In addition, the amplitudes of the 
radiative sidebands increase as the number of channels increases 
(see also Ref. \cite{CPN2016} for similar behavior), and as a result, 
the stable propagation distance $z_{s}$ decreases with increasing $N$. 
The growth of radiative sidebands and pulse distortion  
with increasing $z$ eventually leads  to the destruction of the soliton sequences.       
We point out that when each soliton sequence propagates through the fiber  
on its own, no radiative sidebands develop and no instability is observed 
up to distances as large as $z=20000$ \cite{CPN2016}. 
The latter finding is also in accordance with results of single-channel soliton transmission 
experiments, which demonstrated stable soliton propagation over distances 
as large as $10^{6}$ km \cite{Nakazawa91}.  
Based on these observations we conclude that 
transmission instability in the multichannel optical fiber system 
is caused by the Kerr-induced interaction in interchannel soliton collisions, that is, 
it is associated with the terms $2|\psi_{k}|^2\psi_{j}$ in Eq. (\ref{raman1}).

\begin{figure}[ptb]
\begin{tabular}{cc}
\epsfxsize=7.0cm  \epsffile{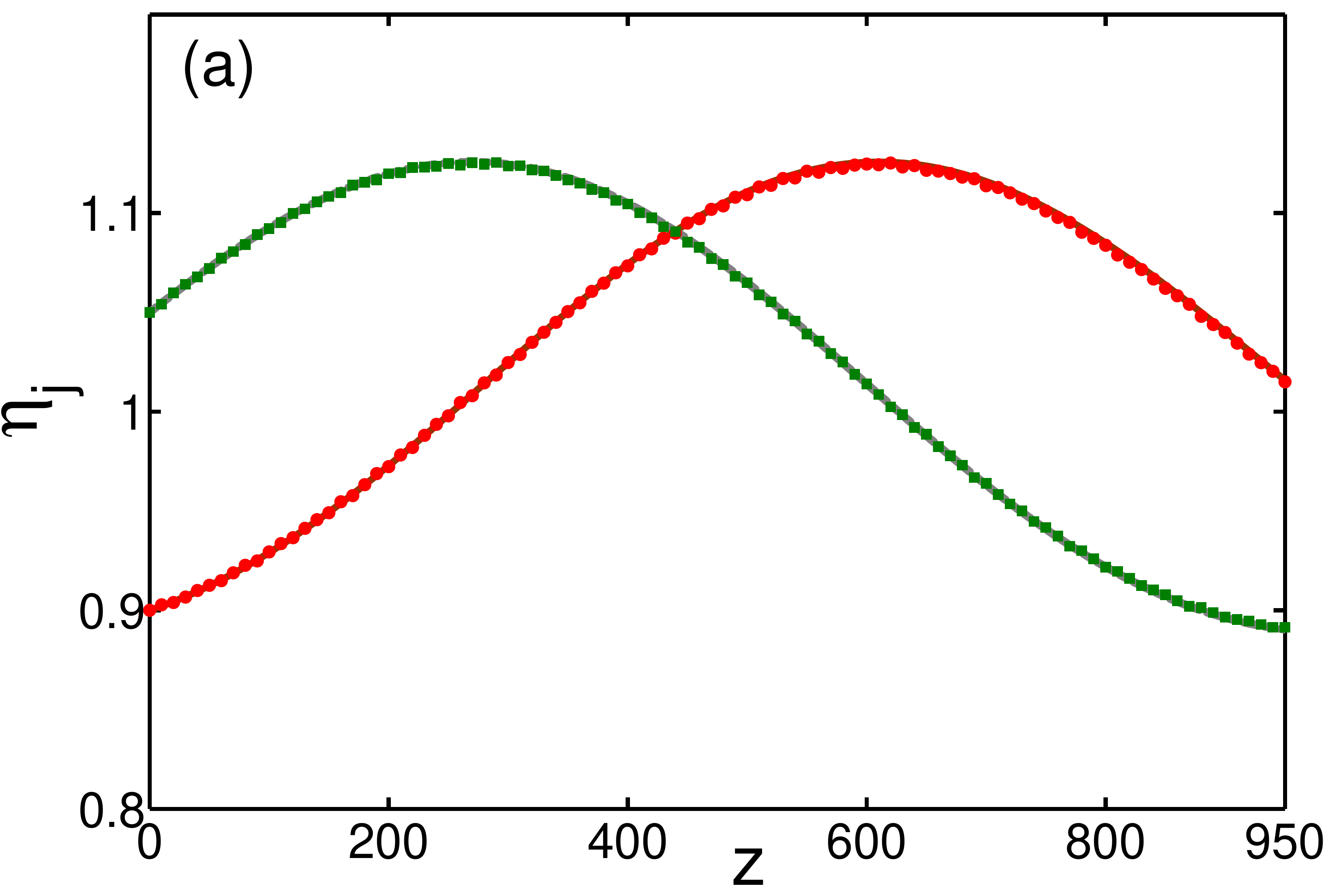} \\
\epsfxsize=7.0cm  \epsffile{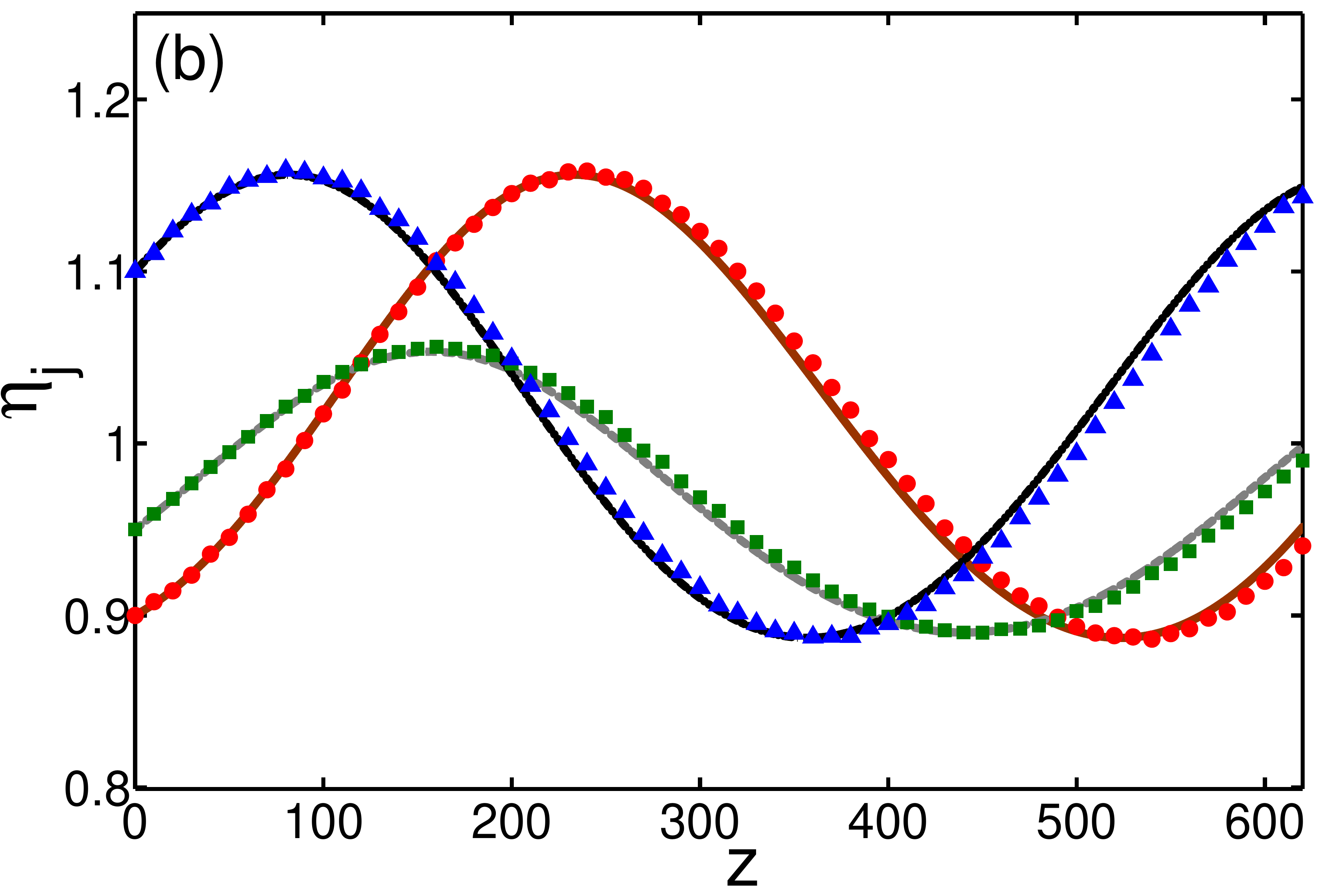} \\
\epsfxsize=7.0cm  \epsffile{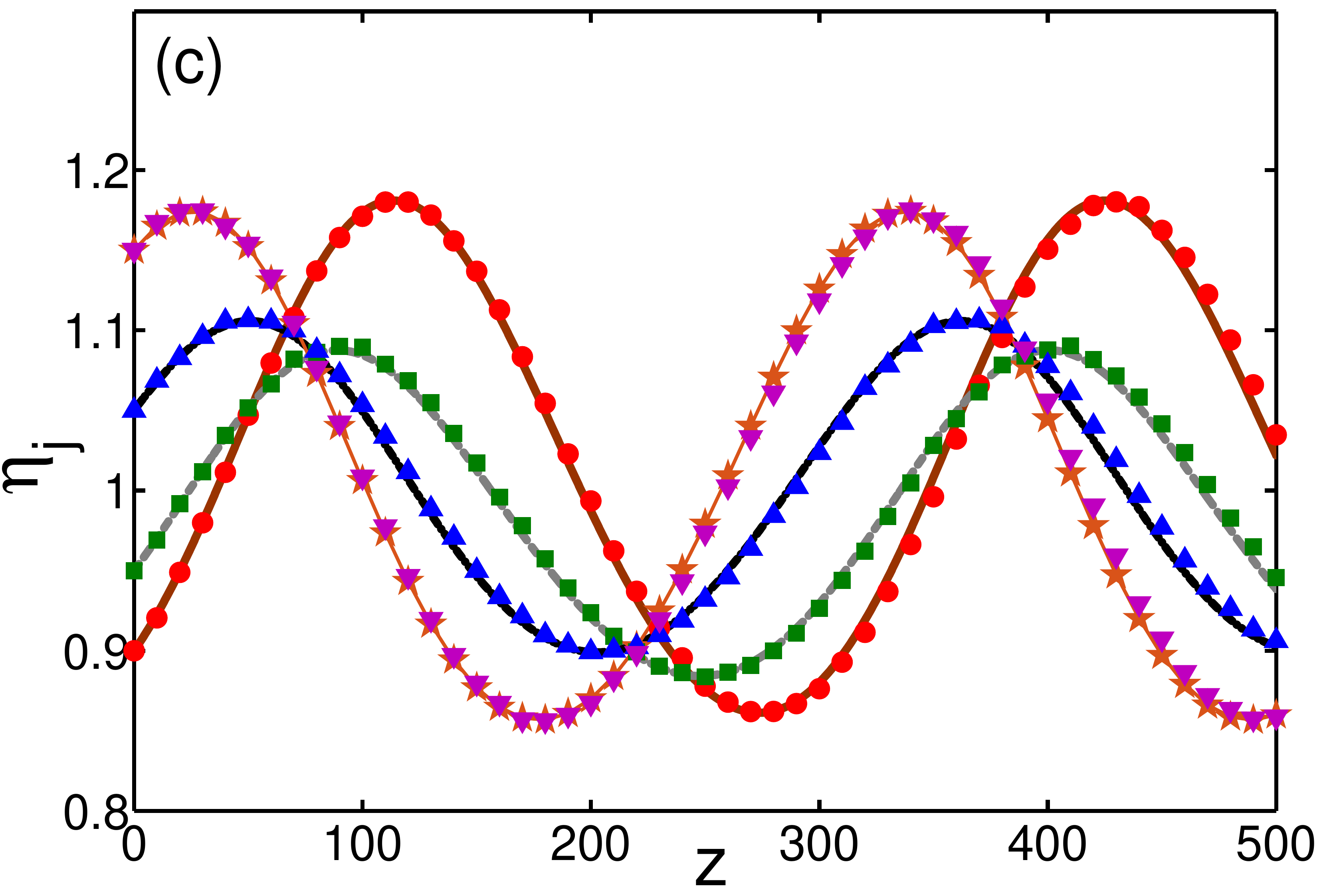} \\
\end{tabular}
\caption{(Color online) The $z$ dependence of soliton amplitudes $\eta_{j}$ 
for two-channel (a), three-channel (b), and four-channel (c) transmission 
in the presence of delayed Raman response 
and the frequency dependent linear gain-loss (\ref{raman2}).   
The physical parameter values are listed in rows 1, 5, and 8 of Table 1. 
The red circles, green squares, blue up-pointing triangles, 
and magenta down-pointing triangles represent 
$\eta_{1}(z)$, $\eta_{2}(z)$, $\eta_{3}(z)$, and $\eta_{4}(z)$,  
obtained by numerical solution of Eqs. (\ref{raman1}) and (\ref{raman2}). 
The solid brown, dashed gray, dashed-dotted black, and solid-starred orange curves 
correspond to $\eta_{1}(z)$, $\eta_{2}(z)$, $\eta_{3}(z)$, and $\eta_{4}(z)$, 
obtained by the predator-prey model (\ref{raman5}).}
 \label{fig3}
\end{figure}

We now take into account the effects of delayed Raman response 
and frequency dependent linear gain-loss on the propagation. 
Our first objective is to check the validity of the predator-prey 
model's predictions for collision-induced dynamics of soiton 
amplitudes in the presence of delayed Raman response.  
For this purpose, we carry out numerical simulations with the full coupled-NLS 
model (\ref{raman1}) with the linear gain-loss function (\ref{raman2})  
for two, three, and four frequency channels with the 
physical parameter values listed in rows 1, 5, and 8 of Table 1. 
To enable comparison with the results presented in Figs. \ref{fig2} 
and \ref{mag_fig2}, we discuss the results of simulations with the same sets 
of initial soliton amplitudes as the ones used in Figs. \ref{fig2} and \ref{mag_fig2}.
The numerical simulations are carried out up to the onset of 
transmission instability, which occurs at $z_{s_4}=950$ for $N=2$, 
at $z_{s_5}=620$ for $N=3$, and at $z_{s_6}=500$ for $N=4$.  
The $z$ dependence of soliton amplitudes obtained by numerical solution 
of the full coupled-NLS model (\ref{raman1}) with the gain-loss  
(\ref{raman2}) is shown in Fig. \ref{fig3} 
along with the prediction of the predator-prey model (\ref{raman5}). 
In all three cases the soliton amplitudes oscillate about 
their equilibrium value $\eta=1$, i.e., the dynamics of 
soliton amplitudes is stable up to the distance $z_{s}$. 
Furthermore, the agreement between the coupled-NLS simulations and 
the predator-prey model's predictions is excellent throughout the propagation. 
Thus, our coupled-NLS simulations validate the predictions of the 
predator-prey model (\ref{raman5}) for collision-induced amplitude 
dynamics in the presence of delayed Raman response at distances 
$0 \le z \le z_{s}$. This is a very important observation, because of 
the major simplifying assumptions that were made in the 
derivation of the model (\ref{raman5}). 
In particular, we conclude that the effects of radiation emission, 
modulation instability, intrachannel interaction, and other high-order perturbations 
can indeed be neglected for distances smaller than $z_{s}$.

\begin{figure}[ptb]
\begin{tabular}{cc}
\epsfxsize=5.8cm  \epsffile{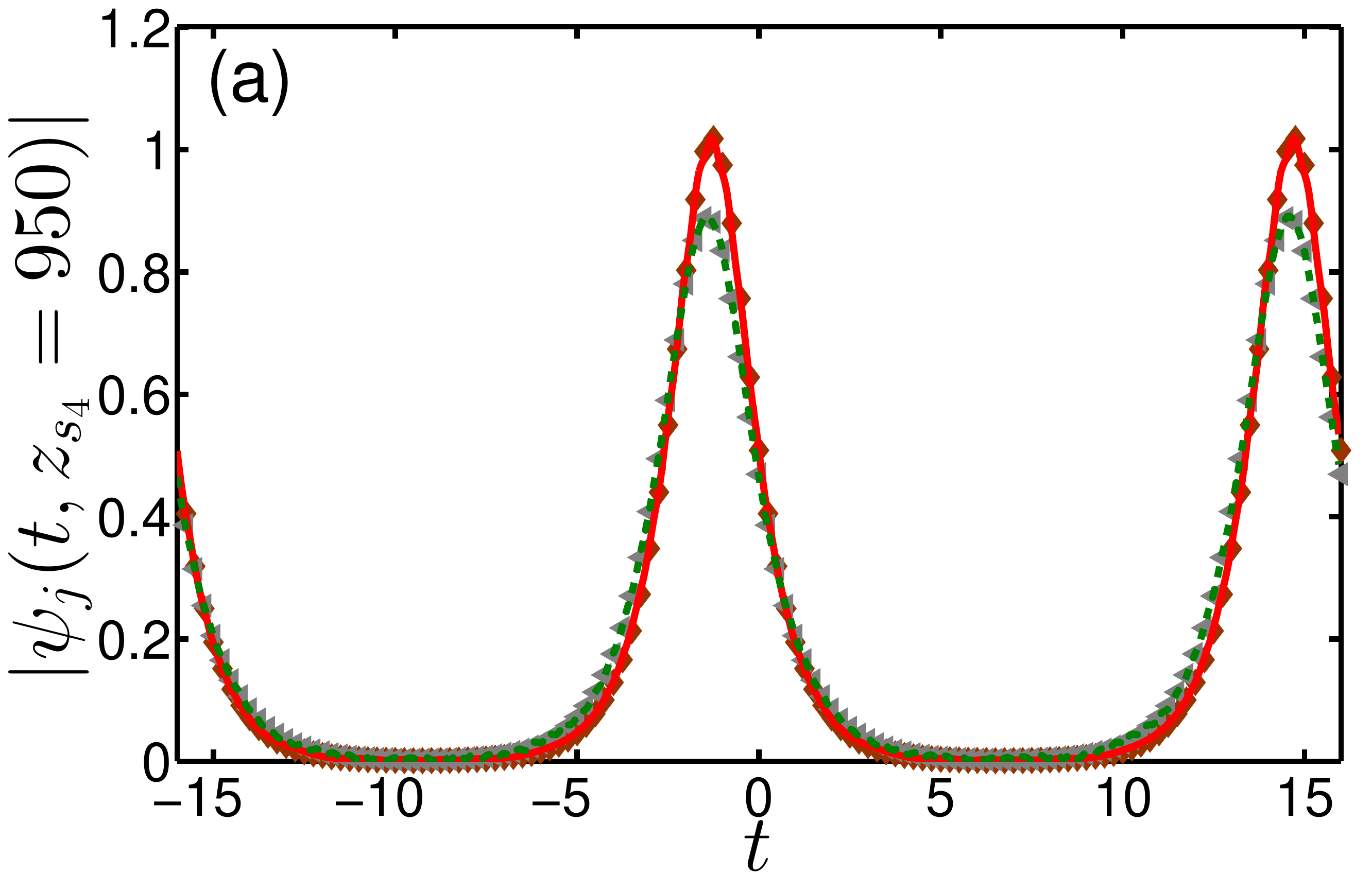} &
\epsfxsize=5.8cm  \epsffile{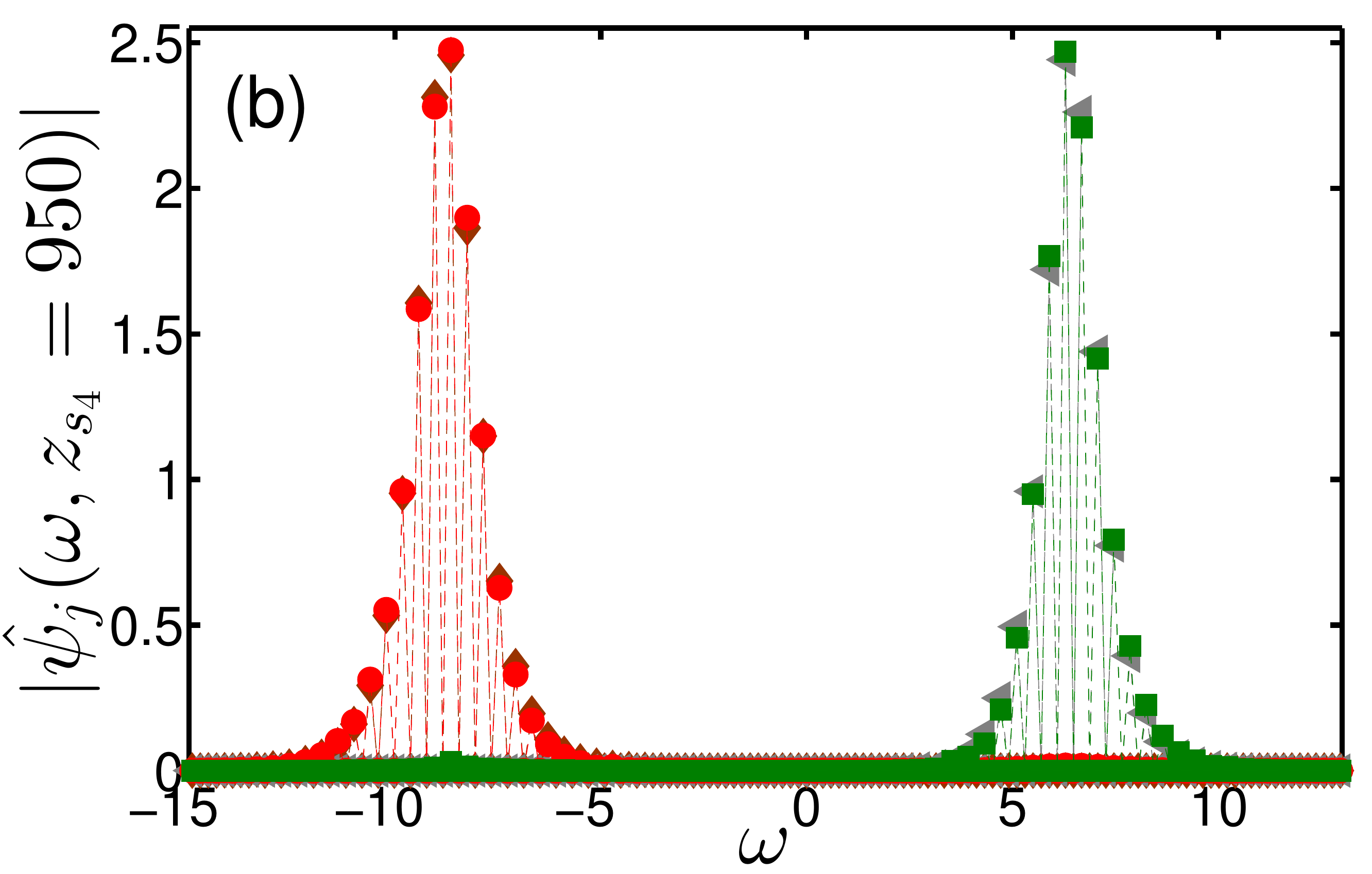} \\
\epsfxsize=5.8cm  \epsffile{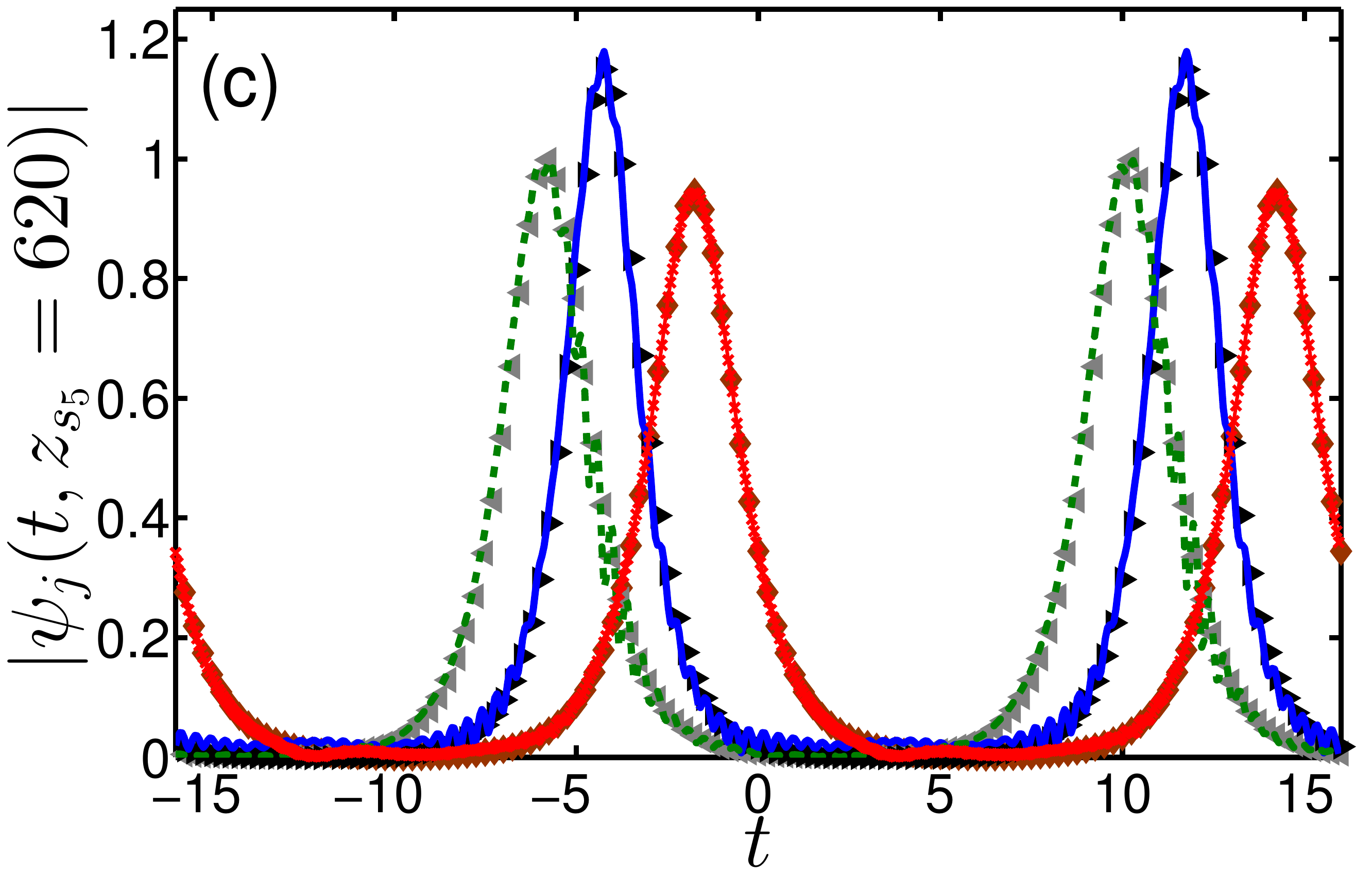}&
\epsfxsize=5.8cm  \epsffile{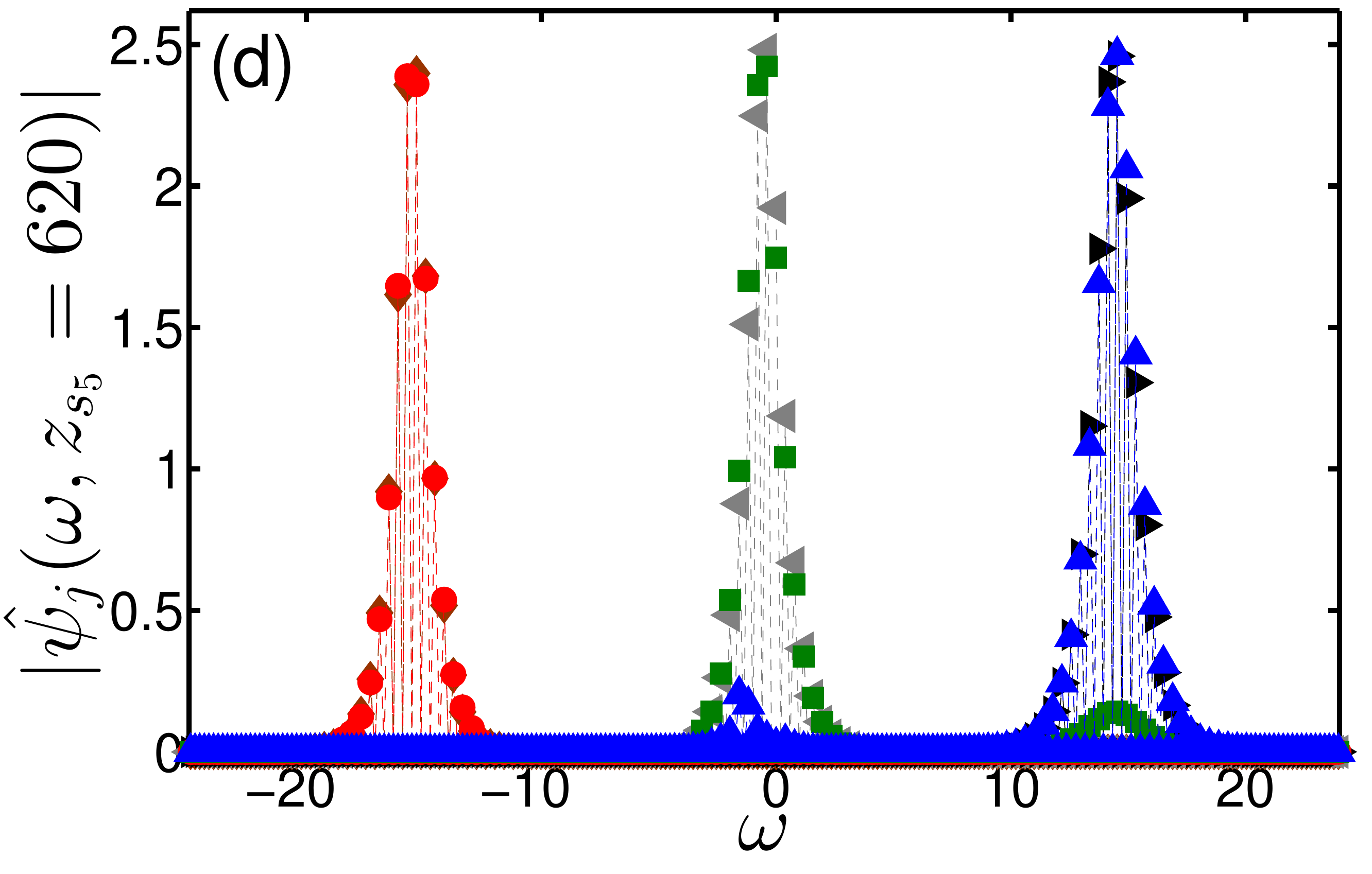} \\
\epsfxsize=5.8cm  \epsffile{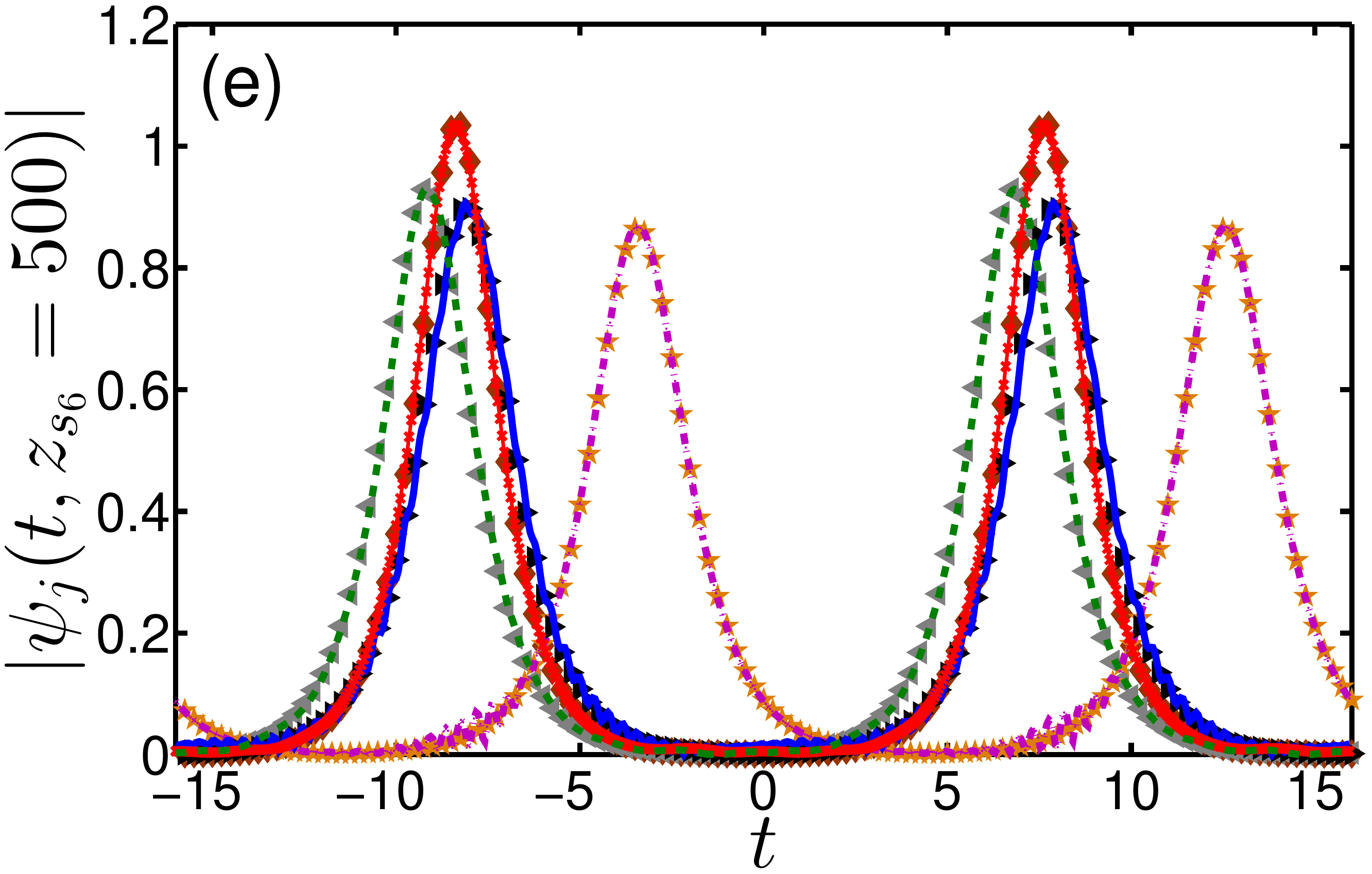}&
\epsfxsize=5.8cm  \epsffile{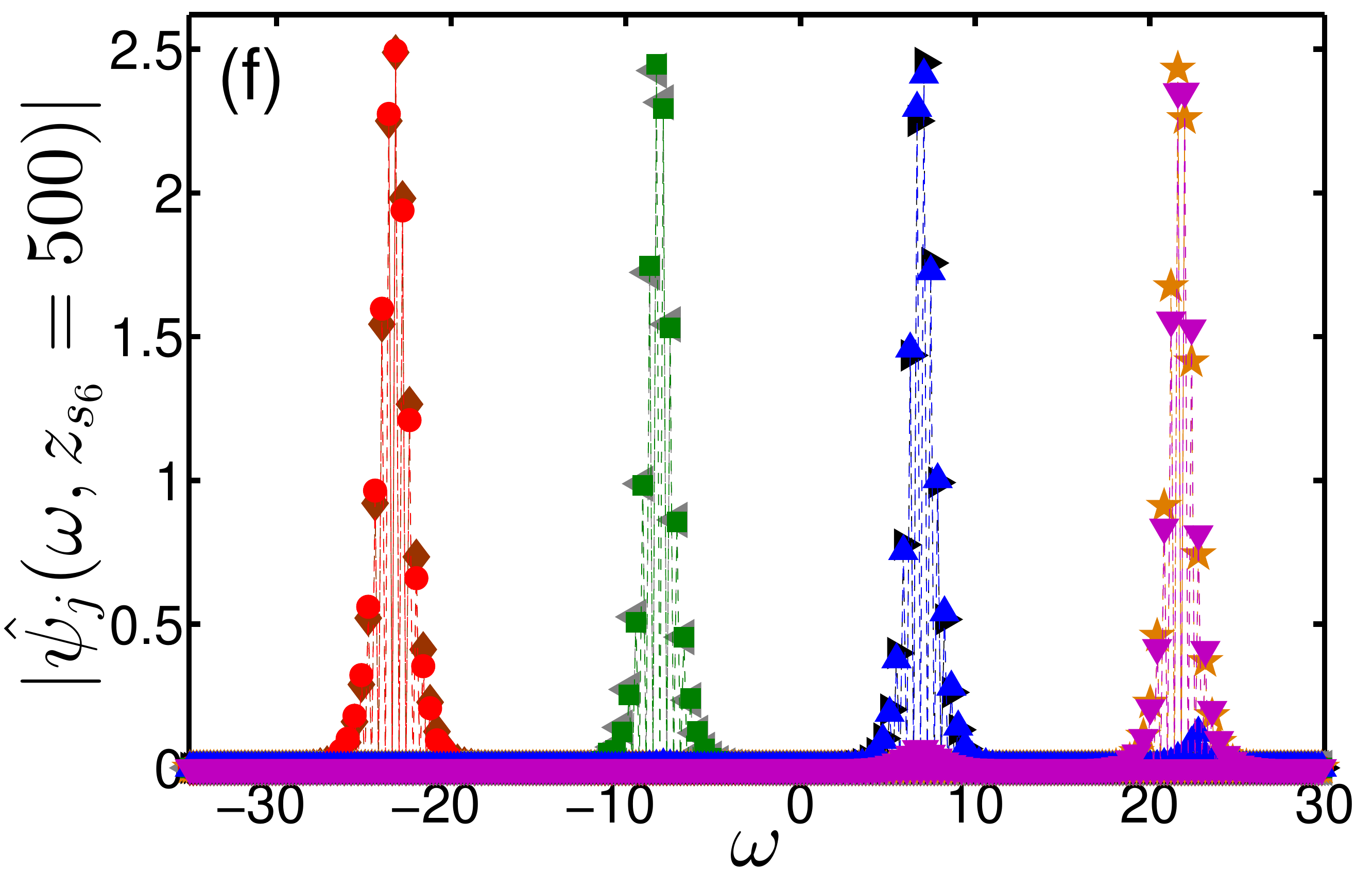} \\
\end{tabular}
\caption{(Color online) The pulse patterns at the onset of transmission instability  
$|\psi_{j}(t,z_{s})|$ and their Fourier transforms  
$|\hat{\psi_{j}}(\omega,z_{s})|$ for two-channel [(a)-(b)], 
three-channel [(c)-(d)], and four-channel [(e)-(f)] transmission 
in the presence of delayed Raman response and the 
linear gain-loss (\ref{raman2}). 
The physical parameter values are listed in rows 1, 5, and 8 of Table 1. 
The stable transmission distances are 
$z_{s_4}=950$ for $N=2$, $z_{s_5}=620$ for $N=3$, 
and $z_{s_6}=500$ for $N=4$.
The solid-crossed red curve [solid red curve in (a)], dashed green curve, 
solid blue curve, and dashed-dotted magenta curve represent 
$|\psi_{j}(t,z_{s})|$ with $j=1,2,3,4$, obtained by 
numerical simulations with Eqs. (\ref{raman1}) and (\ref{raman2}). 
The red circles, green squares, blue up-pointing triangles, 
and magenta down-pointing triangles represent 
$|\hat{\psi_{j}}(t,z_{s})|$ with $j=1,2,3,4$, obtained by the simulations. 
The brown diamonds, gray left-pointing triangles, black right-pointing triangles, 
and orange stars represent the theoretical prediction for $|\psi_{j}(t,z_{s})|$ 
or $|\hat{\psi_{j}}(\omega,z_{s})|$ with $j=1,2,3,4$, respectively.}
 \label{fig4}
\end{figure}

\begin{figure}[ptb]
\begin{tabular}{cc}
\epsfxsize=5.8cm  \epsffile{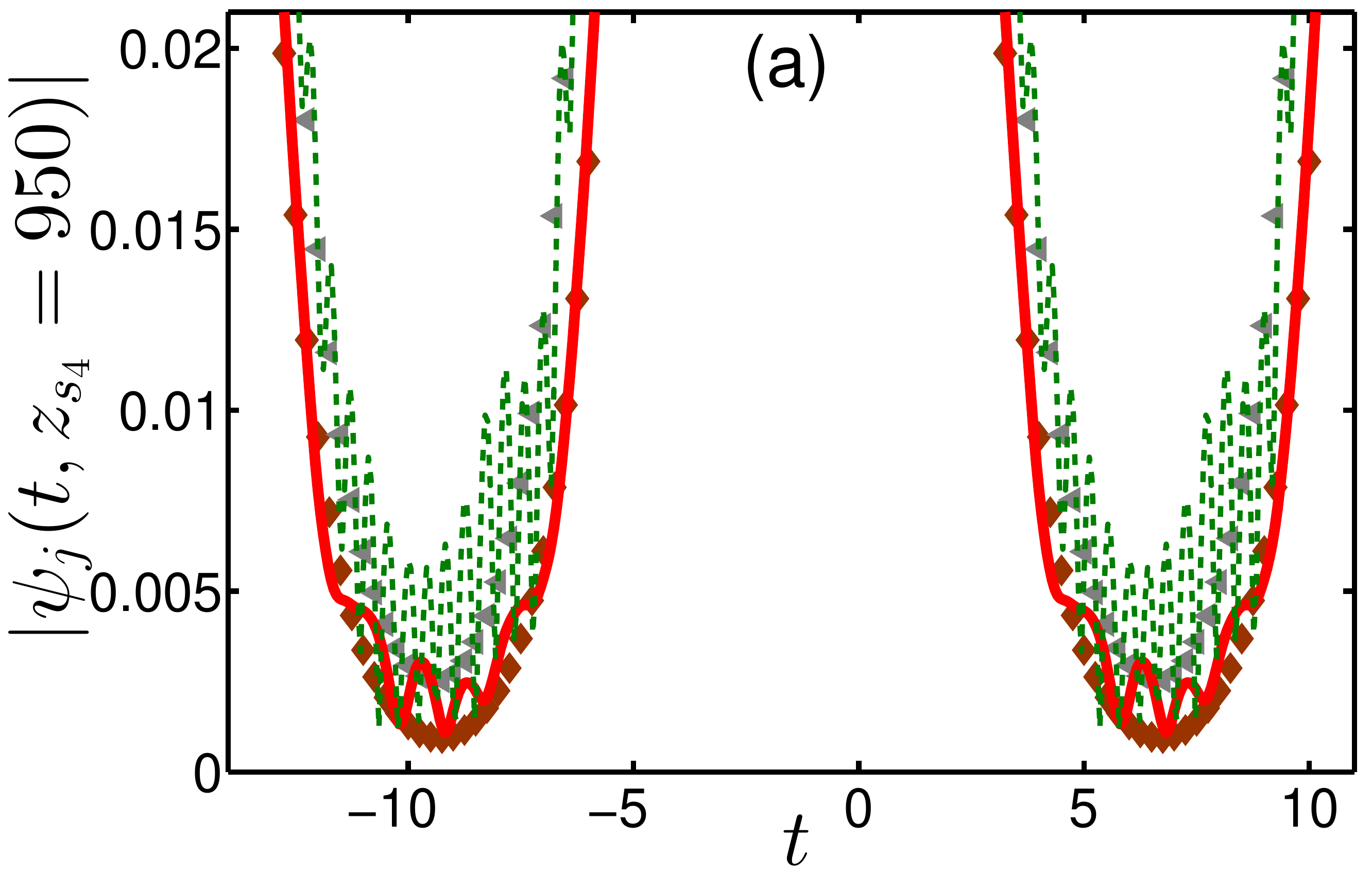} &
\epsfxsize=5.8cm  \epsffile{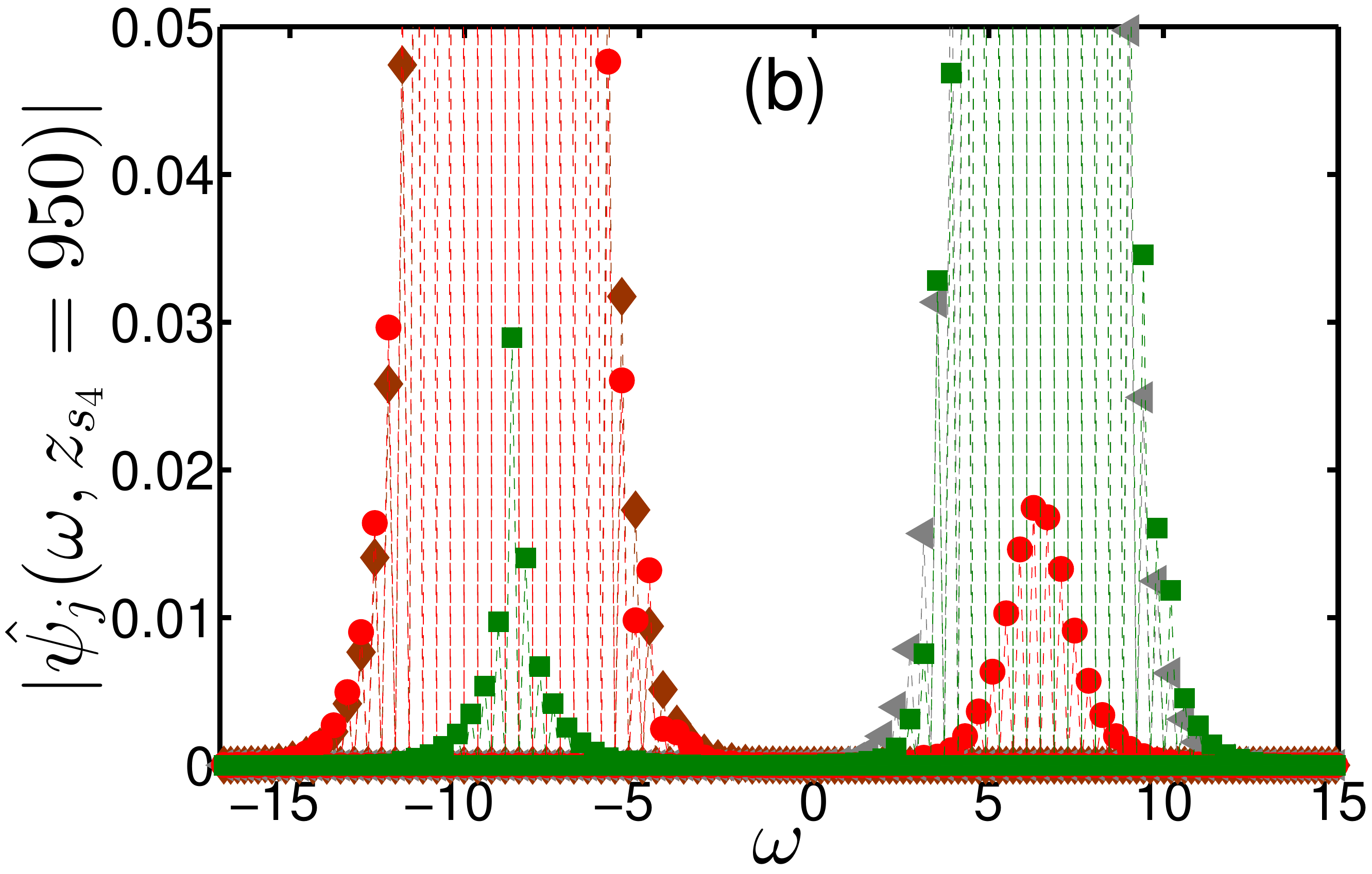} \\
\epsfxsize=5.8cm  \epsffile{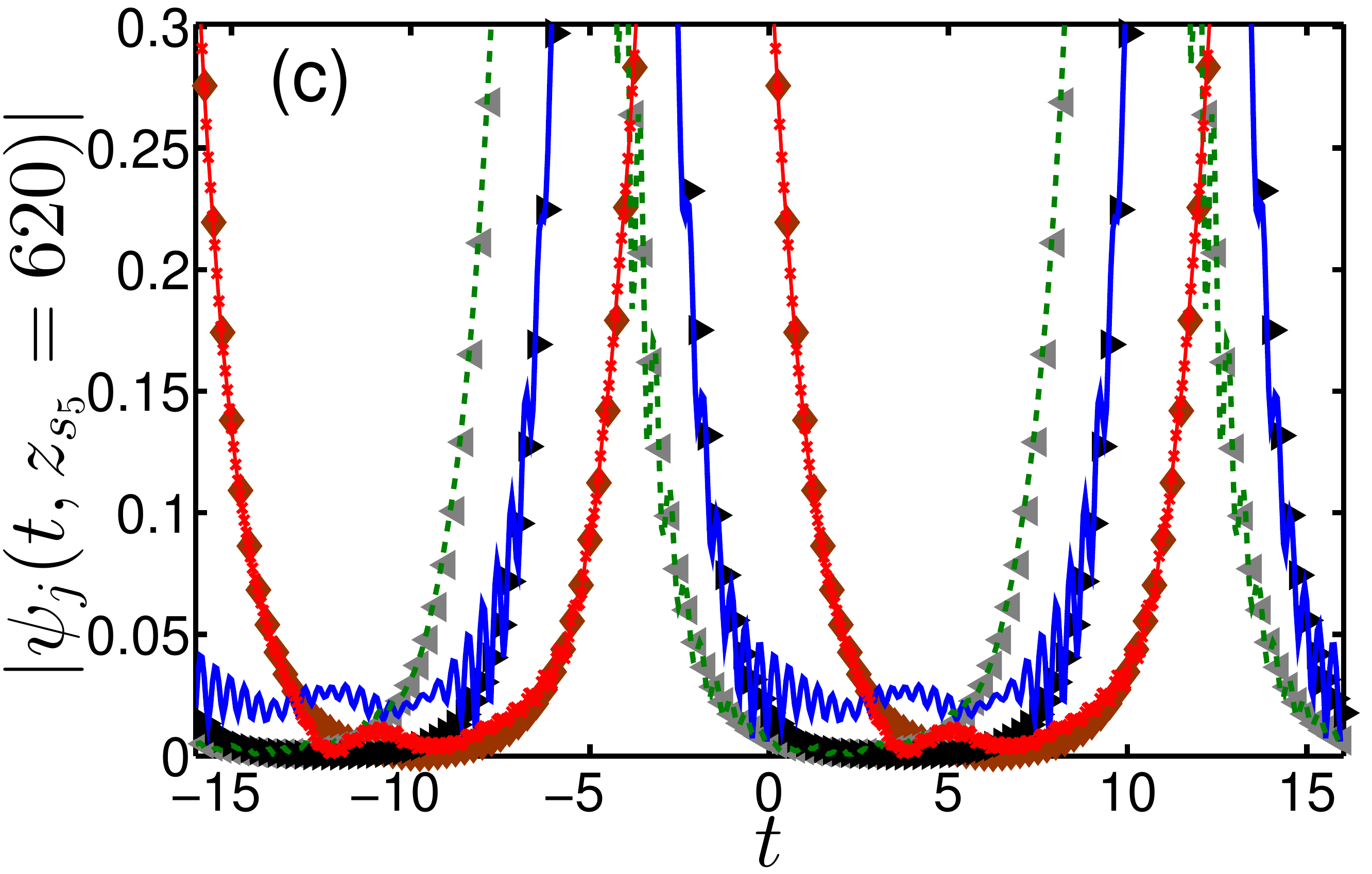}&
\epsfxsize=5.8cm  \epsffile{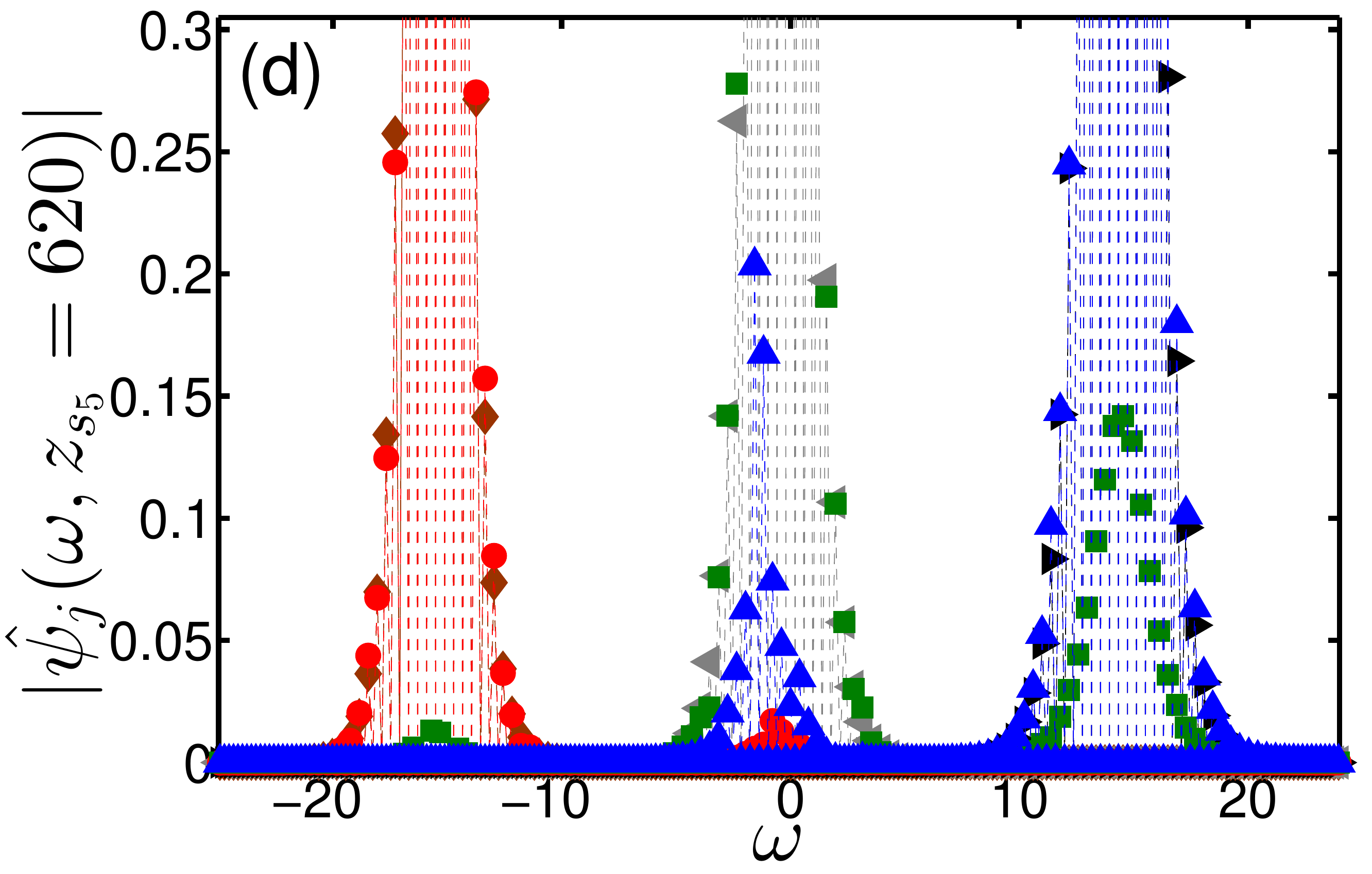} \\
\epsfxsize=5.8cm  \epsffile{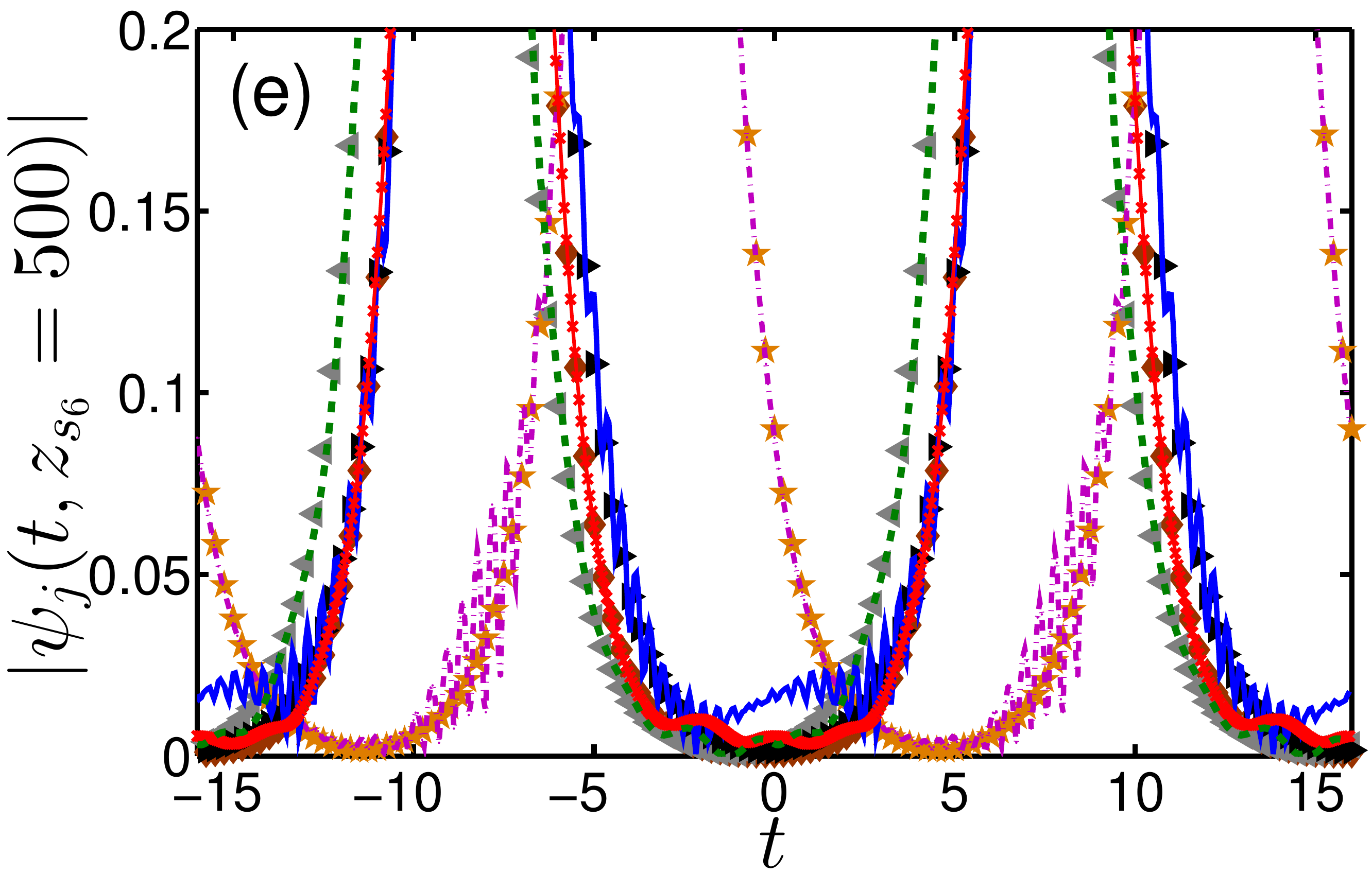}&
\epsfxsize=5.8cm  \epsffile{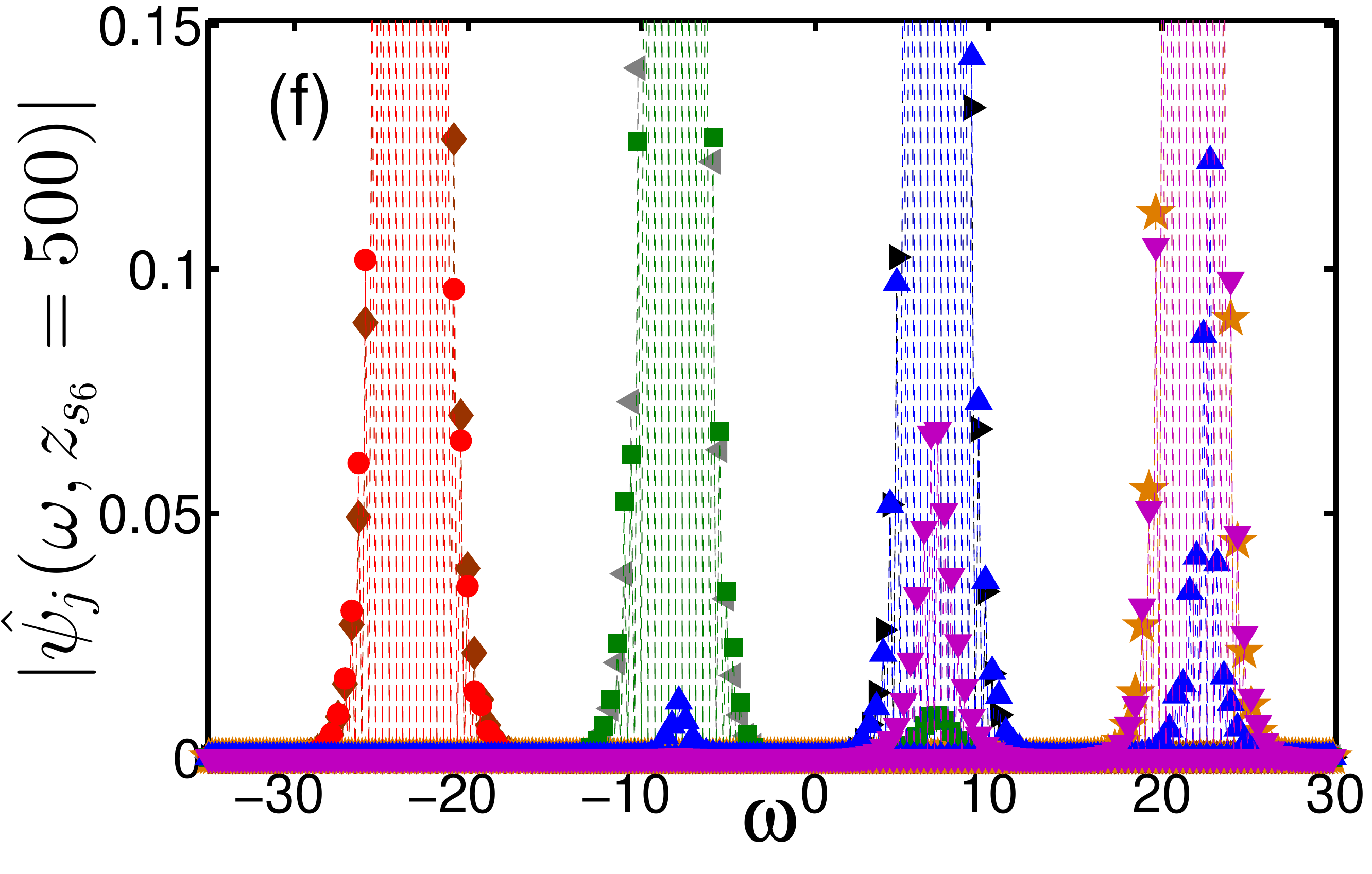} \\
\end{tabular}
\caption{(Color online) Magnified versions of the graphs in Fig. \ref{fig4} for 
small $|\psi_{j}(t,z_{s})|$ and $|\hat{\psi_{j}}(\omega,z_{s})|$ values. 
The symbols are the same as in Fig. \ref{fig4}.}
 \label{mag_fig4}
\end{figure}

Further insight about transmission stability and about the processes 
leading to transmission destabilization is gained by an analysis of the 
soliton patterns at the onset of instability.   
Figure \ref{fig4} shows the pulse patterns at the onset of instability 
$|\psi_{j}(t,z_{s})|$ and their Fourier transforms 
$|\hat{\psi_{j}}(\omega,z_{s})|$, obtained by the numerical simulations 
that are described in the preceding paragraph. 
The theoretical predictions for the pulse patterns and their Fourier transforms, 
which are calculated in the same manner as in Fig. \ref{fig2}, 
are also shown. In addition, Fig. \ref{mag_fig4} shows magnified versions 
of the graphs in Fig. \ref{fig4} for small $|\psi_{j}(t,z_{s})|$ 
and $|\hat{\psi_{j}}(\omega,z_{s})|$ values. 
We observe that the soliton patterns are almost intact at $z_{s}$. 
Based on this observation and the observation that dynamics 
of soliton amplitudes is stable for $0 \le z \le z_{s}$ we conclude 
that the multichannel soliton-based transmission is stable at distances 
smaller than $z_{s}$.  
However, as seen in Figs. \ref{mag_fig4}(a), \ref{mag_fig4}(c), and \ref{mag_fig4}(e), 
the soliton patterns are actually slightly distorted at $z_{s}$, 
and the distortion appears as fast oscillations in the solitons tails. 
Moreover, as seen in Figs. \ref{mag_fig4}(b), \ref{mag_fig4}(d), and \ref{mag_fig4}(f), 
pulse-pattern distortion is caused by resonant formation of radiative sidebands, where the 
largest sidebands for the $j$th sequence form near  
the frequencies $\beta_{j-1}(z)$ and/or $\beta_{j+1}(z)$ of the 
neighboring soliton sequences.       
Thus, the mechanisms leading to deterioration of the soliton sequences 
in multichannel transmission in a single fiber  
in the presence of delayed Raman response 
and linear gain-loss are very similar to the ones observed in the absence 
of delayed Raman response and linear gain-loss. This also indicates that 
the dominant cause for transmission destabilization in the full coupled-NLS 
simulations for multichannel transmission in a single fiber  
is due to the effects of Kerr-induced interaction in interchannel 
soliton collisions. As explained earlier, the latter effects are 
represented by the $2|\psi_{k}|^2\psi_{j}$ terms in Eq. (\ref{raman1}).

It is interesting to note that the stable propagation distances $z_{s}$ 
in the presence of delayed Raman response and linear gain-loss are 
larger compared with the distances obtained in the absence of these 
two processes by factors of 1.7 for $N=2$, 1.2 for $N=3$, 
and 1.5 for $N=4$. We attribute this moderate increase in $z_{s}$ values to the 
introduction of frequency dependent linear gain-loss 
with strong loss $g_{L}$ outside the frequency intervals 
$\beta_{j}(0)- W/2 < \omega \le \beta_{j}(0)+ W/2$, where $1 \le j \le N$,  
which leads to partial suppression of radiative sideband generation. 
However, the suppression of radiative instability in a single fiber is quite limited, 
since the radiative sidebands for a given sequence form near the frequencies 
$\beta_{k}(z)$ of the other soliton sequences. As a result, in a single fiber, 
one cannot employ strong loss at the latter frequencies, as this would lead 
to the decay of the propagating solitons. Better suppression of radiative instability 
and significantly larger $z_{s}$ values can be realized in nonlinear waveguide 
couplers with frequency dependent linear gain-loss. 
This subject is discussed in detail in Section \ref{coupler}.

\begin{figure}[ptb]
\begin{tabular}{cc}
\epsfxsize=5.8cm  \epsffile{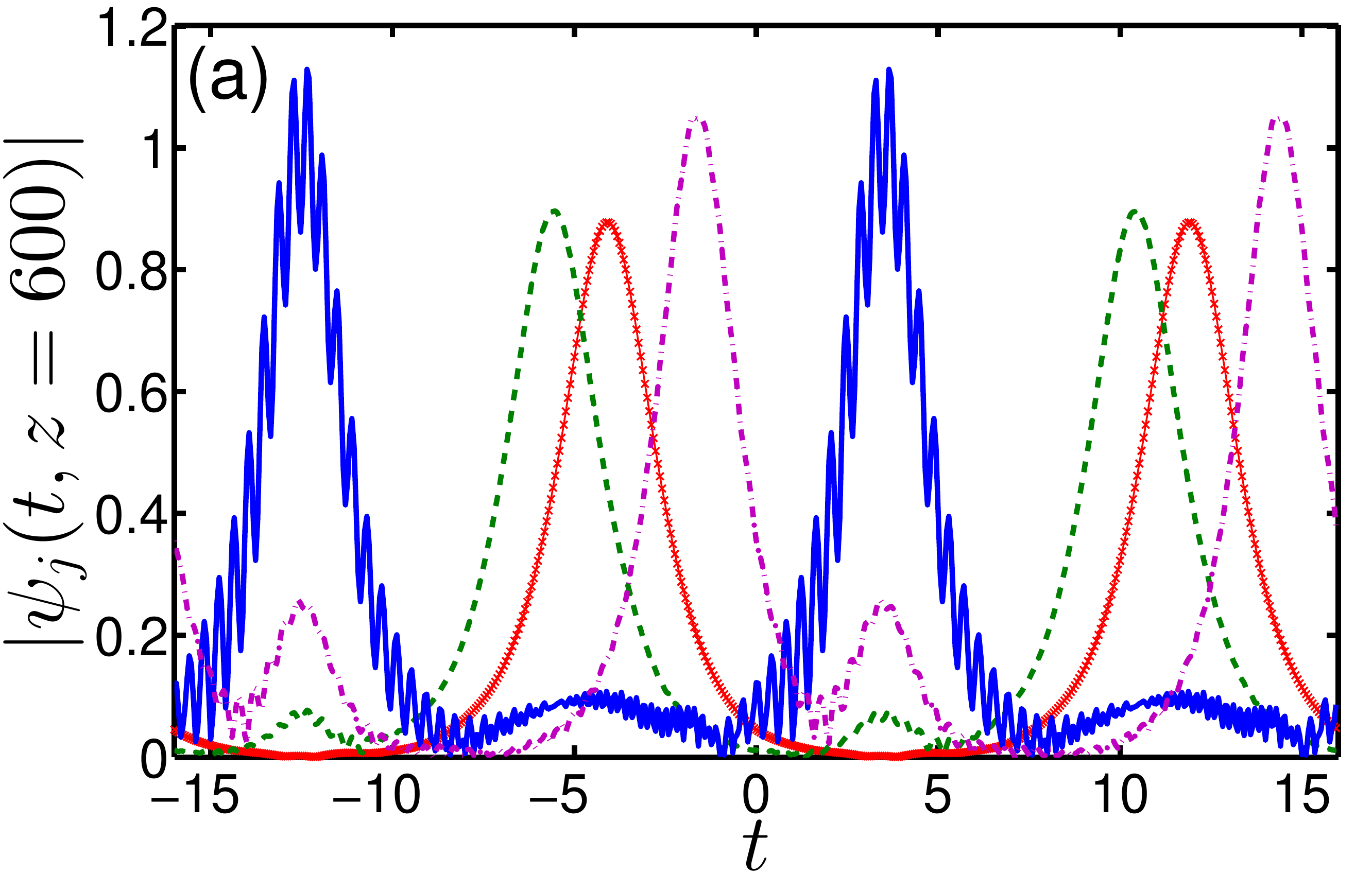} &
\epsfxsize=5.8cm  \epsffile{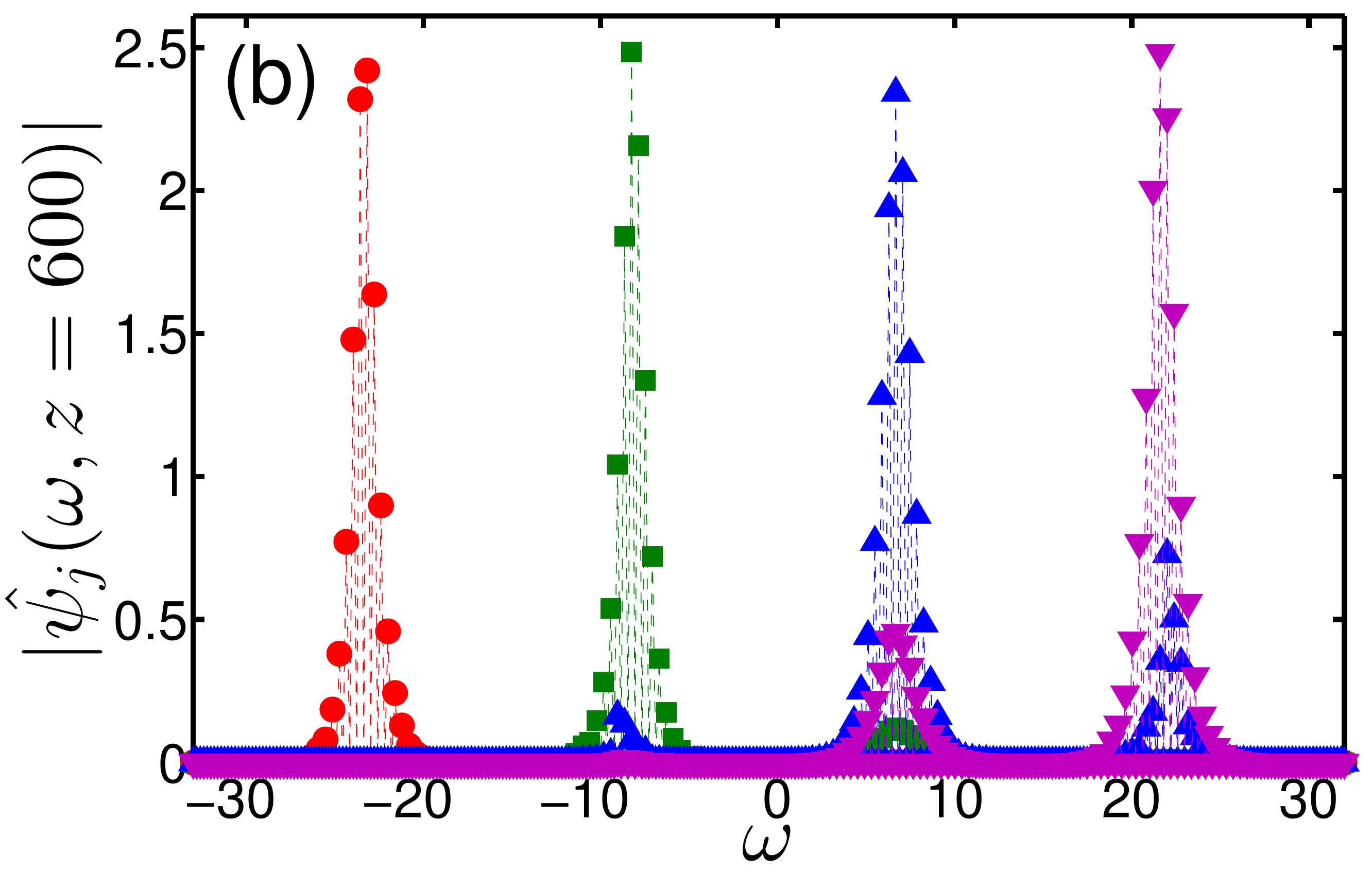} \\
\epsfxsize=5.8cm  \epsffile{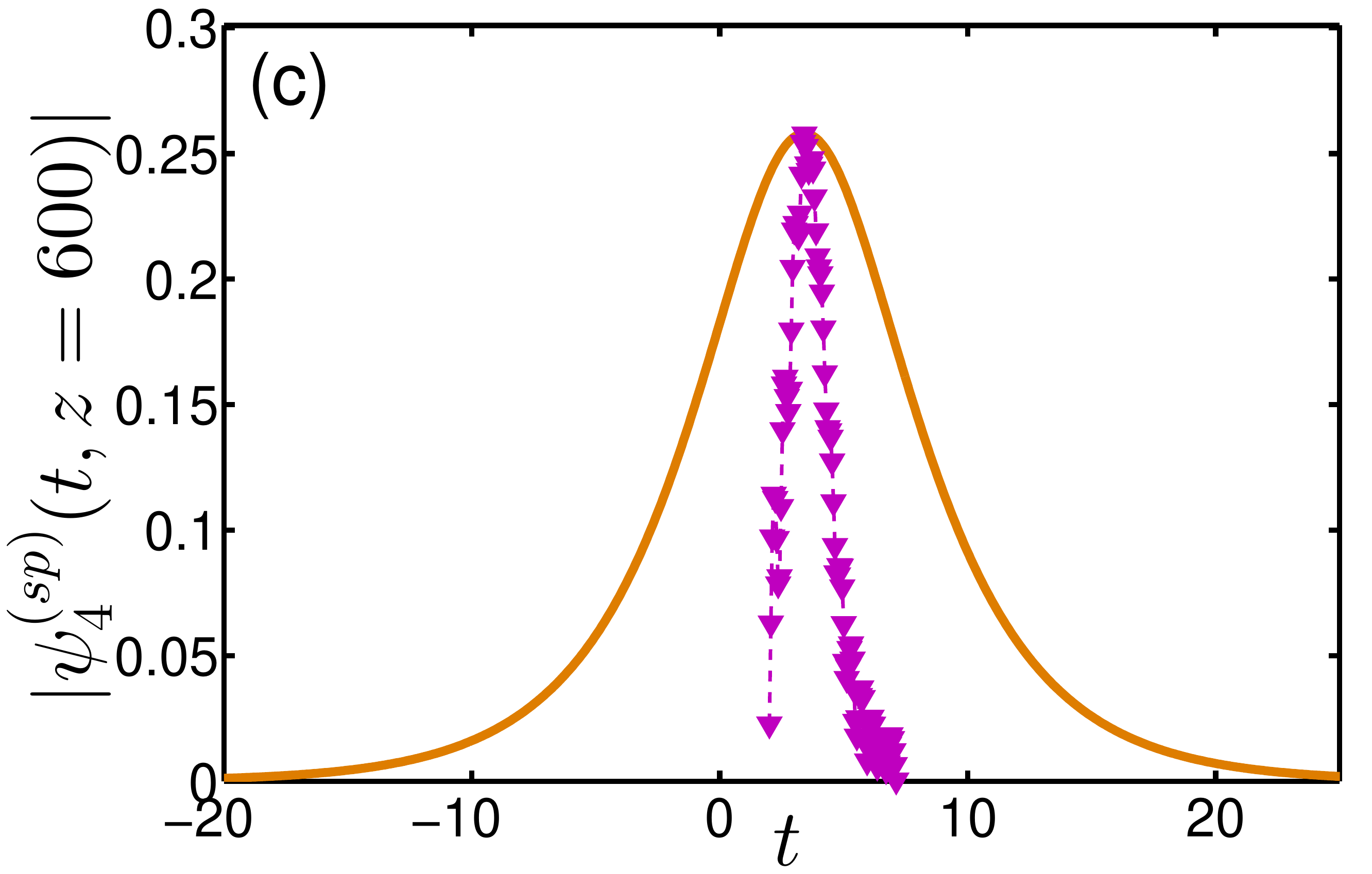} &
\epsfxsize=5.8cm  \epsffile{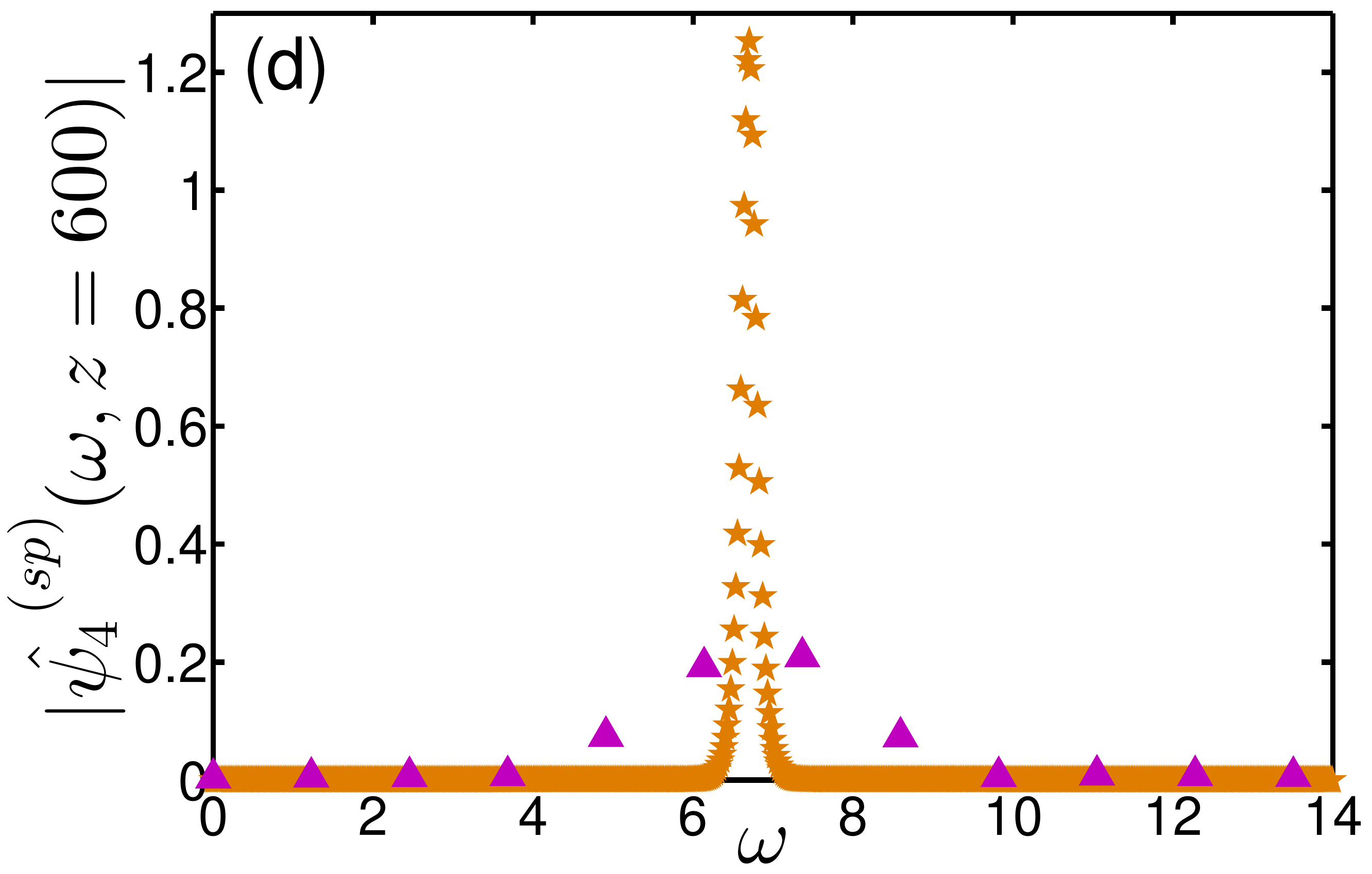} \\ 
\epsfxsize=5.8cm  \epsffile{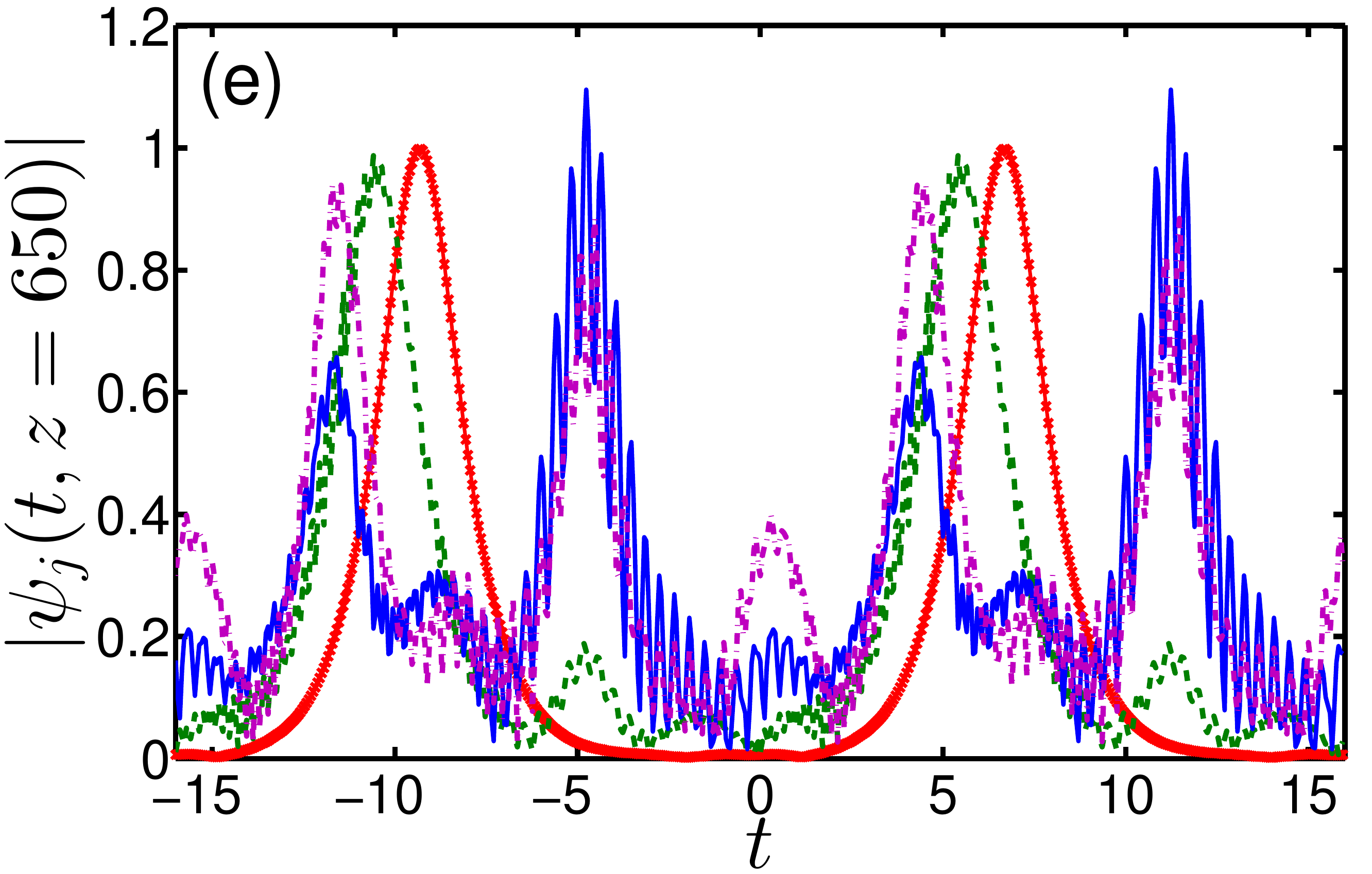} &
\epsfxsize=5.8cm  \epsffile{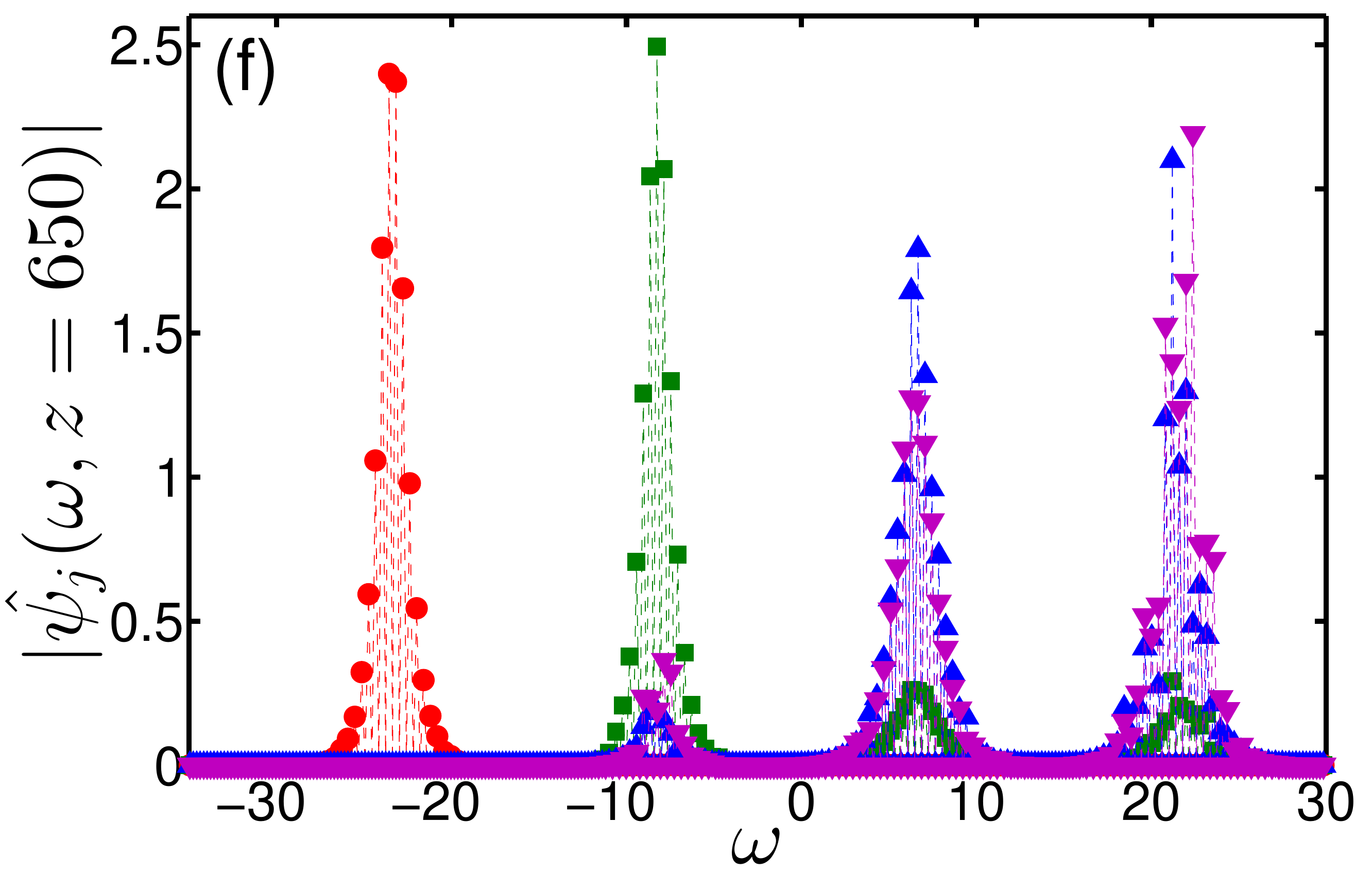} \\  
\end{tabular}
\caption{(Color online)  
The pulse patterns $|\psi_{j}(t,z)|$ and their Fourier transforms 
$|\hat{\psi_{j}}(\omega,z)|$ at $z=600$ [(a) and (b)] 
and at $z=650$ [(e) and (f)], obtained by numerical simulations  
with Eqs. (\ref{raman1}) and (\ref{raman2}) 
for the four-channel system considered in Figs. \ref{fig3} and \ref{fig4}. 
The symbols in (a), (b), (e), and (f) are the same as in Fig. \ref{fig4}.  
(c) The shape of a radiation-induced pulse $|\psi_{4}^{(sp)}(t,z)|$,   
generated due to radiation emitted by the $j=4$ soliton sequence 
at $z=600$. The magenta down-pointing triangles represent 
the numerically obtained $|\psi_{4}^{(sp)}(t,z)|$ with $z=600$, 
while the solid orange curve corresponds to the shape 
of a single NLS soliton with the same amplitude.
(d) The Fourier transform of the radiation-induced pulse in (c). 
The magenta down-pointing triangles represent 
the numerically obtained $|\hat\psi_{4}^{(sp)}(\omega,z)|$ with $z=600$, 
while the solid orange curve corresponds to the Fourier transform 
of a single NLS soliton with the same amplitude and central frequency.} 
\label{fig6}
\end{figure}

We now turn to discuss the later stages of pulse pattern deterioration, 
i.e., the evolution of the soliton sequences in a single fiber 
for distances $z> z_{s}$. As a concrete example, we discuss 
the four-channel setup considered in Figs. \ref{fig3}(c), 
\ref{fig4}(e), and \ref{fig4}(f), for which $z_{s_{6}}=500$ \cite{parameter_values}. 
Figures \ref{fig6}(a) and \ref{fig6}(b) show the  
pulse patterns $|\psi_{j}(t,z)|$ and their Fourier transforms  
$|\hat{\psi_{j}}(\omega,z)|$ at $z=600>z_{s_{6}}$,    
as obtained by numerical solution of 
Eqs. (\ref{raman1}) and (\ref{raman2}). 
It is seen that the largest sidebands at $z=600$ form near the frequency 
$\beta_{3}(z)$ for the $j=2$ and $j=4$ soliton sequences, 
and near the frequencies $\beta_{2}(z)$ and $\beta_{4}(z)$ 
for the $j=3$ soliton sequence. These larger sidebands 
lead to significantly stronger pulse distortion at $z=600$ compared with 
$z_{s_{6}}=500$. In particular, at $z=600$, the $j=3$ pulse sequence 
is strongly distorted, where the distortion is in the form of fast oscillations 
in the main body of the solitons. In contrast, at $z_{s_{6}}=500$, 
the $j=3$ sequence is only weakly distorted, and the distortion 
is in the form of fast oscillations, which are significant only in the solitons tails. 
Additionally, radiation emitted by the solitons in the $j=2$, $j=3$, and $j=4$ 
frequency channels develops into small pulses at $z=600$. The largest radiation-induced 
pulses are generated due to radiation emitted by solitons in the 
$j=4$ channel near the frequency $\beta_{3}(z)$.  
Figures \ref{fig6}(c) and \ref{fig6}(d) show a comparison of the shape 
and Fourier transform of the latter pulses with the shape and 
Fourier transform expected for a single NLS soliton with the same 
amplitude and frequency. It is clear that these radiation-induced 
pulses do not posses the soliton form. Similar conclusion holds 
for the other radiation-induced pulses.   
The amplitudes of the radiative sidebands generated by the 
$j=2$, $j=3$, and $j=4$ pulse sequences continue to increase 
with increasing propagation distance and this leads to further 
pulse pattern degradation. Indeed, as seen in Fig. \ref{fig6}(f), 
at $z=650$, the radiative sidebands generated by the $j=4$ sequence  
near $\beta_{3}(z)$ and by the $j=3$ sequence near $\beta_{4}(z)$      
are comparable in magnitude to the Fourier transforms of the 
$j=3$ and $j=4$ pulse sequences, respectively. Additional strong radiative 
sidebands are observed for the $j=2$ sequence near frequencies 
$\beta_{3}(z)$ and $\beta_{4}(z)$ and for the $j=4$ sequence 
near frequency $\beta_{2}(z)$.  
As a result, the $j=2$, $j=3$, and $j=4$ pulse sequences 
are strongly degraded due to pulse distortion at $z=650$. 
More specifically, distortion due to fast oscillations in both the 
main body and the tail of the pulses is observed for these 
three pulse sequences [see  Fig. \ref{fig6}(e)]. 
In addition, the number and amplitudes of the radiation-induced 
pulses are much larger at $z=650$ compared with the corresponding 
number and amplitudes of these pulses at $z=600$.

\section{Nonlinear waveguide coupler transmission}
\label{coupler}
The results of the numerical simulations in Section \ref{simu} show that in a single fiber, 
radiative instabilities can be partially mitigated by employing the frequency dependent 
linear gain-loss (\ref{raman2}). However, as described in Section \ref{simu},  
suppression of radiation emission in a single fiber 
is still quite limited, and generation of radiative sidebands leads 
to severe pulse pattern degradation at large distances. 
The limitation of the single-fiber setup is explained by noting that 
the radiative sidebands for each pulse sequence form near the frequencies 
$\beta_{k}(z)$ of the other pulse sequences. As a result, in a single fiber, 
one cannot employ strong loss at or near the frequencies $\beta_{k}(z)$, 
as this would lead to the decay of the propagating pulses.
It is therefore interesting to look for other waveguide setups that 
can significantly enhance transmission stability. A very promising 
approach for enhancing transmission stability is based on employing a nonlinear waveguide coupler, 
consisting of $N$ very close waveguides \cite{CPN2016}. 
In this case each pulse sequence propagates through its own waveguide 
and each waveguide is characterized by its own frequency dependent linear 
gain-loss function $\tilde g_{j}(\omega,z)$ \cite{linear_gain_loss}. 
This enables better suppression of radiation emission, 
since the linear gain-loss of each waveguide can be set equal to the required $g_{j}$ value   
within a certain $z$-dependent bandwidth $(\beta_{j}(z)-W/2, \beta_{j}(z)+W/2]$
around the central frequency $\beta_{j}(z)$ of the solitons in that waveguide, 
and equal to a relatively large negative value $g_{L}$ outside of that bandwidth. 
This leads to enhancement of transmission stability compared with the 
single fiber setup, since generation of all radiative sidebands outside of 
the interval $(\beta_{j}(z)-W/2, \beta_{j}(z)+W/2]$ 
is suppressed by the relatively strong linear loss $g_{L}$.

In the current section, we investigate the possibility to significantly enhance 
transmission stability in multichannel soliton-based systems 
by employing $N$-waveguide couplers with frequency dependent linear gain-loss. 
The enhanced transmission stability is also expected to enable observation 
of the stable oscillatory dynamics of soliton amplitudes, predicted by the 
predator-prey model (\ref{raman5}), along significantly larger distances 
compared with the distances observed in single-fiber transmission.    
Similar to the single-fiber setup considered in 
Section \ref{simu}, we take into account the effects of second-order dispersion,  
Kerr nonlinearity, delyaed Raman response, and linear gain-loss. 
The main difference between the waveguide coupler setup and the 
single-fiber setup is that the single linear gain-loss function  $\tilde g(\omega)$
of Eq. (\ref{raman2}) is now replaced by $N$ $z$-dependent linear gain-loss 
functions $\tilde g_{j}(\omega,z)$, where $1 \le j \le N$.  
Thus, the propagation of the pulse sequences through the waveguide coupler 
is described by the following coupled-NLS model:   
\begin{eqnarray} &&
i\partial_z\psi_{j}+\partial_{t}^2\psi_{j}+2|\psi_{j}|^2\psi_{j}
+4\sum_{k=1}^{N}(1-\delta_{jk})|\psi_{k}|^2\psi_{j}=
i{\cal F}^{-1}(\tilde g_{j}(\omega,z) \hat\psi_{j})/2
\nonumber \\&&
-\epsilon_{R}\psi_{j}\partial_{t}|\psi_{j}|^2 
-\epsilon_{R}\sum_{k=1}^{N}(1-\delta_{jk})
\left[\psi_{j}\partial_{t}|\psi_{k}|^2
+\psi_{k}\partial_{t}(\psi_{j}\psi_{k}^{\ast})\right],  
\label{raman11}
\end{eqnarray}     
where $1 \le j \le N$. The linear gain-loss function of the $j$th waveguide $\tilde g_{j}(\omega,z)$, 
appearing on the right-hand side of Eq. (\ref{raman11}), is defined by: 
\begin{eqnarray} &&
\tilde g_{j}(\omega,z) = \left\{ 
\begin{array}{l l}
g_{j} &  \mbox{ if $\beta_{j}(z)-W/2 < \omega \le \beta_{j}(z)+W/2,$}\\
g_{L} &  \mbox{ if $\omega \le \beta_{j}(z) - W/2$, 
or $\omega > \beta_{j}(z) + W/2,$}\\
\end{array} \right. 
\label{raman13}
\end{eqnarray}    
where the $g_{j}$ coefficients are determined by Eq. (\ref{raman4}), 
the $z$ dependence of the frequencies $\beta_{j}(z)$ is determined 
from the numerical solution of the coupled-NLS 
model (\ref{raman11}), and $g_{L}<0$. 
Notice the following important properties of the gain-loss  (\ref{raman13}).  
First, the gain-loss $g_{j}$ inside the central frequency interval 
$(\beta_{j}(z)-W/2, \beta_{j}(z)+W/2]$ is expected to compensate 
for amplitude shifts due to Raman crosstalk and by this, lead to  
stable oscillatory dynamics of soliton amplitudes. 
Second, the relatively strong linear loss $g_{L}$ outside the interval 
$(\beta_{j}(z)-W/2, \beta_{j}(z)+W/2]$ should enable efficient 
suppression of radiative sideband generation for {\it any frequency} 
outside of this interval.
Third, the end points of the central frequency interval are shifting with $z$, 
such that the interval is centered around $\beta_{j}(z)$ throughout the propagation.
This shifting of the central amplification interval is introduced to compensate for 
the significant Raman-induced frequency shifts experienced by the solitons 
during the propagation \cite{Sliding_Filter}.  
The combination of the three properties of $\tilde g_{j}(\omega,z)$ should 
lead to a significant increase of the stable transmission distances in the nonlinear 
$N$-waveguide coupler compared with the single-fiber system considered in Section \ref{simu}.     
As a result, one can expect that the stable oscillatory dynamics of soliton 
amplitudes, predicted by the predator-prey model (\ref{raman5}), 
will also hold along significantly larger distances.

\begin{figure}[ptb]
\begin{tabular}{cc}
\epsfxsize=7.2cm  \epsffile{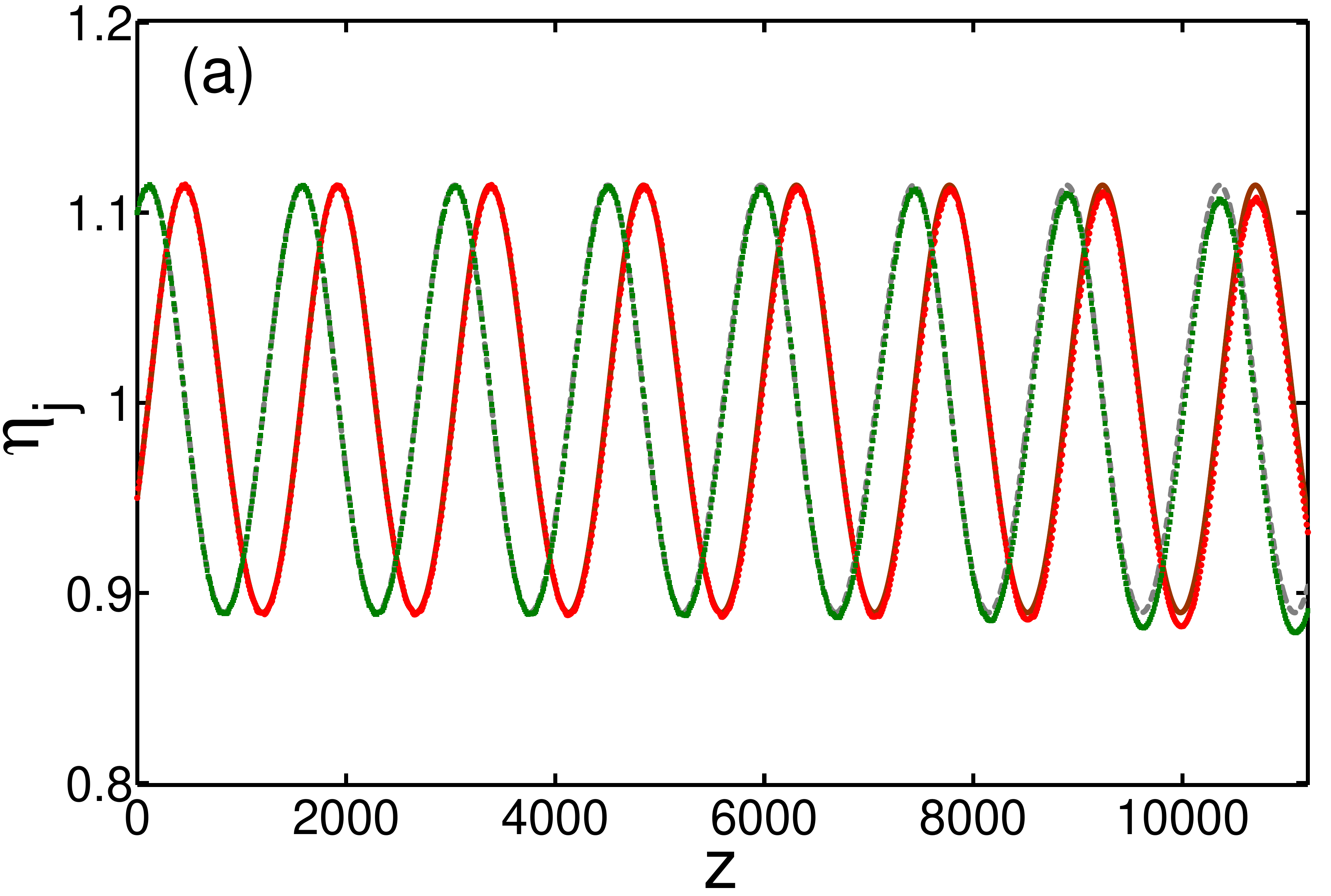} \\
\epsfxsize=7.5cm  \epsffile{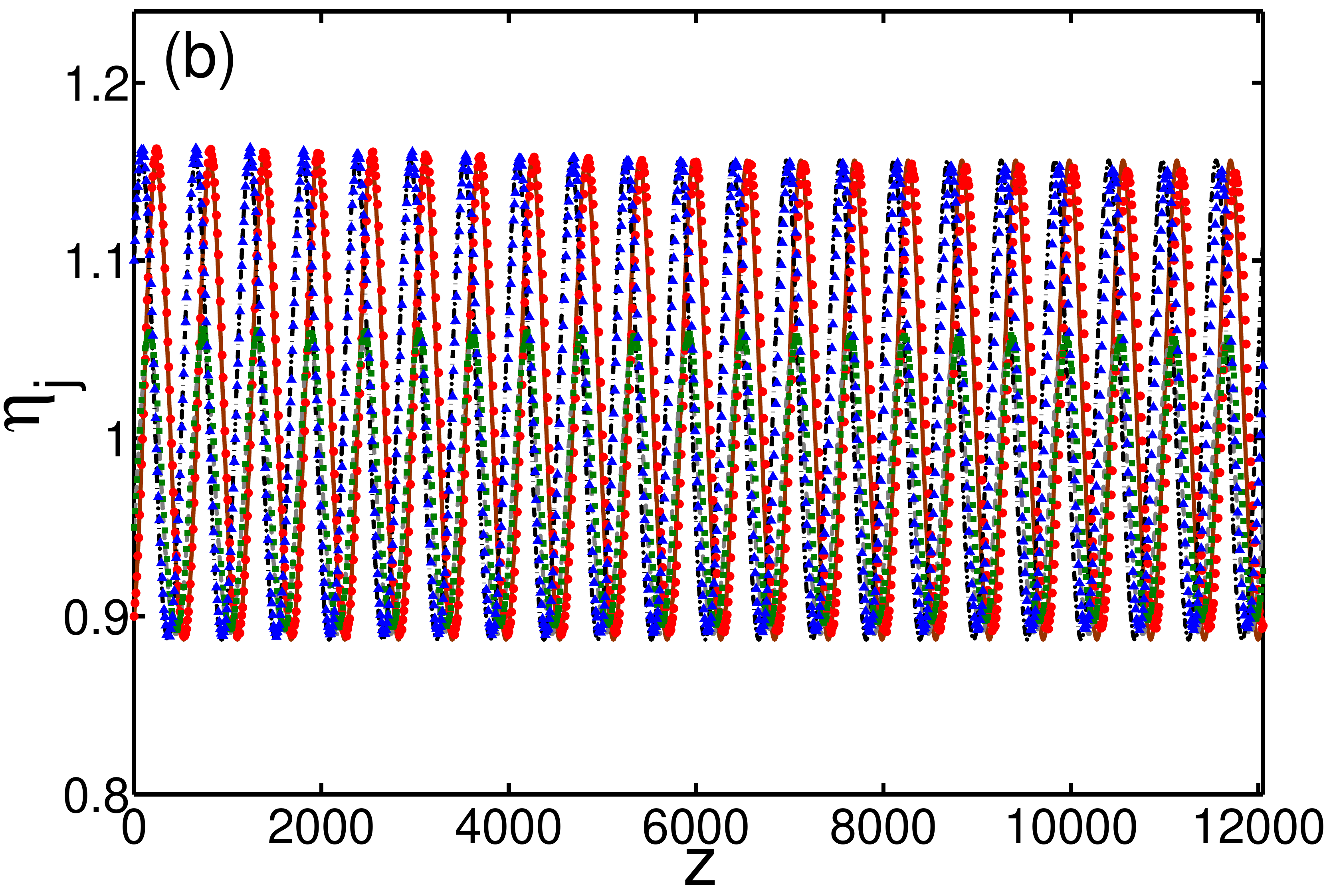}\\
\epsfxsize=7.2cm  \epsffile{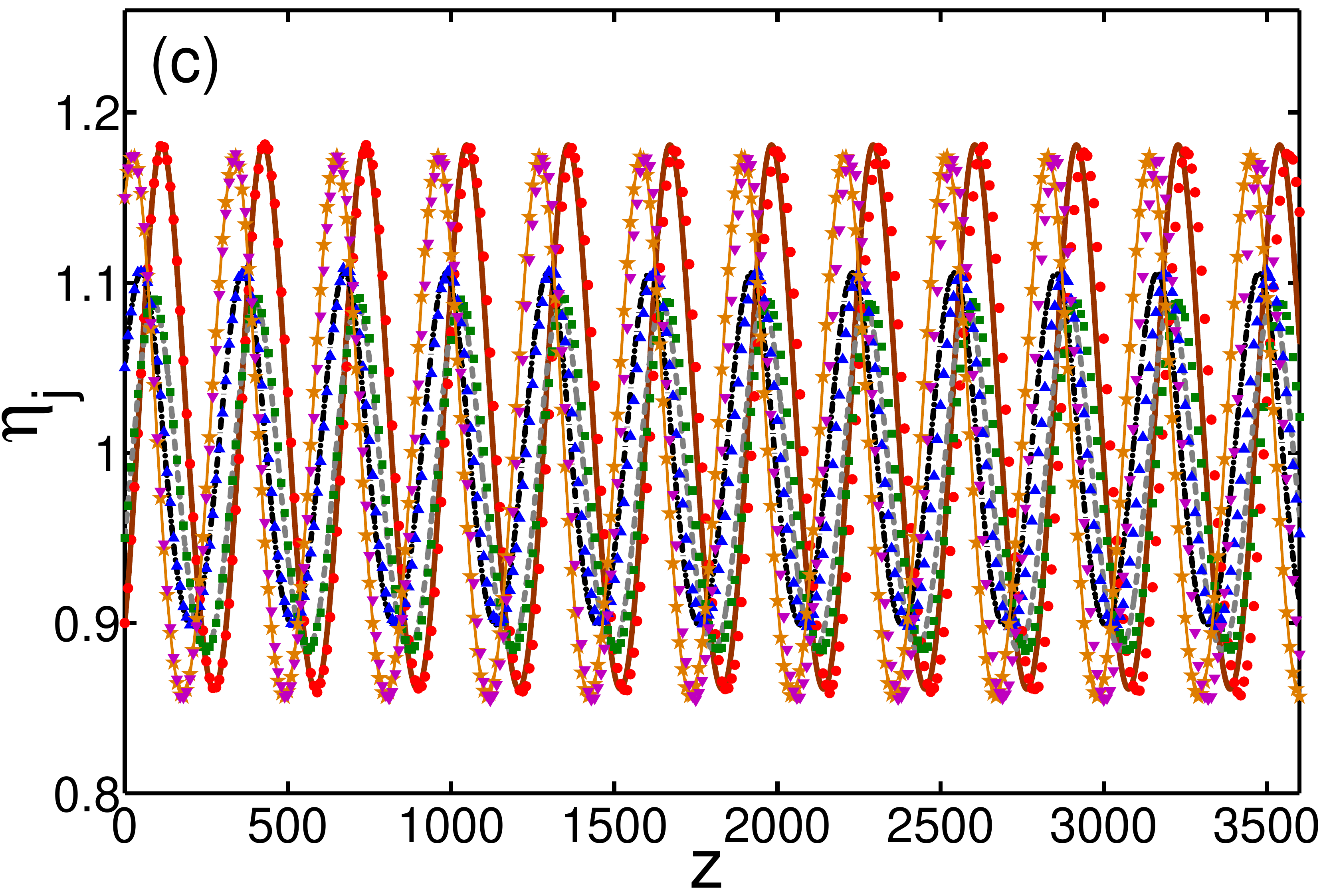}
\end{tabular}
\caption{(Color online) The $z$ dependence of soliton amplitudes $\eta_{j}$ in 
a two-channel (a), a three-channel (b), and a four-channel (c) 
nonlinear waveguide coupler with linear gain-loss (\ref{raman13}). 
The values of the physical parameters are 
the same as the ones used in Fig. \ref{fig3}. 
The red circles, green squares, blue up-pointing triangles, and magenta down-pointing triangles 
represent $\eta_{1}(z)$, $\eta_{2}(z)$, $\eta_{3}(z)$, and $\eta_{4}(z)$ obtained 
by numerical solution of the coupled-NLS model (\ref{raman11}) with the gain-loss (\ref{raman13}). 
The solid brown, dashed gray, dashed-dotted black, and solid-starred orange curves correspond to 
$\eta_{1}(z)$, $\eta_{2}(z)$, $\eta_{3}(z)$, and $\eta_{4}(z)$ obtained 
by the predator-prey model (\ref{raman5}).}
\label{fig7}
\end{figure}

In order to check whether the $N$-waveguide coupler setup leads to enhancement 
of transmission stability, we numerically solve Eq. (\ref{raman11})
with the gain-loss (\ref{raman13}) for two, three, and four channels. 
The comparison with results obtained for single-fiber transmission is enabled by  
using the same values of the physical parameters that were used  
in Figs. \ref{fig3}(a), \ref{fig3}(b) and \ref{fig3}(c). 
The numerical simulations are carried out up to the onset of transmission 
instability, which occurs at $z_{s_7}=11200$ for $N=2$, $z_{s_8}=12050$ for $N=3$, 
and $z_{s_{9}}=3600$ for $N=4$. 
Figure \ref{fig7} shows the $z$ dependence of soliton amplitudes 
as obtained by the coupled-NLS simulations along with the prediction of the predator-prey model (\ref{raman5}). 
It is seen that the amplitudes exhibit stable oscillations about 
the equilibrium value $\eta=1$ for $N=2,3$, and 4. 
Furthermore, the agreement between the coupled-NLS simulations 
and the predictions of the predator-prey model  are excellent throughout 
the propagation. Thus, both transmission stability and the validity of the predator-prey 
model's predictions are extended to distances that are larger by factors of 
11.8 for $N=2$, 19.4 for $N=3$, and 7.2 for $N=4$ compared with the 
distances obtained with the single-fiber WDM system in Section \ref{simu}.

\begin{figure}[ptb]
\begin{tabular}{cc}
\epsfxsize=6.0cm  \epsffile{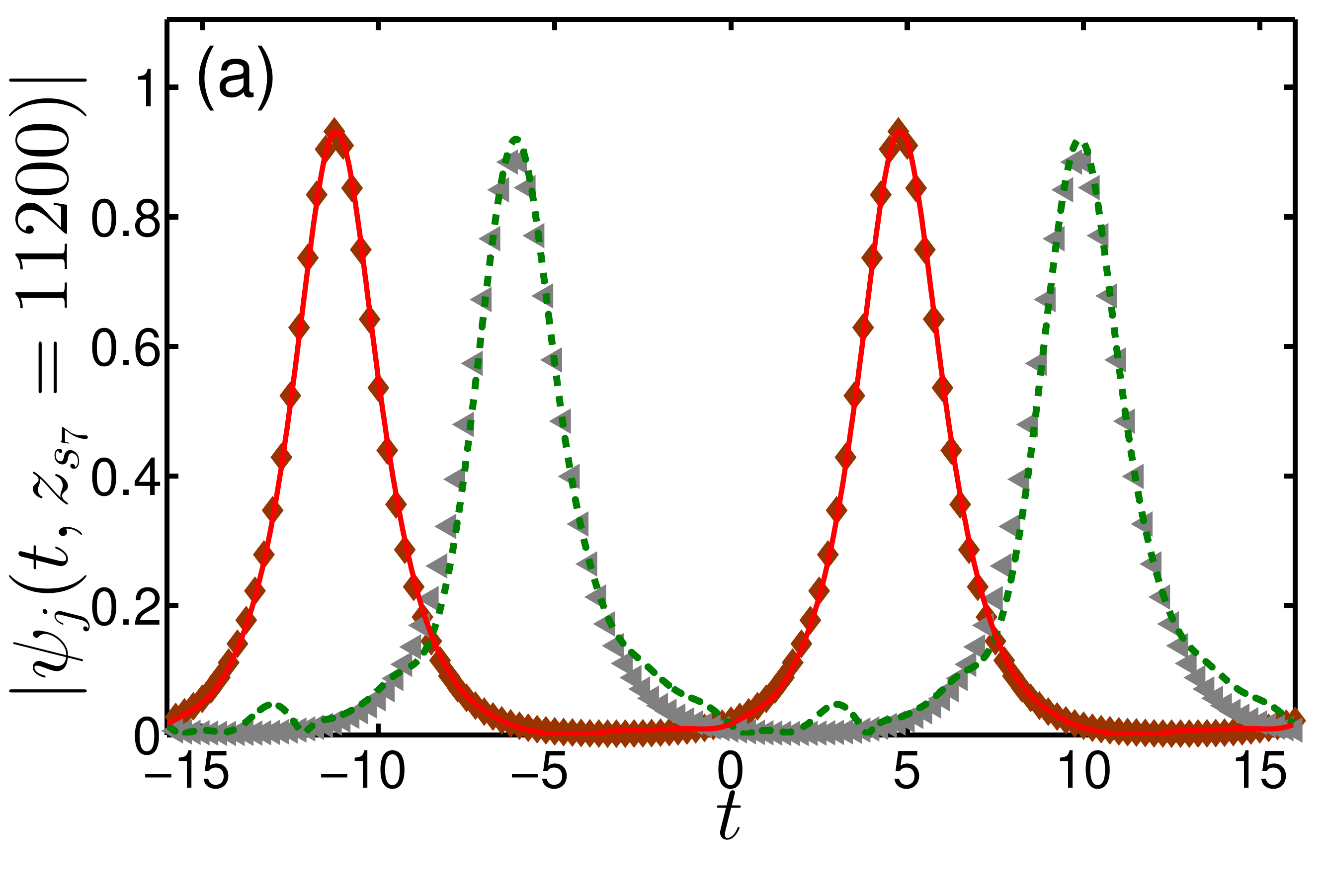} &
\epsfxsize=6.0cm  \epsffile{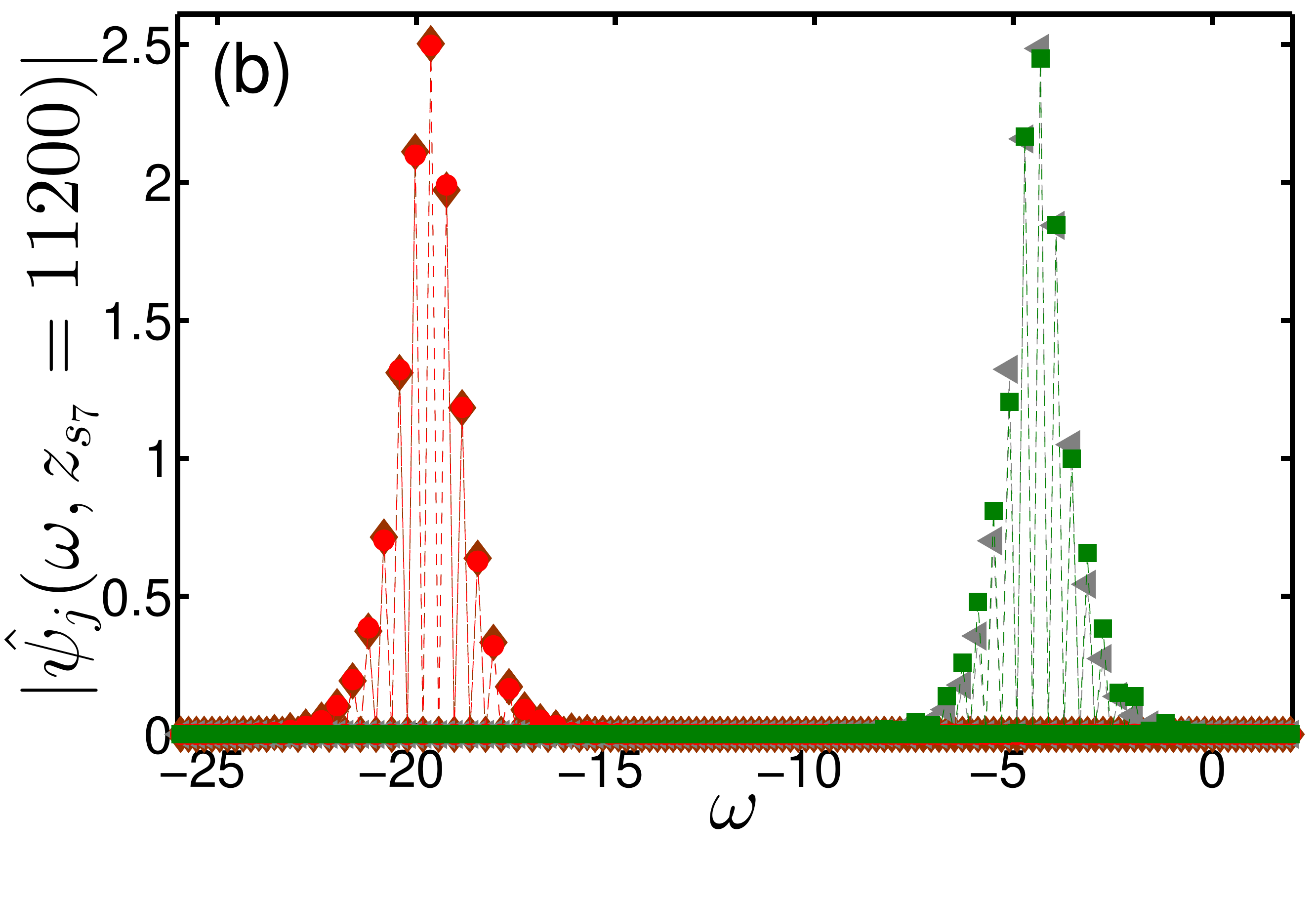} \\
\epsfxsize=6.0cm  \epsffile{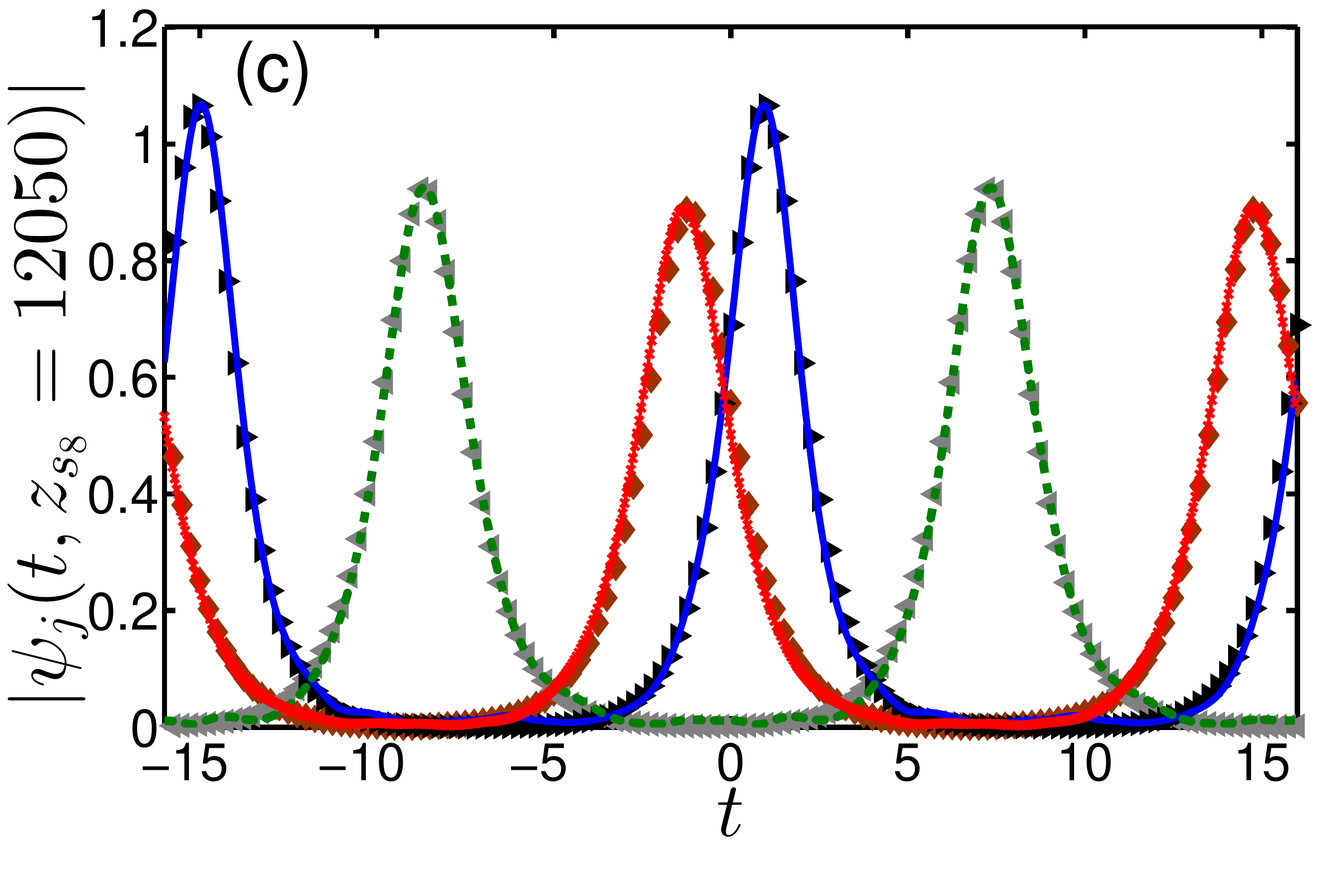} &
\epsfxsize=6.0cm  \epsffile{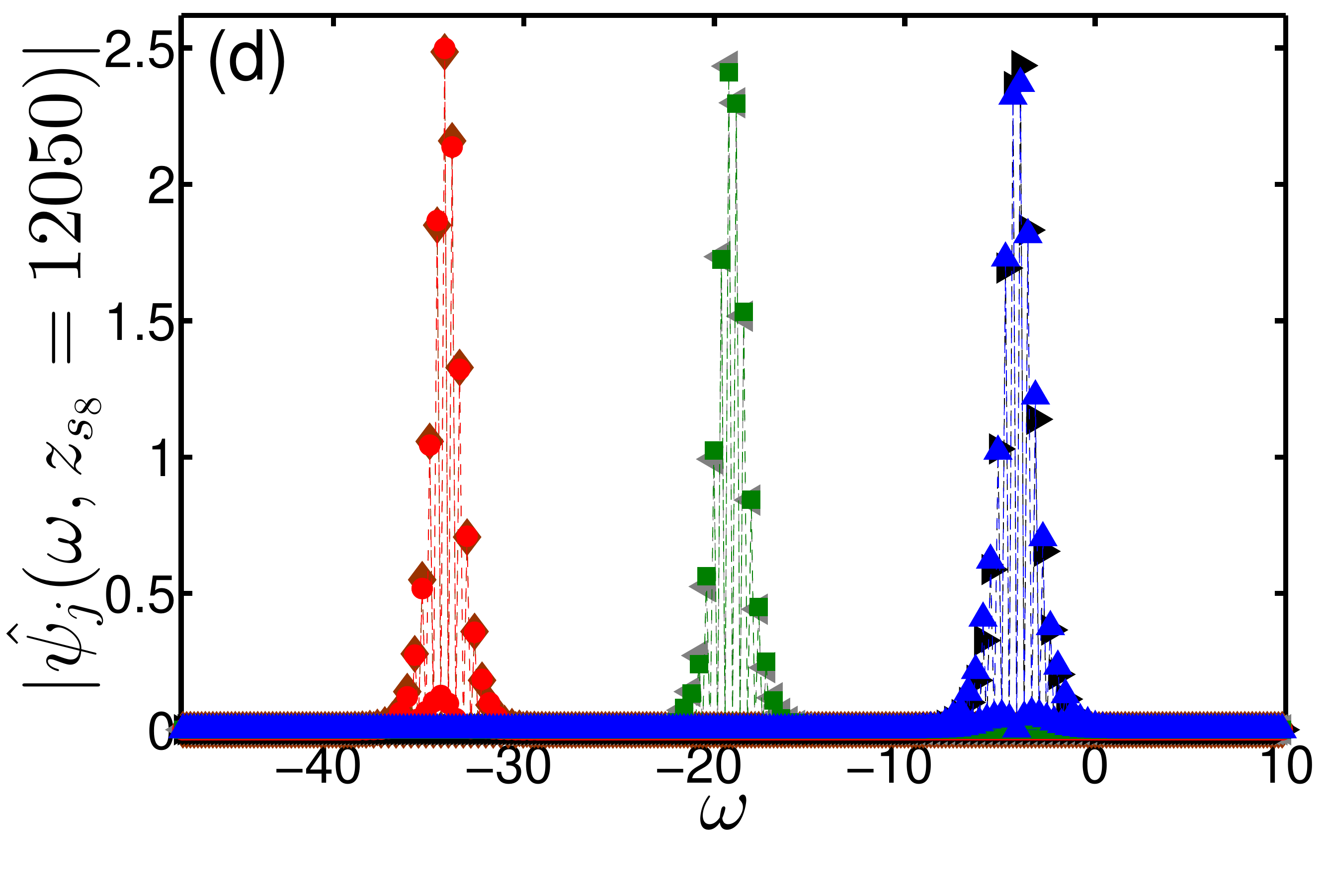} \\
\epsfxsize=6.0cm  \epsffile{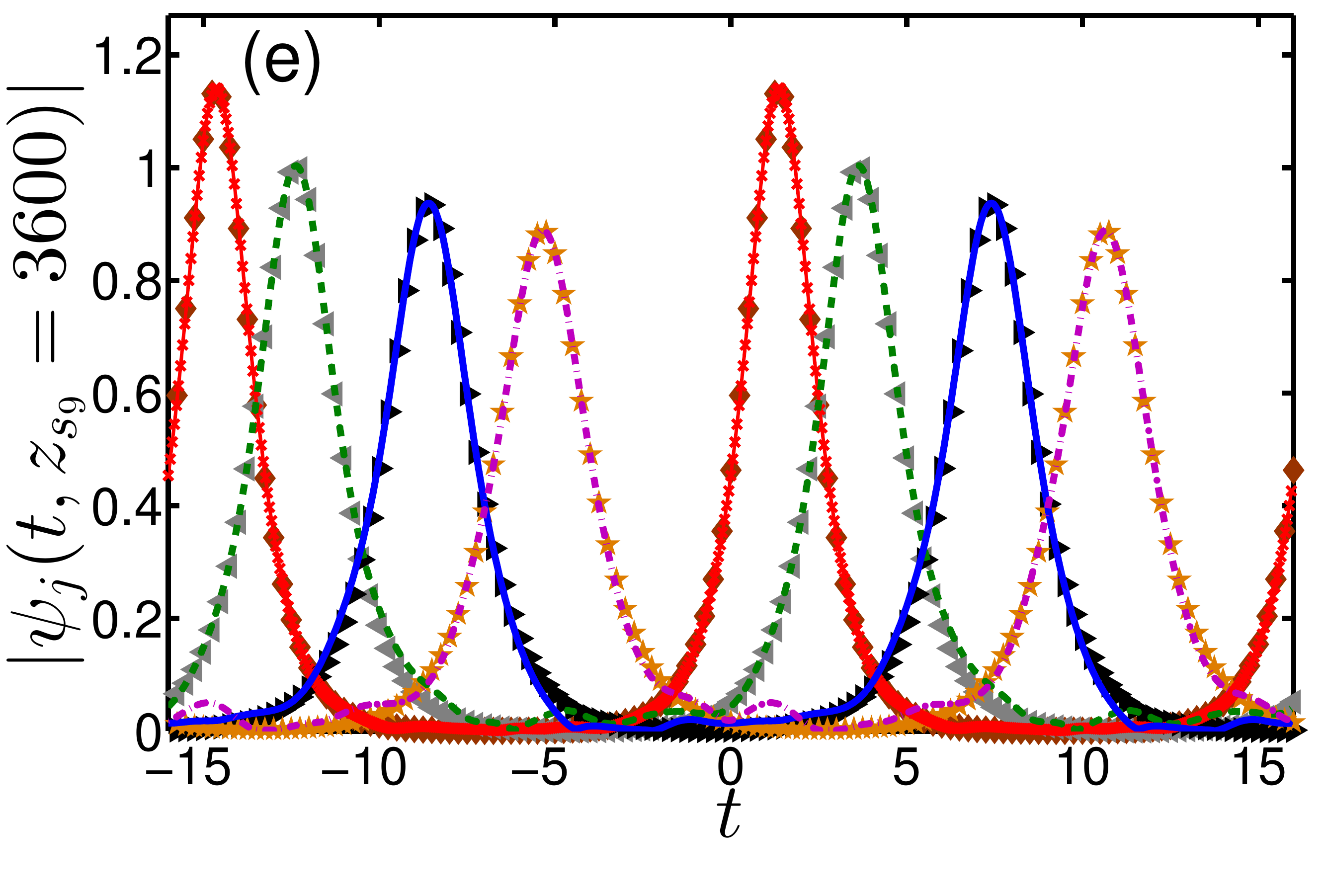} &
\epsfxsize=6.0cm  \epsffile{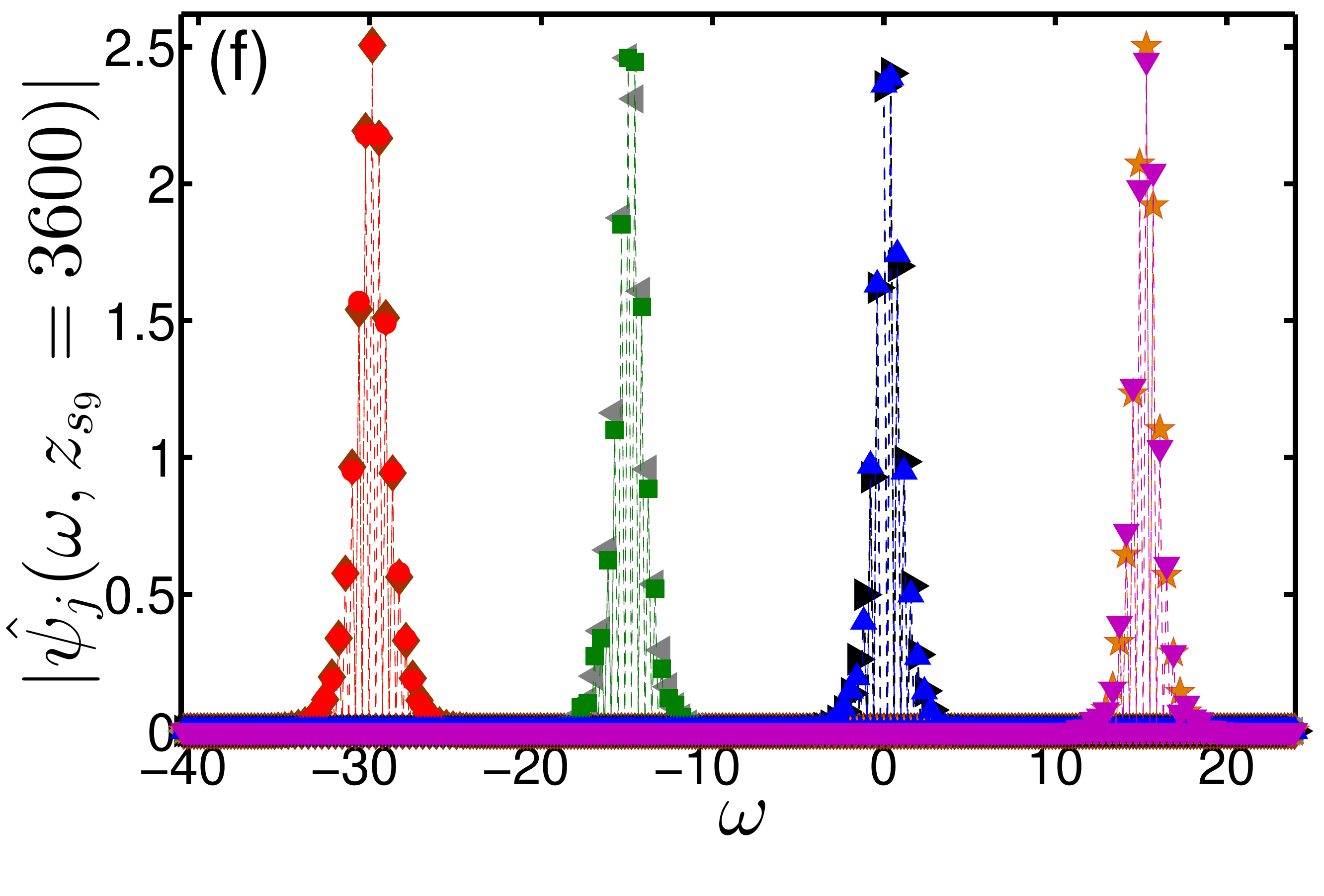}
\end{tabular}
\caption{(Color online) 
The pulse patterns at the onset of instability $|\psi_{j}(t,z_{s})|$ and their Fourier transforms  
$|\hat{\psi_{j}}(\omega,z_{s})|$ for the two-channel [(a)-(b)], 
the three-channel  [(c)-(d)], and the four-channel [(e)-(f)] 
waveguide couplers of Fig. \ref{fig7}.  
The final distances are $z_{s_7}=11200$ in (a)-(b), 
$z_{s_8}=12050$ in (c)-(d), and $z_{s_9}=3600$ in (e)-(f).
The solid-crossed red curve [solid red curve in (a)], dashed green curve, 
solid blue curve, and dashed-dotted magenta curve represent 
$|\psi_{j}(t,z_{s})|$ with $j=1,2,3,4$, obtained by 
numerical simulations with Eqs. (\ref{raman11}) and (\ref{raman13}). 
The red circles, green squares, blue up-pointing triangles, 
and magenta down-pointing triangles represent 
$|\hat{\psi_{j}}(t,z_{s})|$ with $j=1,2,3,4$, obtained by the simulations. 
The brown diamonds, gray left-pointing triangles, black right-pointing triangles, 
and orange stars represent the theoretical prediction for $|\psi_{j}(t,z_{s})|$ 
or $|\hat{\psi_{j}}(\omega,z_{s})|$ with $j=1,2,3,4$, respectively.}
 \label{fig8}
\end{figure}

\begin{figure}[ptb]
\begin{tabular}{cc}
\epsfxsize=5.8cm  \epsffile{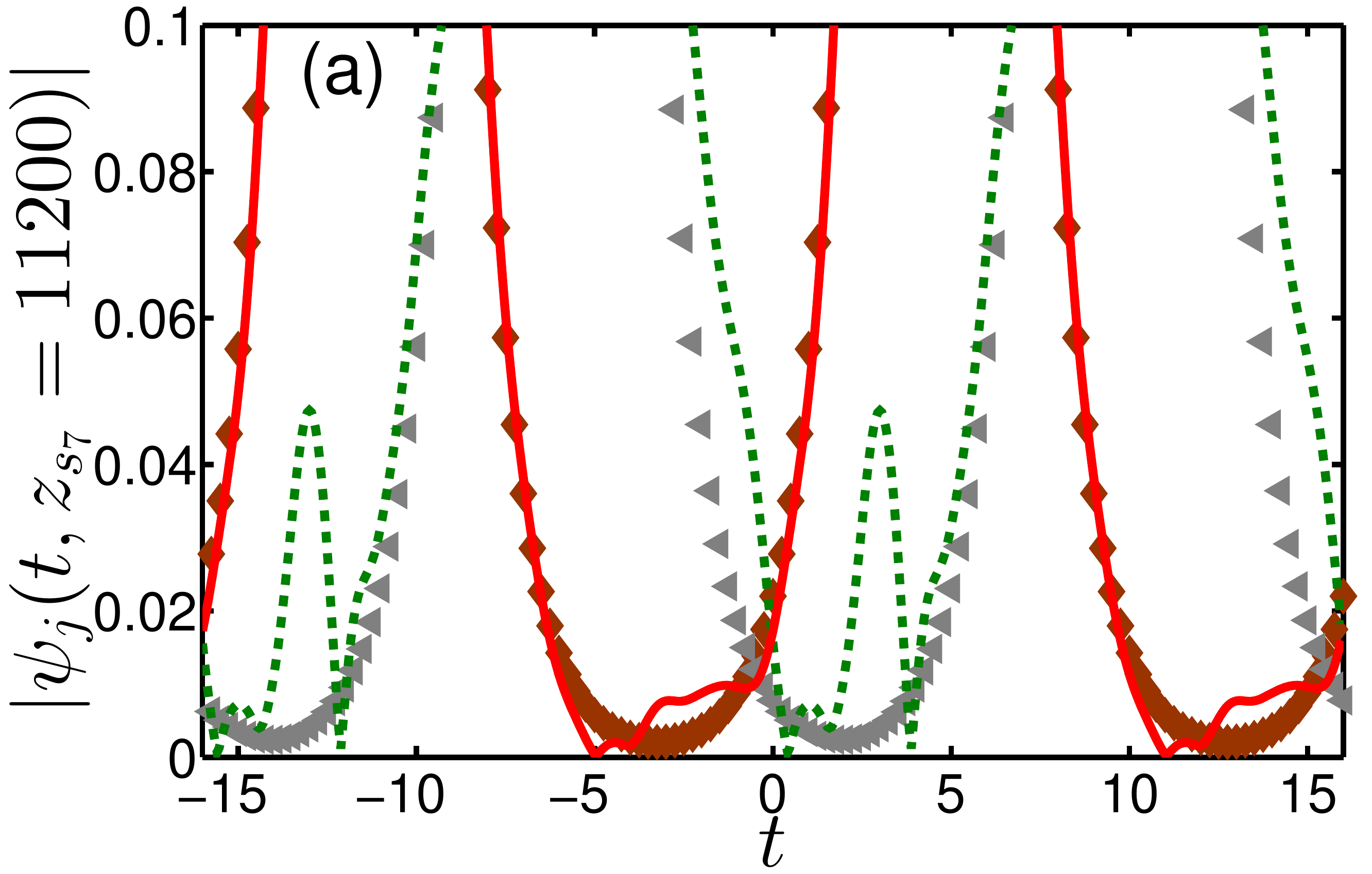} &
\epsfxsize=5.8cm  \epsffile{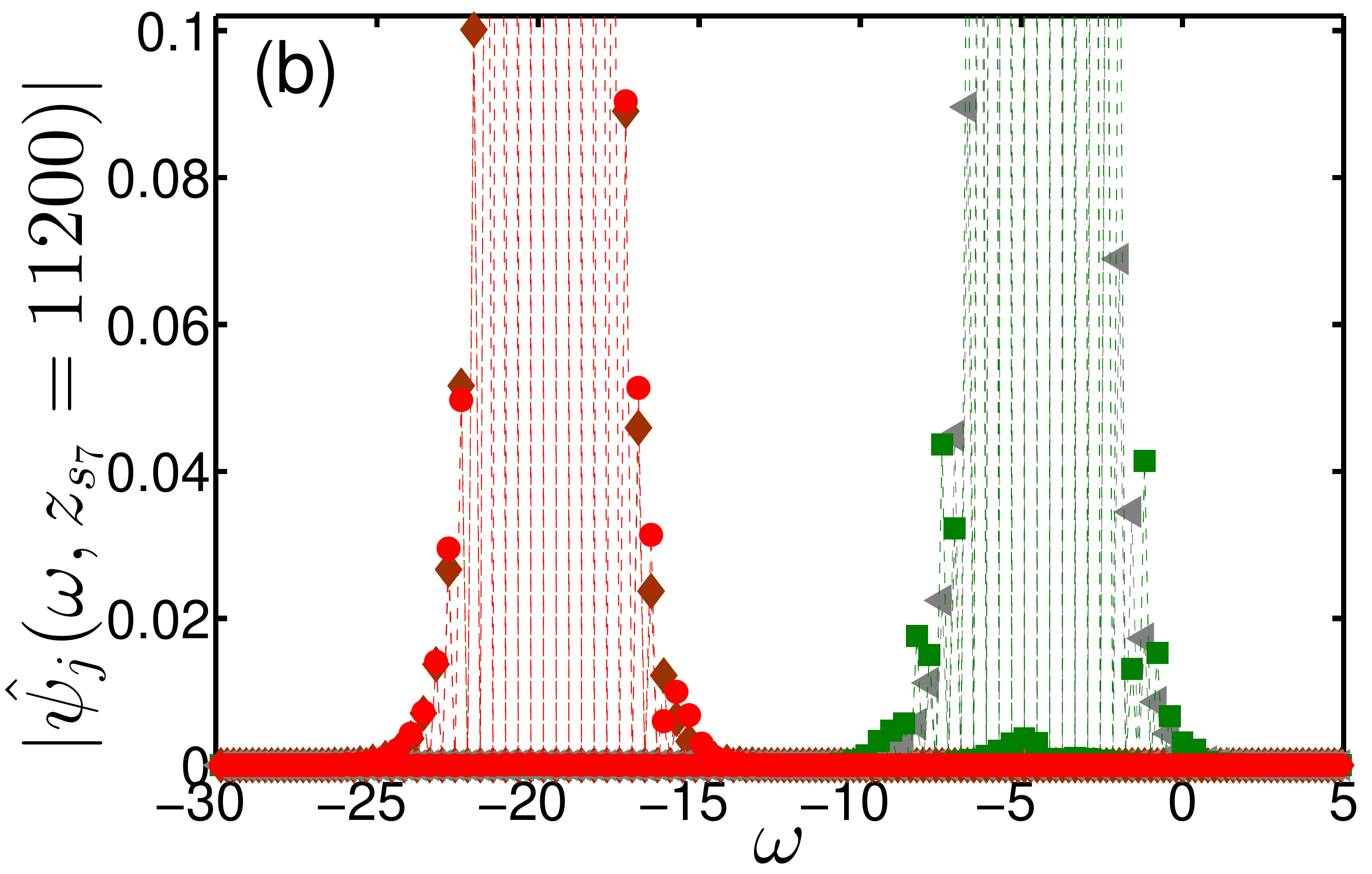} \\
\epsfxsize=5.8cm  \epsffile{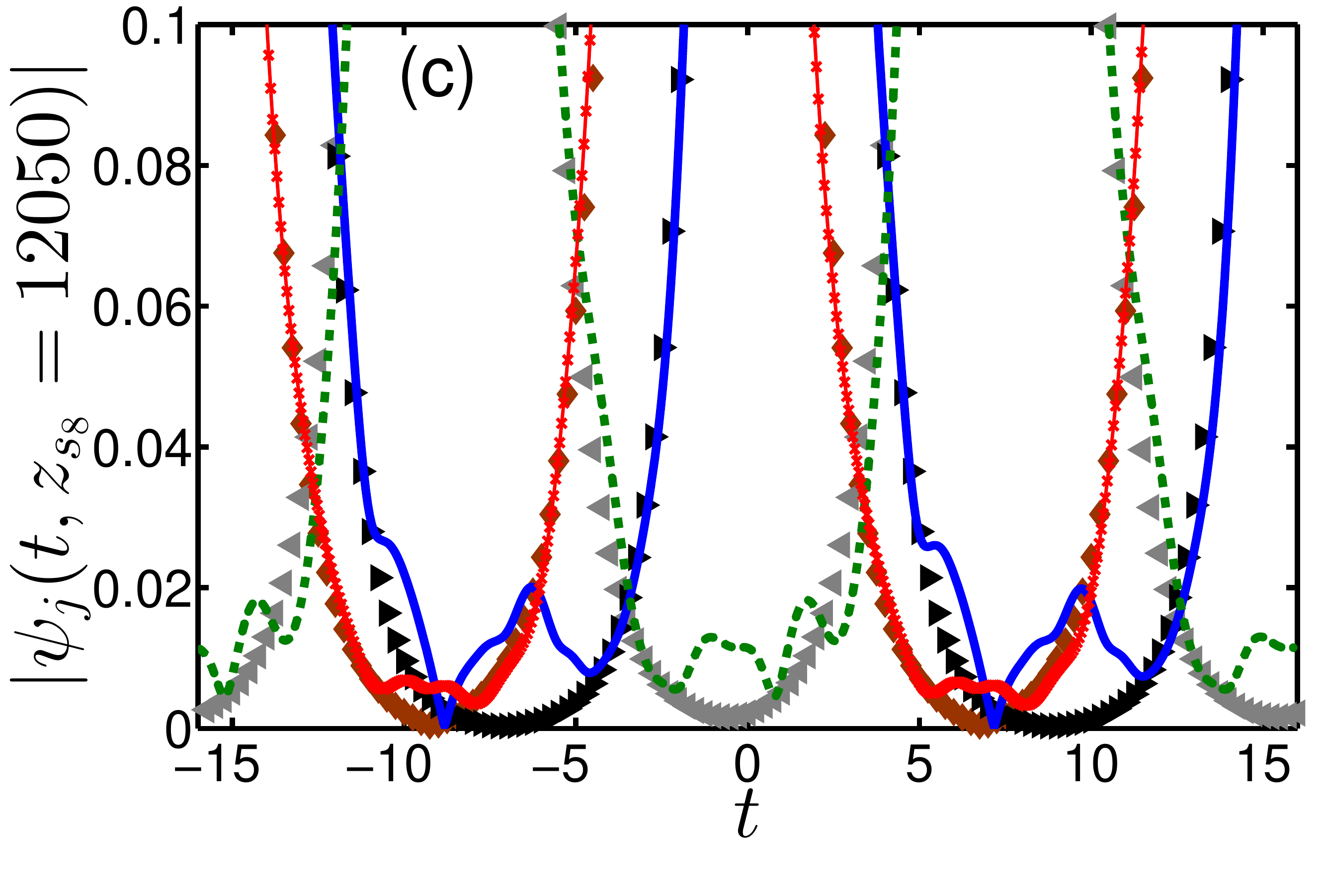}&
\epsfxsize=5.8cm  \epsffile{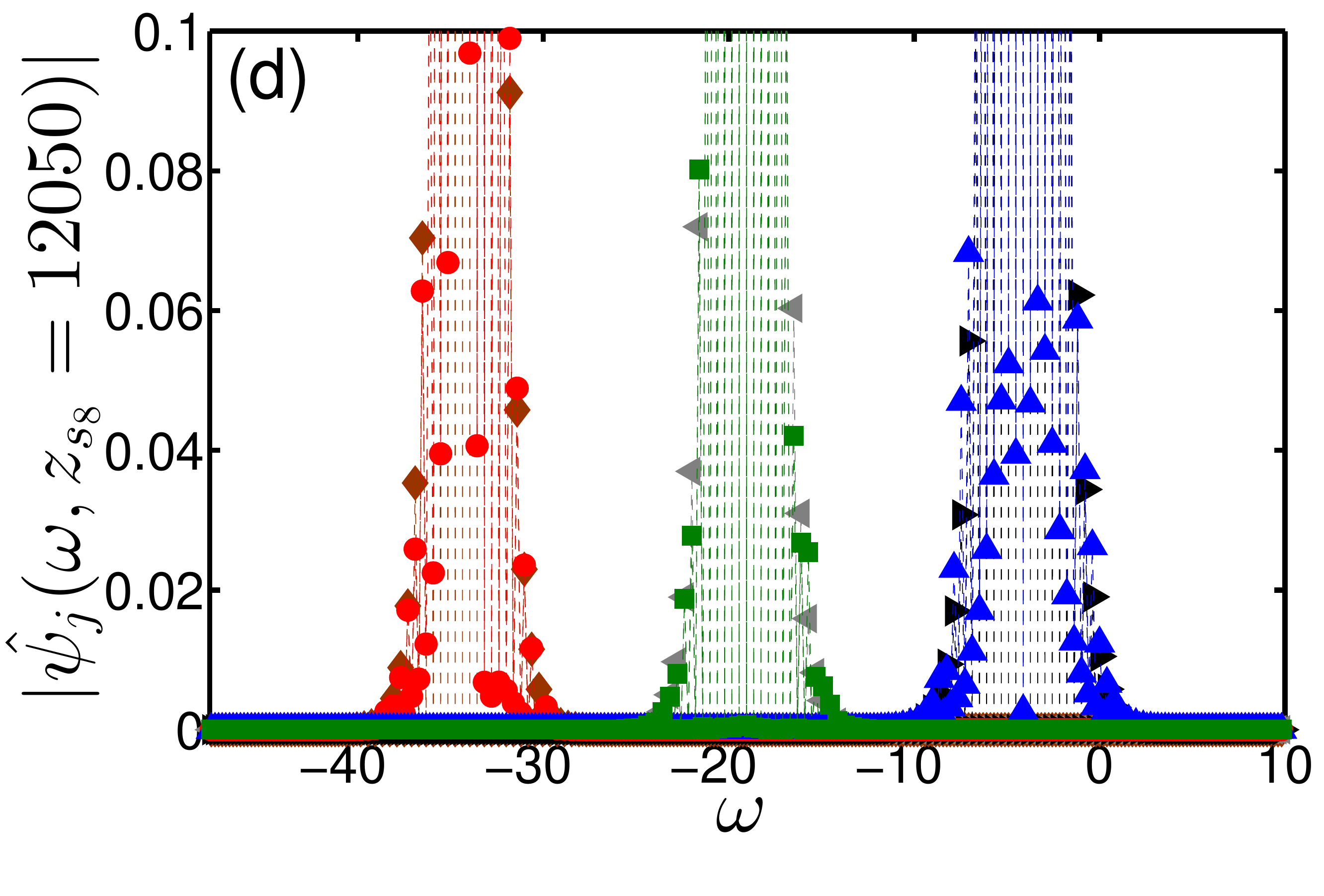} \\
\epsfxsize=5.8cm  \epsffile{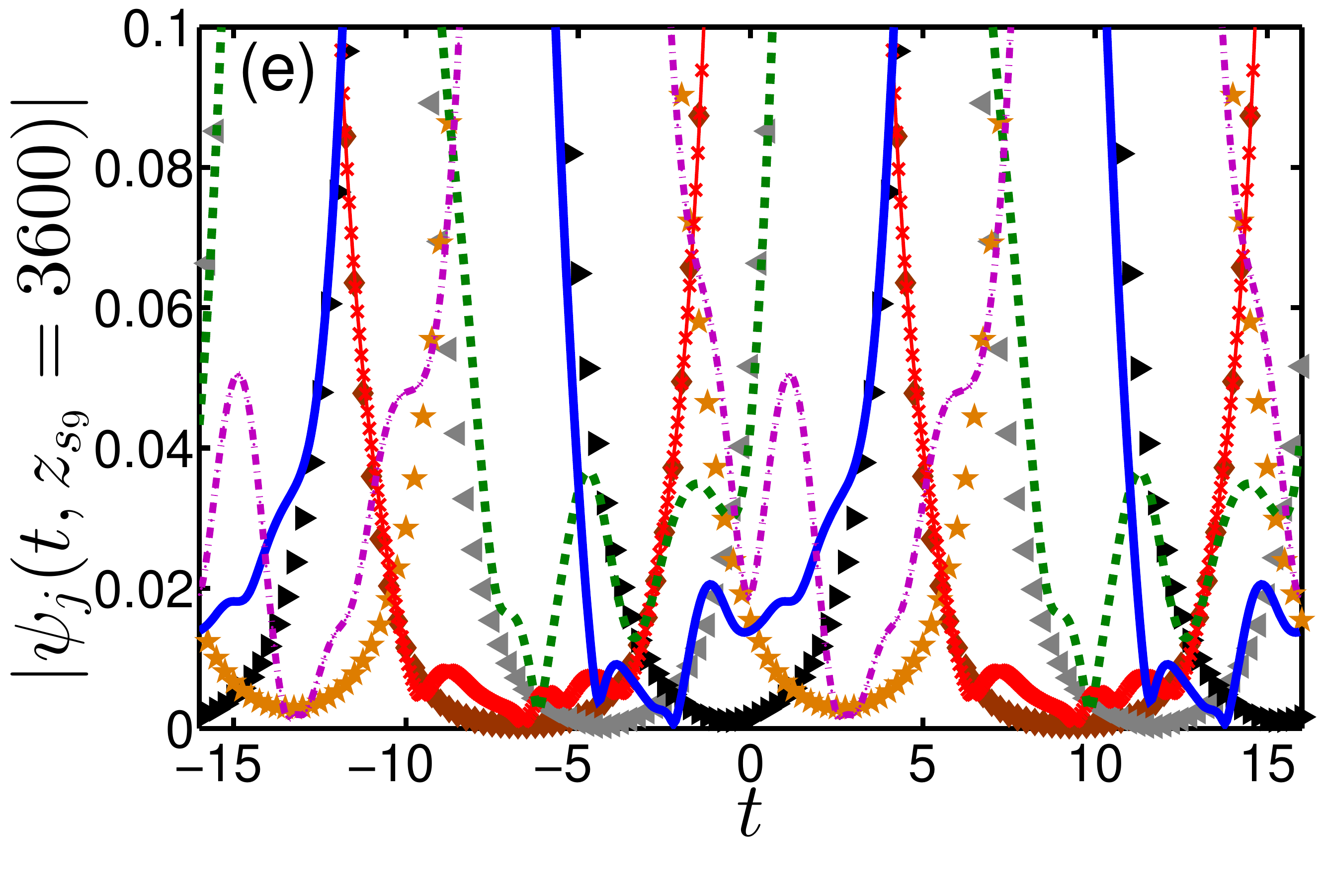}&
\epsfxsize=5.8cm  \epsffile{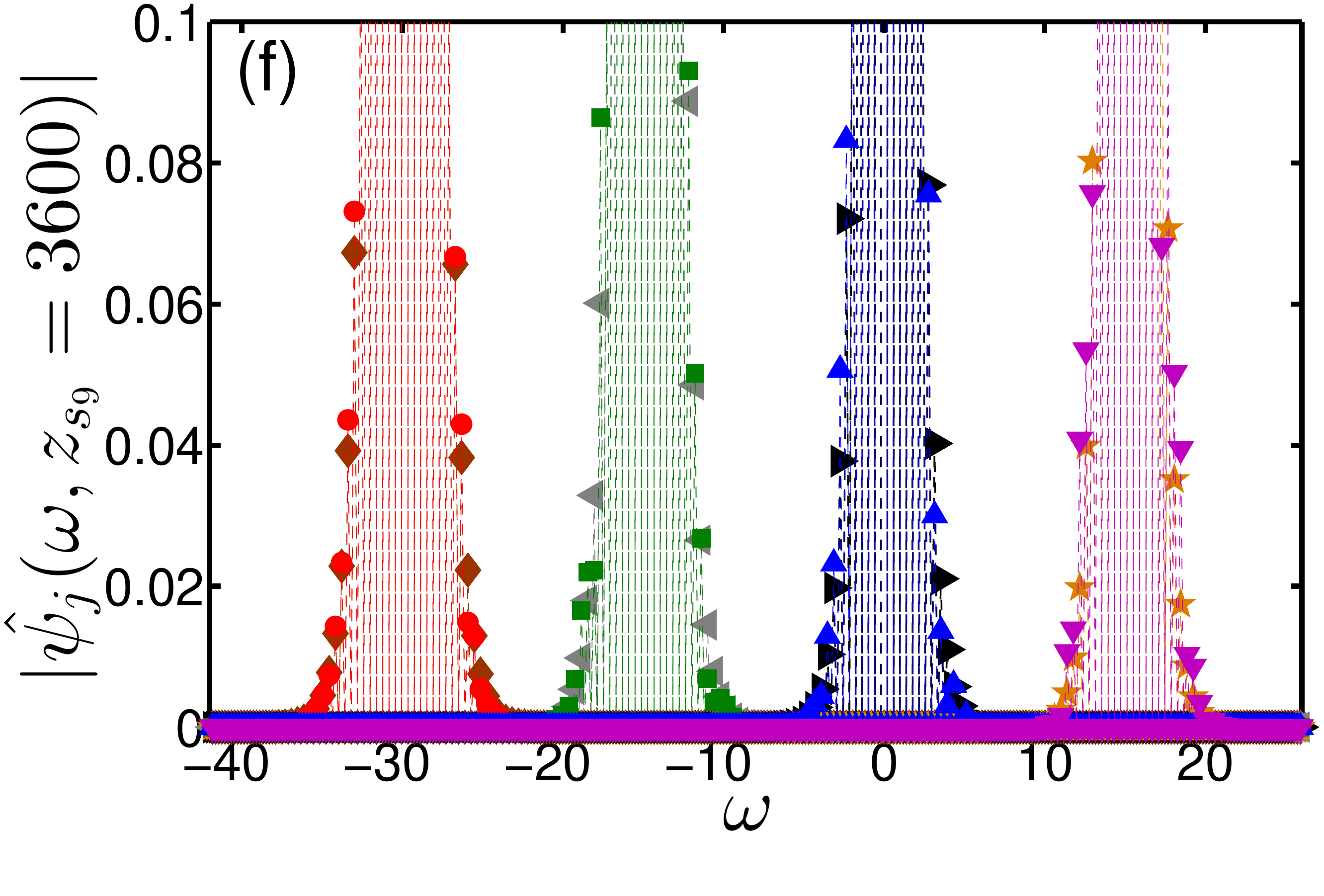} \\
\end{tabular}
\caption{(Color online) Magnified versions of the graphs in Fig. \ref{fig8} for 
small $|\psi_{j}(t,z_{s})|$ and $|\hat{\psi_{j}}(\omega,z_{s})|$ values. 
The symbols are the same as in Fig. \ref{fig8}.}
 \label{mag_fig8}
\end{figure}

Further insight into the enhanced transmission stability in waveguide couplers is gained 
by analyzing the pulse patterns at the onset of instability $|\psi_{j}(t,z_{s})|$ 
and their Fourier transforms $|\hat{\psi_{j}}(\omega,z_{s})|$. 
Figure \ref{fig8} shows the results obtained by numerical solution 
of Eqs. (\ref{raman11}) and (\ref{raman13}) together with the theoretical prediction. 
Figure \ref{mag_fig8} shows magnified versions of the graphs in Fig. \ref{fig8} 
for small $|\psi_{j}(t,z_{s})|$ and $|\hat{\psi_{j}}(\omega,z_{s})|$ values. 
It is seen that the soliton patterns are almost intact at $z=z_{s}$, 
in accordance with our conclusion about transmission stability for $0 \le z \le z_{s}$.      
Moreover, as seen from Fig. \ref{mag_fig8}, no radiative sidebands and no 
fast oscillations in the solitons shapes are observed at $z=z_{s}$. 
Instead, soliton distortion at $z=z_{s}$ is due to slow variations in the shape at 
the pulse tails, formation of small radiative pulses, and position shifts 
of the solitons from the same sequence relative to one another. 
Thus, instability due to formation of radiative sidebands is completely 
suppressed in $N$-waveguide couplers with the gain-loss (\ref{raman13}). 
This finding explains the significant increase in the values of the stable 
propagation distance in $N$-waveguide coupler transmission compared with 
the distances achieved in single-fiber transmission in Section \ref{simu}.

\begin{figure}[ptb]
\begin{tabular}{cc}
\epsfxsize=10.0cm  \epsffile{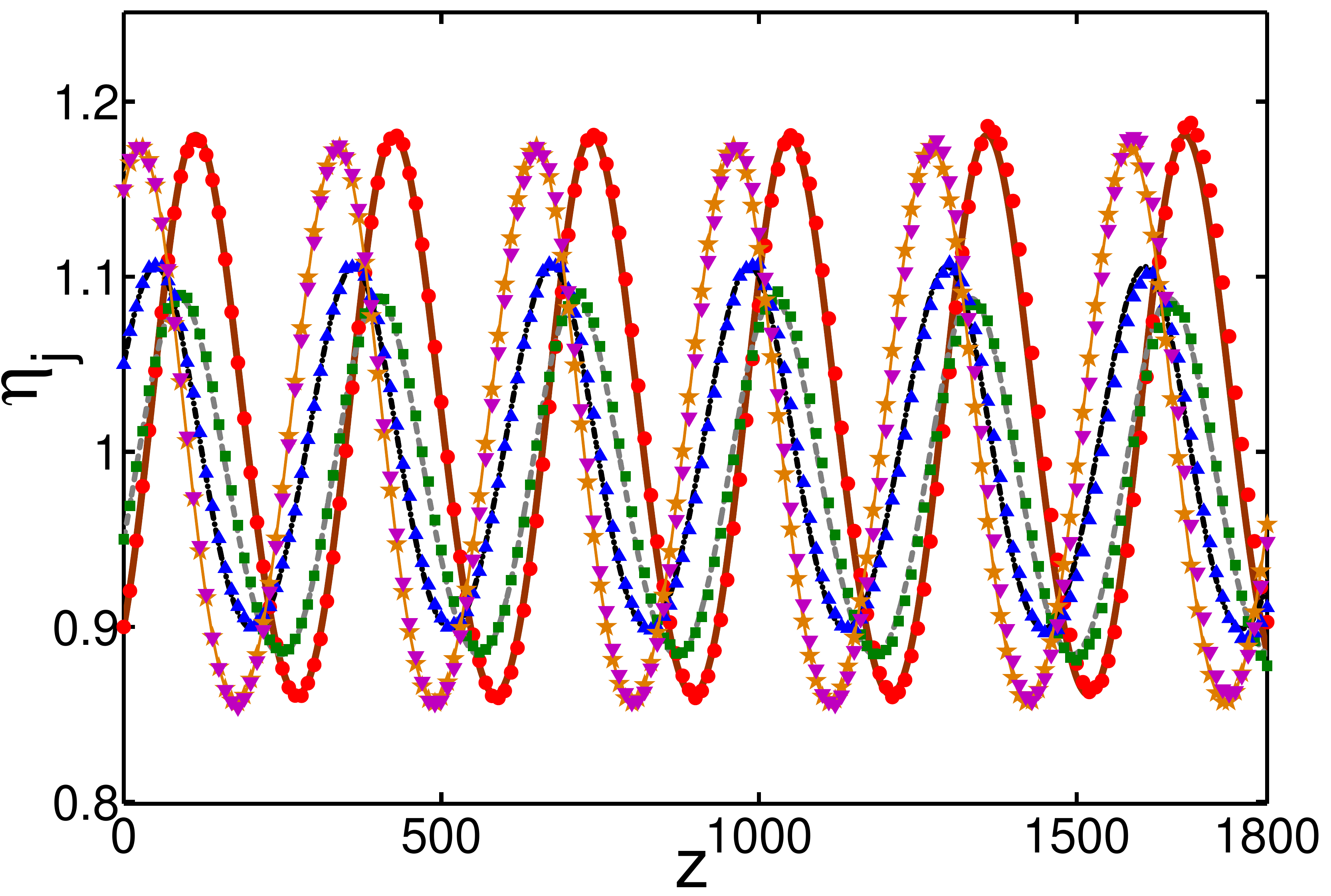} 
\end{tabular}
\caption{(Color online) The $z$ dependence of soliton amplitudes $\eta_{j}$ in 
a four-channel nonlinear waveguide coupler with linear gain-loss (\ref{raman12}) and $W=15$. 
The values of the other physical parameters are 
the same as the ones used in Fig. \ref{fig7} (c). 
The red circles, green squares, blue up-pointing triangles, and magenta down-pointing triangles 
represent $\eta_{1}(z)$, $\eta_{2}(z)$, $\eta_{3}(z)$, and $\eta_{4}(z)$ obtained 
by numerical solution of the coupled-NLS model (\ref{raman11}) with the gain-loss (\ref{raman12}). 
The solid brown, dashed gray, dashed-dotted black, and solid-starred orange curves correspond to 
$\eta_{1}(z)$, $\eta_{2}(z)$, $\eta_{3}(z)$, and $\eta_{4}(z)$ obtained 
with the predator-prey model (\ref{raman5}).}
\label{fig9}
\end{figure}

We note that the frequency shifts experienced by the propagating solitons at 
large propagation distances are quite large for both the single-fiber systems of 
Section \ref{simu} and the $N$-waveguide coupler systems of the current section. 
For example, the total frequency shifts measured at $z_{s_9}=3600$ from 
the coupled-NLS simulations for the four-channel waveguide coupler of Fig. \ref{fig7}(c) are $\Delta\beta_1(z_{s_{9}})=-7.354$, $\Delta\beta_2(z_{s_{9}})=-7.296$,
$\Delta\beta_3(z_{s_{9}})=-7.284$, and $\Delta\beta_4(z_{s_{9}})=-7.2204$. 
These Raman-induced frequency shifts make the gain-loss functions with 
fixed frequency intervals [such as the function in Eq. (\ref{raman2})] less effective in 
stabilizing soliton amplitude dynamics. In order to compensate for the effects 
of these frequency shifts, a shifting of the central amplification interval was 
introduced into the gain-loss function (\ref{raman13}). 
We now complete the analysis of transmission stabilization in the waveguide coupler, 
by evaluating the impact of shifting of the amplification interval in the gain-loss 
function (\ref{raman13}). For this purpose, we consider the following alternative 
gain-loss functions with {\it fixed} amplification intervals: 
\begin{eqnarray} &&
\tilde g_{j}(\omega) = \left\{ 
\begin{array}{l l}
g_{j} &  \mbox{ if $\beta_{j}(0)-W/2 < \omega \le \beta_{j}(0)+W/2,$}\\
g_{L} &  \mbox{ if $\omega \le \beta_{j}(0) - W/2$, or $\omega > \beta_{j}(0) + W/2,$}\\
\end{array} \right. 
\label{raman12}
\end{eqnarray}    
where $1 \le j \le N$. We carry out numerical simulations with Eq. (\ref{raman11}) 
and the gain-loss functions (\ref{raman12}) with $W=15$ for a four-channel waveguide coupler. 
The values of the other physical parameters are the same as the ones used in Fig. \ref{fig7}(c). 
The simulations are carried out up to the onset of transmisison instability at $z_{s_{10}}=1800$. 
Figure \ref{fig9} shows the $z$ dependence 
of soliton amplitudes obtained by these coupled-NLS simulations along with the prediction of the 
predator-prey model (\ref{raman5}). We observe stable oscillatory dynamics and good agreement 
with the predator-prey model's prediction throughout the propagation. 
We note that the stable propagation distance $z_{s_{10}}=1800$ 
is larger by a factor of 3.6 compared with the stable propagation distance for 
the corresponding single-fiber system [see Fig. \ref{fig3}(c)], but smaller by a factor of 0.5
compared with the distance obtained with the waveguide coupler and the gain-loss (\ref{raman13}). 
Based on this comparison we conclude that the introduction of shifting of the central amplification interval 
does lead to an enhancement of transmission stability. On the other hand, we also observe 
that even in the absence of shifting of the amplification interval, the waveguide coupler setup 
enables stable propagation along significantly larger distances compared with the single-fiber systems 
considered in Section \ref{simu}.

\begin{figure}[ptb]
\begin{tabular}{cc}
\epsfxsize=10.0cm  \epsffile{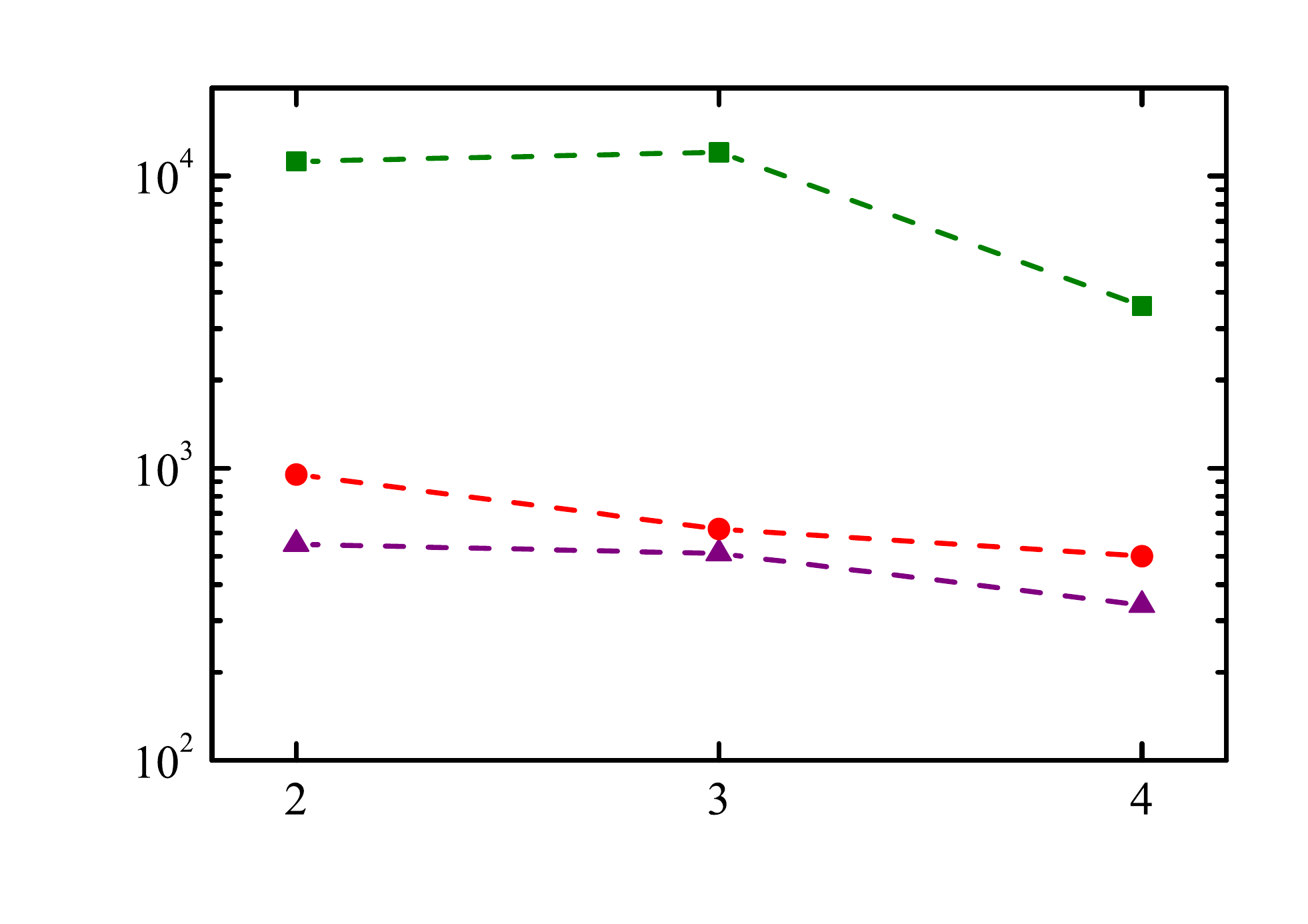} 
\end{tabular}
\caption{(Color online) The dependence of the stable propagation 
distance $z_s$ on the number of channels $N$
in different transmission setups.
The red circles represent the values obtained for single-fiber transmission 
in the presence of delayed Raman response and the linear gain-loss (\ref{raman2})  
by numerical solution of Eqs. (\ref{raman1}) and (\ref{raman2}). 
The purple triangles represent the values obtained for single-fiber transmission 
in the absence of delayed Raman response and linear gain-loss 
by numerical solution of Eq. (\ref{raman1B}). 
The green squares represent the values obtained for $N$-waveguide coupler 
transmission by numerical solution of Eqs. (\ref{raman11}) and (\ref{raman13}).}
\label{fig10}
\end{figure}

We conclude this section by summarizing the dependence of the stable propagation 
distance $z_{s}$ on the number of channels $N$ in the main transmission setups 
considered in the paper.  Figure \ref{fig10} shows the $z_{s}$ values obtained by 
numerical simulations for single-fiber transmission with and without the effects of 
delayed Raman response and the linear gain-loss (\ref{raman2}). 
The $z_{s}$ values obtained by the simulations for $N$-waveguide coupler 
transmission with the linear gain-loss (\ref{raman13}) are also shown. 
We observe that in single-fiber transmission, the introduction of the gain-loss 
(\ref{raman2}) leads to a moderate increase in the $z_{s}$ values, despite of 
the presence of delayed Raman response. Moreover, the $z_{s}$ values 
obtained in $N$-waveguide coupler transmission are significantly larger than 
the ones obtained in single-fiber transmission. As explained earlier, 
the enhanced transmission stability in waveguide couplers can be attributed 
to the more efficient suppression of radiative sideband generation by the linear 
gain-loss (\ref{raman13}).

\section{Conclusions}
\label{conclusions}
We investigated transmission stabilization and destabilization 
and dynamics of pulse amplitudes induced by Raman crosstalk in multichannel 
soliton-based optical waveguide systems with $N$ frequency channels. 
We considered two main transmission setups. In the first setup, 
the $N$ soliton sequences propagate through a single optical fiber, 
while in the second setup, the sequences propagate through a waveguide coupler, 
consisting of $N$ close waveguides. 
We studied the transmission by performing numerical simulations with coupled-NLS models, 
which take into account second-order dispersion, Kerr nonlinearity, delayed Raman response, 
and frequency dependent linear gain-loss. The simulations were carried out for two, three, 
and four frequency channels in both single-fiber and waveguide coupler setups. 
The results of the coupled-NLS simulations were compared with the predictions of a 
simplified predator-prey model for dynamics of pulse amplitudes \cite{NP2010}, 
which incorporates amplitude shifts due to linear gain-loss and delayed Raman response, 
but neglects radiation emission and intrachannel interaction. 
The predator-prey model in Ref. \cite{NP2010} predicts that stable dynamics 
of soliton amplitudes can be realized by a suitable choice of amplifier gain in different 
frequency channels. One major goal of our study was to validate this prediction. 
A second major goal was to characterize the processes that lead to 
transmission destabilization and to develop waveguide setups, 
which lead to significant enhancement of transmission stability.

We first studied soliton-based multichannel transmission in a single fiber 
in the absence of delayed Raman response and linear gain-loss. 
We found that in this case, transmission destabilization is caused 
by resonant formation of radiative sidebands, where the largest sidebands 
for the $j$th soliton sequence form at frequencies $\beta_{j-1}(0)$ 
and / or $\beta_{j+1}(0)$ of the solitons in the neighboring frequency channels. 
Additionally, the amplitudes of the radiative sidebands increase with increasing 
number of channels $N$, and as a result, the stable propagation distances 
decrease with increasing $N$. Furthermore, the stable propagation distances 
obtained in our numerical simulations are significantly smaller compared with 
the distances achieved in Ref. \cite{CPN2016} for single-channel transmission 
with the same values of the physical parameters. 
Based on these findings we conclude that destabilization of multichannel 
soliton-based transmission in a single fiber is caused by Kerr-induced interaction 
in interchannel soliton collisions.

We then carried out numerical simulations for multichannel transmission in a single fiber, 
taking into account the effects of delayed Raman response 
and frequency dependent linear gain-loss. 
We assumed that the gain-loss function $g(\omega)$ 
for single-fiber transmission is given by Eq. (\ref{raman2}). 
That is, $g(\omega)$ is equal to the constants $g_{j}$, 
determined by the predator-prey model, in frequency intervals 
$(\beta_{j}(0)- W/2, \beta_{j}(0)+ W/2]$ of constant width $W$  
centered about the initial soliton frequencies $\beta_{j}(0)$, 
and is equal to a negative value $g_{L}$ outside of these intervals.
Numerical simulations with the full coupled-NLS model showed 
that at distances smaller than the stable propagation distance $z_{s}$, 
soliton amplitudes exhibit stable oscillatory dynamics, 
in excellent agreement with the predictions of the predator-prey model of Ref. \cite{NP2010}. 
These findings are very important because of the major simplifying assumptions made 
in the derivation of the predator-prey model. In particular, based on these findings, 
we conclude that the effects of radiation emission and intrachannel interaction 
can indeed be neglected at distances smaller than $z_{s}$. 
However, at distances $z \simeq z_{s}$, we observed transmission destabilization 
due to formation of radiative sidebands. 
The destabilization process is very similar to destabilization in the absence 
of delayed Raman response and linear gain-loss, i.e., the largest sidebands 
for the $j$th soliton sequence form at frequencies $\beta_{j-1}(z)$ 
and / or $\beta_{j+1}(z)$ of the solitons in the neighboring frequency channels.
At distances larger than $z_{s}$, the continued growth of the radiative sidebands 
leads to fast oscillations in the main body of the solitons and to generation of new pulses, 
which do not possess the soliton sech form. As a result, the pulse patterns 
at this late stage of the propagation are strongly distorted.

We note that the stable propagation distances $z_{s}$ 
for single-fiber multichannel transmission  
in the presence of delayed Raman response and linear gain-loss are 
larger compared with the distances obtained in the absence of these 
processes. We attribute this increase in $z_{s}$ values to the 
introduction of frequency dependent linear gain-loss 
with relatively strong loss $g_{L}$ outside the frequency intervals 
$(\beta_{j}(0)- W/2, \beta_{j}(0)+ W/2]$, 
which leads to partial suppression of radiative sideband generation. 
However, the suppression of radiative instability in single-fiber transmission is quite limited, 
since the radiative sidebands for each sequence form near the frequencies 
$\beta_{k}(z)$ of the other soliton sequences. As a result, in a single fiber, 
one cannot employ strong loss at the latter frequencies, as this would lead 
to the decay of the propagating solitons.

A more promising approach for achieving significant enhancement of 
transmission stability is based on employing a nonlinear waveguide coupler, 
consisting of N close waveguides. In this case, each soliton sequence 
propagates through its own waveguide, and each waveguide 
is characterized by its own frequency dependent linear gain-loss function. We assumed that the linear 
gain-loss for the $j$th waveguide $\tilde g_{j}(\omega,z)$ is equal to the constant $g_{j}$, 
determined by the predator-prey model, inside a  
$z$-dependent frequency interval centered about the soliton frequency $\beta_{j}(z)$, 
and is equal to a negative value $g_{L}$ outside of this interval. 
This waveguide coupler setup is expected to lead to enhanced transmission stability, 
since generation of all radiative sidebands outside of the central 
amplification interval is suppressed by the relatively strong linear loss $g_{L}$ 
for each of the $N$ waveguides.  
To test this prediction, we carried out numerical simulations with a new 
coupled-NLS model, which takes into account the effects of 
second-order dispersion, Kerr nonlinearity, delayed Raman response, 
and the $N$ frequency-dependent linear gain-loss functions $\tilde g_{j}(\omega,z)$.    
The simulations with the new coupled-NLS model showed that transmission stability 
and the validity of the predator-prey model's predictions in waveguide coupler transmission  
are extended to distances that are larger by factors of 11.8 for two channels, 
19.4 for three channels, and 7.2 for four channels, compared with the distances in the single-fiber system. 
Additionally, the simulations showed that the solitons retain their shape at distances smaller than $z_{s}$ 
and no radiative sidebands appear throughout the propagation. 
Based on these observations we conclude that transmission stability in the 
waveguide coupler system is indeed significantly enhanced compared with the single-fiber system. 
Furthermore, the enhanced transmission stability in the waveguide coupler is enabled by the efficient 
suppression of radiative sideband generation due to the presence of the linear 
gain-loss $\tilde g_{j}(\omega,z)$.

To complete the analysis of transmission stabilization in the waveguide coupler, we evaluated the impact 
of the shifting of the central amplification intervals of the gain-loss  functions $\tilde g_{j}(\omega,z)$. 
For this purpose, we replaced each $\tilde g_{j}(\omega,z)$ by a 
$z$-independent gain-loss function $\tilde g_{j}(\omega)$ 
that is equal to $g_{j}$ inside a {\it fixed} frequency interval centered about 
the initial soliton frequency $\beta_{j}(0)$, and is equal to a negative value $g_{L}$ 
outside of this interval. Numerical simulations with the coupled-NLS model for a four-channel 
system showed that the stable propagation distance 
and the distance along which the predator-prey model's predictions are valid are larger 
by a factor of 3.6 compared with the distance achieved in the single-fiber system, but smaller 
by a factor of 0.5 compared with the distance obtained in the waveguide coupler 
with shifting of the central amplification intervals of the linear gain-loss. 
Based on these findings we conclude that the introduction of shifting of the central amplification intervals 
does lead to enhanced transmission stability. On the other hand, even in the absence 
of shifting of the central amplification intervals, 
the waveguide coupler setup enables stable propagation along 
significantly larger distances compared with the single-fiber setup.

\appendix
\section{The method for determining the stable propagation distance}
\label{appendA}

In this Appendix, we present the method that we used for determining 
the value of the stable propagation distance $z_{s}$ from the results of the coupled-NLS simulations. 
In addition, we present the theoretical predictions for the soliton patterns 
and their Fourier transforms, which were used in the analysis of transmission stability.

We consider propagation of the soliton sequence in the $j$th frequency channel 
through an optical waveguide, where the envelope of the electric field 
of this sequence at $z=0$ is given by Eq. (\ref{raman8}). 
We are interested in the envelope of the pulse sequence at distance $z$.  
We assume that the pulse sequence is only weakly distorted at this distance. 
Thus, by the standard adiabatic perturbation technique for the NLS soliton, 
one can write the envelope of the electric field of the $j$th sequence at distance $z$ as: 
$\psi_{j}(t,z)=\psi_{tj}(t,z)+v_{rj}(t,z)$, where $\psi_{tj}(t,z)$ is the soliton 
part and $v_{rj}(t,z)$ is the radiation part \cite{Kaup91,Chertkov2003}. 
Since $T \gg 1$, we neglect intrasequence interaction, 
in accordance with the assumptions of the predator-prey model 
of Ref. \cite{NP2010} (see also Section \ref{models}). 
Under this assumption, the soliton part of the electric field can be expressed as: 
\begin{eqnarray} &&
\psi_{tj}(t,z) = \eta_{j}(z)e^{i\theta_{j}(z)}
\sum_{k = -J}^{J-1}\frac{\exp\{ i\beta_{j}(z) 
\left[ t-y_{j}(z)-kT \right] \}}
{\mathrm{cosh}\{\eta_{j}(z)\left[t-y_{j}(z)-kT\right]\}}, 
\label{criterion1}
\end{eqnarray} 
where $\eta_{j}(z)$ is the amplitude, $\beta_{j}(z)$ is the frequency, 
$\theta_{j}(z)$ is the common overall phase, 
$y_{j}(z)=\Delta y_{j}(z) + T/2 + \delta_{j}$, and $\Delta y_{j}(z)$ 
is the common overall position shift. The Fourier transform of $\psi_{tj}(t,z)$ 
with respect to time is given by: 
\begin{eqnarray} &&
\!\!\!\!\!
\hat{\psi}_{tj}(\omega,z)=\left(\frac{\pi}{2}\right)^{1/2}
\mbox{sech}\left\{\frac{\pi\left[\omega - \beta_{j}(z)\right]}{2\eta_{j}(z)}\right\}
e^{i\theta_{j}(z) - i\omega y_{j}(z)}{\sum_{k=-J}^{J-1}e^{-ikT\omega}}.
\;\;\;
\label{criterion2}
\end{eqnarray}                         
Our theoretical prediction for the $j$th pulse pattern at distance $z$, $|\psi_{j}^{(th)}(t,z)|$, 
is calculated by using Eq. (\ref{criterion1}) with values of $\eta_{j}(z)$, $\beta_{j}(z)$, 
and $y_{j}(z)$, which are measured from the numerical simulations. Similarly, our theoretical 
prediction for the Fourier transform of the $j$th pulse pattern, $|\hat{\psi}_{j}^{(th)}(\omega,z)|$, 
is obtained by using Eq. (\ref{criterion2}) with the numerically obtained values of 
$\eta_{j}(z)$ and $\beta_{j}(z)$ \cite{eta_and_beta}.

The method for determining the stable propagation distance is based on a comparison 
of the theoretical predictions for the pulse patterns $|\psi_{j}^{(th)}(t,z)|$ with the results 
of the numerical simulation $|\psi_{j}^{(num)}(t,z)|$ for $1 \le j \le N$. 
More specifically, we calculate the following normalized integrals, which measure the deviation of the 
numerically obtained pulse patterns from the corresponding theoretical predictions: 
\begin{eqnarray} &&
I_{j}(z)=\tilde I_{j}^{(dif)}(z)/\tilde I_{j}(z), 
\label{criterion3}
\end{eqnarray}   
where 
\begin{eqnarray} &&
\tilde I_{j}^{(dif)}(z)=     
\left\{ \int\limits_{-JT}^{JT} 
\left[\;\left| \psi _j^{\left( {th} \right)}\left( {t,z} \right) \right| - 
\left| \psi _j^{\left( {num} \right)}\left( {t,z} \right) \right| \; \right]^2 dt 
\right\}^{1/2},   
\label{criterion4}
\end{eqnarray}           
\begin{eqnarray} &&
\tilde I_{j}(z)=     
\left[ \int\limits_{-JT}^{JT} 
\left| \psi _j^{\left( {th} \right)}\left( {t,z} \right) \right|^2 dt \right]^{1/2},   
\label{criterion5}
\end{eqnarray}       
and $1 \le j \le N$. We then define the stable propagation distance $z_{s}$ 
as the largest distance at which the values of $I_{j}(z)$ are still smaller than 
a constant $C$ for $1 \le j \le N$. In practice, we used the value $C=0.05$ in 
the numerical simulations. We emphasize, however, that the values of the stable 
propagation distance obtained by this method are not very sensitive to the choice of 
the constant $C$. That is, we found that small changes in the value of $C$ 
lead to small changes in the measured $z_{s}$ values.



\begin{thebibliography}{}
\bibitem{NLS_model} Note that similar perturbed coupled-NLS models with second-order 
dispersion, Kerr nonlinearity, and delayed Raman response, but without frequency 
dependent linear gain-loss, were used in Refs. \cite{Kumar98,Kaup99}.
\bibitem{dimensions} 
The dimensionless distance $z$ in Eq. (\ref{raman1})  is 
$z=X/(2L_{D})$, where $X$ is the dimensional distance, 
$L_{D}=\tau_{0}^{2}/|\tilde\beta_{2}|$ is the dimensional dispersion length,    
$\tau_{0}$ is the soliton width, and $\tilde\beta_{2}$ is the second-order
dispersion coefficient. The dimensionless time is
$t=\tau/\tau_{0}$, where $\tau$ is the time. 
The solitons spectral width is $\nu_{0}=1/(\pi^{2}\tau_{0})$ and the 
interchannel frequency difference is $\Delta\nu=(\pi\Delta\beta\nu_{0})/2$. 
$\psi_{j}=E_{j}/\sqrt{P_{0}}$, where $E_{j}$ is proportional to the 
electric field of the $j$th pulse sequence and $P_{0}$ is the peak power. 
The dimensionless second order dispersion coefficient is 
$d=-1=\tilde\beta_{2}/(\gamma P_{0}\tau_{0}^{2})$, 
where $\gamma$ is the Kerr nonlinearity coefficient.
The dimensionless Raman coefficient is  
$\epsilon_{R}=0.006/\tau_{0}$, where $\tau_{0}$ is in picoseconds. 
The dimensionless linear gain-loss coefficients $g_{j}$ are given by $g_{j}=2\tau_{0}^{2}\tilde g_{j} /|\tilde\beta_{2}|$, 
where $\tilde g_{j}$ are the dimensional linear gain-loss coefficients.
\bibitem{parameter_values} We point out that similar results are obtained 
with other choices of the physical parameter values. That is, our results 
are not very sensitive to the values of $J$, $T$, and $\Delta\beta$,    
as long as $T>10$ and $\Delta\beta > 10$.
\bibitem{Sliding_Filter} The shifting of the central 
amplification bandwidth of the linear gain-loss 
(\ref{raman13}) is somewhat similar to the shifting 
of the central frequency in sliding frequency filters, 
which were widely studied in the context of soliton-based transmission. 
See Ref. \cite{Iannone98} and references therein.
\bibitem{eta_and_beta} In the absence of delayed Raman response 
and linear gain-loss, we employ Eqs. (\ref{criterion1}) and (\ref{criterion2})  
with $\eta_{j}(z)=\eta_{j}(0)$ and $\beta_{j}(z)=\beta_{j}(0)$.
\end{thebibliography}
\end{document}